\documentclass{jfm}
\usepackage[utf8]{inputenc}
\usepackage{newtxtext}  
\usepackage{newtxmath}  
\usepackage{amsmath,amssymb}
\usepackage{bm}
\usepackage{graphicx}
\usepackage{xcolor}
\usepackage[justification=justified,labelfont=sc]{caption}
\usepackage[labelfont=normalfont]{subcaption}
\usepackage{natbib}
\usepackage{hyperref}  
\usepackage{cleveref}

\graphicspath{{./}}

\newcommand*{\four}[1]{\hat{#1}}
\newcommand*{\mean}[1]{\overline{#1}}
\newcommand*{\avgxz}[1]{\left \langle #1 \right \rangle_{xz}}
\newcommand*{\fluct}[1]{{#1}^{\prime}}
\newcommand*{\phaseShift}{\angle \hat{A}_d}
\newcommand*{\bkappa}{\bm{\kappa}}
\newcommand*{\blambda}{\bm{\lambda}}

\newcommand*{\reBulk}{\mathrm{Re}_b}
\newcommand*{\reTau}{\mathrm{Re}_{\tau}}
\newcommand*{\reGeneral}{\mathrm{Re}}
\newcommand*{\pd}[2]{\frac{\partial #1}{\partial #2}}
\newcommand*{\td}[2]{\frac{\mathrm{d} #1}{\mathrm{d} #2}}
\newcommand*{\ppdSeparate}[3]{\frac{\partial^2 #1}{\partial #2 \, \partial #3}}
\newcommand*{\absVal}[1]{\lvert #1 \rvert}
\newcommand{\extraCaptionWidth}{0.8\linewidth}
\newcommand*{\dimensional}[1]{{#1}^{\star}}

\title{Coupled dynamics of wall pressure and transpiration, with implications for the modeling of tailored surfaces and turbulent drag reduction}

\author{Simon Toedtli\aff{1}
\corresp{\email{stoedtli@ucar.edu}},
Anthony Leonard\aff{2}
\and Beverley McKeon\aff{3}}

\affiliation{\aff{1}National Center for Atmospheric Research, Boulder, Colorado 80301, USA
\aff{2}Graduate Aerospace Laboratories, California Institute of Technology, Pasadena, California 91125, USA
\aff{3}Stanford University, Department of Mechanical Engineering, Palo Alto, California 94305, USA}

\begin{document}

\maketitle

\abstract{
Wall-based active and passive flow control for drag reduction in low Reynolds number ($\reGeneral$) turbulent flows can lead to three typical phenomena: i) attenuation or ii) amplification of the near-wall cycle, and iii) generation of spanwise rollers.
The present study conducts direct numerical simulations (DNS) of a low $\reGeneral$ turbulent channel flow and demonstrates that each flow response can be generated with a wall transpiration at two sets of spatial scales, termed ``streak’’ and ``roller’’ scales.
The effect of the transpiration is controlled by its relative phase to the background flow, which can be parametrized by the wall pressure.
Streak scales i) attenuate the near-wall cycle if transpiration and wall-pressure are approximately in-phase or ii) amplify it otherwise, and iii) roller scales energize spanwise rollers when transpiration and wall pressure are out-of-phase.
The dynamics of the wall pressure and transpiration are coupled and robust relative phase relations, which are required to trigger the flow responses, can result if the source term of the linear fast or nonlinear slow pressure correlates with the wall transpiration over a scale-dependent height or if the temporal frequency content of the wall transpiration is approximately sparse.
The importance of each condition depends on the relative magnitude of the pressure components, which is significantly altered by the transpiration.
The analogy in flow response suggests that transpiration with the two scale families and their phase relations to the wall pressure represent fundamental building blocks for flows over tailored surfaces including riblets, porous, and permeable walls.
}

\section{Introduction}
Active and passive flow control techniques promise reduced skin friction drag and vast energetic and monetary savings in practical applications \citep[see, for example,][]{Kim2011Physicscontrolwall}.
However, as pointed out by \citet{BechertBruseHageEtAl1997Experimentsdragreducing}, the development of effective drag reduction techniques is challenging and current engineering solutions fall short of delivering even a fraction of the projected savings.
A key challenge is predicting the flow response to control in advance, which remains elusive and limits the effectiveness of controllers.
Current control techniques are typically informed by the physics of canonical flows and target coherent flow structures as a proxy for suppressing turbulent fluctuations.
However, there is no guarantee that this approach will be successful and there is growing evidence that these controllers can modify the dynamics of a low Reynolds number flow in at least three ways:
i) attenuation or ii) amplification of the near-wall cycle, and iii) generation of spanwise rollers \citep[see, for example,][]{ChoiMoinKim1994Activeturbulencecontrol,ToedtliYuMcKeon2020origindragincrease,Garcia-MayoralJimenez2011Hydrodynamicstabilitybreakdown,ChavarinLuhar2020ResolventAnalysisTurbulent,JimenezUhlmannPinelliEtAl2001Turbulentshearflow,BreugemBoersmaUittenbogaard2006influencewallpermeability,Gomez-de-SeguraGarcia-Mayoral2019Turbulentdragreduction,KimChoi2014Spacetimecharacteristicscompliant}.
Of the three possible modifications, only the first one leads to drag reduction, which underscores the difficulty of designing effective controllers.
The present work examines the three aforementioned flow modifications in a low Reynolds number turbulent channel flow with active wall transpiration.
Using data from direct numerical simulations, we study how specific phase relations between wall pressure and transpiration at distinct spatial scales correlate with each flow response.

\subsection{Active flow control: opposition control}
Active feedback flow control, specifically techniques based on the opposition control scheme of \citet{ChoiMoinKim1994Activeturbulencecontrol}, are one way to generate the flow phenomena described in the previous section.
Opposition control measures the wall-normal velocity $\dimensional{v}$ (the superscript $\star$ labels a dimensional quantity) at a distance $\dimensional{y}_d$ from the wall (located at $\dimensional{y}_w$) and generates a wall transpiration that is proportional to the sensor measurement
\begin{equation}  \label{eq:classicalOc}
    \dimensional{v}(\dimensional{y}_w) = - A_d \, \dimensional{v}(\dimensional{y}_d)
\end{equation}
\Cref{eq:classicalOc} will be referred to as ``classical opposition control'' hereafter and has two parameters, the real-valued controller gain $A_d$ and the sensor location $\dimensional{y}_d$.
There is broad consensus that optimal drag reduction is achieved when the gain is set to $A_d = 1$ and the sensors are placed at $y_d^+ \approx 15$, which roughly corresponds to the center of the quasi-streamwise vortices \citep{ChoiMoin1994EffectsComputationalTime,HammondBewleyMoin1998Observedmechanismsturbulence,ChungTalha2011Effectivenessactiveflow,LuharSharmaMcKeon2014Oppositioncontrolresolvent}.
A superscript $+$ as in the above expression indicates normalization by the friction velocity $\dimensional{u}_{\tau} = (\dimensional{\tau}_w / \dimensional{\rho})^{(1/2)}$ and viscous length $\dimensional{l}_v = \dimensional{\nu} / \dimensional{u}_{\tau}$, where $\dimensional{\tau}_w$ is the wall shear stress, $\dimensional{\rho}$ the fluid density and $\dimensional{\nu}$ the kinematic viscosity.
The control scheme can deliver up to 25\% drag reduction in low Reynolds number flows, where the near-wall cycle is the dominant coherent flow motion \citep{ChoiMoinKim1994Activeturbulencecontrol}, but its effectiveness decreases with Reynolds number as large-scale motions become relatively more energetic \citep{HutchinsMarusic2007Evidenceverylong,DengHuangXu2016Origineffectivenessdegradation}.

Numerous generalizations of classical opposition control have since been proposed in the literature.
The relevant generalization for this study is obtained when \cref{eq:classicalOc} is transformed to Fourier domain \citep{LuharSharmaMcKeon2014Oppositioncontrolresolvent,ToedtliLuharMcKeon2019Predictingresponseturbulent}.
The control law then establishes a relation between the Fourier coefficient of the sensor measurement and the transpiration
\begin{equation}  \label{eq:varyingPhaseOcDim}
    \dimensional{\four{v}}(y_w) = - \four{A}_d \, \dimensional{\four{v}}(y_d)
\end{equation}
where the superscript hat labels a complex-valued quantity.
\Cref{eq:varyingPhaseOcDim} will be referred to as ``varying-phase opposition control,'' to make the distinction to \cref{eq:classicalOc} clear.
The formulation in the Fourier domain represents a generalization because the controller gain becomes complex-valued, which introduces the phase $\phaseShift$ as an additional parameter.
Varying-phase opposition control with different $\phaseShift$ can increase the maximum attainable drag reduction or lead to a significant drag increase \citep{ToedtliLuharMcKeon2019Predictingresponseturbulent}.
For positive values of $\phaseShift$, the drag increase coincides with the appearance of spanwise rollers \citep{ToedtliYuMcKeon2020origindragincrease}.
The flow responses therefore likely encompass the three phenomena described initially, but the physical link between their occurrence and $\phaseShift$ remains unclear.

Before moving on to passive control methods, we give a short overview of other relevant opposition control generalizations.
The effectiveness of classical opposition control at low Reynolds number can be improved in multiple ways, for example by using upstream sensor information \citep{Lee2015Oppositioncontrolturbulent} or by adding an integral term to the control law \citep{KimChoi2017Linearproportionalintegralcontrol}.
These approaches are related to \cref{eq:varyingPhaseOcDim}, because spatial shifts in physical domain and integration in time-domain correspond to scale and frequency-dependent phase shifts in Fourier domain.
More recently, opposition control has also been extended to the logarithmic layer, whose coherent structures would be more practical targets at higher Reynolds numbers.
However, the early attempts have not lead to a statistically significant drag reduction and indicate that extending opposition control to other flow regions is non-trivial \citep{AbbassiBaarsHutchinsEtAl2017Skinfrictiondrag,GusevaJimenez2022Linearinstabilityresonance}.
On the other hand, extensions to different flow regimes indicate that opposition control may be fairly robust with regards to additional physical effects.
For example, opposition control can successfully reduce the drag in boundary layers with adverse pressure gradients \citep{WangAtzoriVinuesa2024Oppositioncontrolapplied}, or in compressible turbulent channel flows in the subsonic and supersonic regime \citep{YaoHussain2021Dragreductionvia}.

\subsection{Passive flow control: tailored surfaces}  \label{sec:revTailoredSurf}
A number of passive flow control techniques, which we subsume under the term ``tailored surfaces'' and which include riblets, porous and permeable surfaces and compliant walls, can also induce the three flow responses described initially.
For example, riblets operating in the so-called viscous regime reduce the drag, with drag decreasing as the peak-to-peak spacing increases up to an optimal geometry-dependent spacing \citep[see e.g.][]{BechertBruseHageEtAl1997Experimentsdragreducing,ChoiMoinKim1993Directnumericalsimulation}.
This drag reduction is at least in part due to a suppression of the near-wall cycle \citep{ChavarinLuhar2020ResolventAnalysisTurbulent}.
The drag reduction becomes less effective for riblet spacings past the optimum and eventually turns into a drag increase.
The breakdown of the viscous regime and following drag increase are mainly due to the generation of spanwise rollers, which carry substantial Reynolds stresses \citep{Garcia-MayoralJimenez2011Hydrodynamicstabilitybreakdown}.
The near-wall cycle becomes amplified past the viscous breakdown as well and contributes to the drag increase, but to a lesser extent than the spanwise rollers \citep{ChavarinLuhar2020ResolventAnalysisTurbulent}.
Permeable walls behave similarly to riblets and exhibit a linear regime for small permeabilities, where the drag reduction is approximately proportional to the difference between the streamwise and spanwise permeability.
The linear regime breaks down for sufficiently large wall-normal permeabilities and this breakdown again coincides with the appearance of spanwise rollers that eventually lead to a drag increase \citep{BreugemBoersmaUittenbogaard2006influencewallpermeability,Gomez-de-SeguraGarcia-Mayoral2019Turbulentdragreduction}.
Spanwise rollers have further been reported for flow over
porous \citep{JimenezUhlmannPinelliEtAl2001Turbulentshearflow} and soft compliant walls \citep{KimChoi2014Spacetimecharacteristicscompliant}.
The compliant wall is somewhat different from the other tailored surfaces in that the drag increase and formation of spanwise structures are attributed to a resonance of the compliant wall under forcing of the flow and therefore also depend on the material properties of the compliant wall.

Despite the similarities in flow phenomena, it remains unclear how different tailored surfaces can be related and a unifying framework to analyze them is missing in the literature.
In addition, the link between the properties of a specific surface and the flow response is unknown in many cases.
A dynamic interpretation of tailored surfaces, inspired by the representation of permeable \citep{Gomez-de-SeguraGarcia-Mayoral2019Turbulentdragreduction} and rough walls \citep{KhorasaniLacisPascheEtAl2021wallturbulencealteration} as velocity boundary conditions, may hold clues for how to unify their analysis.
Common to all tailored surfaces is a relaxed no-throughflow condition, which induces a wall-normal velocity in proximity of the bounding surface.
Some configurations, like porous walls, explicitly replace the no-throughflow condition by a non-zero wall-normal velocity \citep[see e.g.][]{JimenezUhlmannPinelliEtAl2001Turbulentshearflow}.
Others, such as riblets, retain an impermeable wall but introduce a complex geometry.
In this case, the notion of relaxed no-throughflow condition does not apply at the wall itself, but perhaps to a suitably chosen plain inside the flow domain that permits a non-zero wall-normal velocity in and out of the riblet groves.
In this interpretation, wall transpiration may provide the essential building blocks to represent the flow response to tailored surfaces.
It remains unclear, however, if and under what conditions this dynamical representation holds.

\subsection{Velocity-pressure phase relations}
The phase relation or, equivalently, relative spatial arrangement between wall pressure and transpiration will be a main focus of this study.
The possible importance of this phase relation was first appreciated by \citet{XuRempferLumley2003Turbulenceovercompliant}, who studied pressure-driven compliant walls and observed no statistically significant reduction in skin friction.
The authors conjectured that the lack of drag reduction might be due to an unfavorable phase relation between wall-normal velocity and pressure.
For example, a sweep event likely correlates with a high wall pressure, which causes the compliant surface to sink in and induce a negative wall-normal velocity in its proximity.
Wall-normal velocity and pressure at the surface therefore have opposite signs and are out-of-phase.
Effective active flow control strategies, such as classical opposition control, likely lead to a different phase relation:
the controller would counter the sweep event with a positive wall transpiration, which leads to a wall-normal velocity and pressure that have the same sign and are in-phase.
To support their hypothesis, the same authors presented a preliminary numerical simulation where the compliant wall was unphysically driven by the negative of the pressure.
This configuration is more likely to induce an in-phase relation between wall-normal velocity and pressure and indeed results in drag reduction.

Simplified analyses based on a rank-1 resolvent model provide additional evidence that compliant surfaces with negative damping coefficients, which lead to an in-phase relation between wall-normal velocity and pressure, are required to suppress the near-wall cycle \citep{LuharSharmaMcKeon2015frameworkstudyingeffect}.
Extensions of the resolvent framework to perforated surfaces and generalized impedance boundary conditions also suggest that an in-phase relation between the two quantities is required to suppress energetic coherent structures \citep{JafariMcKeonArjomandi2023Frequencytunedsurfaces,JafariMcKeonCazzolatoEtAl2024resolventanalysiseffect}.
However, it is important to keep in mind that the resolvent operator, which represents the linear system dynamics, mainly captures the fast pressure component \citep{LuharSharmaMcKeon2014structureoriginpressure}.
The slow and Stokes pressure are missing in these simplified flow models and it remains unclear if and how they may change the phase relation.
In addition, it is unknown if the three flow phenomena described initially can be parameterized in terms of velocity-pressure phase relations.

\subsection{Outline and contributions}
This study aims to address some of the gaps identified in the literature.
We perform DNS of a low Reynolds number turbulent channel flow with varying-phase opposition control and analyze datasets for various $\phaseShift$.
\Cref{sec:methods} introduces the problem formulation, outlines the details of the wall transpiration, and describes the numerical methods used in this study.
\Cref{sec:dr} reviews previous varying-phase opposition control results, which indicate that the structure of the wall transpiration changes with $\phaseShift$ and motivate the definition of two scale families termed ``streak'' and ``roller'' scales.
The structure of the wall pressure for various controlled flows is presented in \cref{sec:wallPressure}, along with an analysis of the relative importance of the pressure components.
\Cref{sec:pvPhaseWall} considers the phase difference between the wall pressure and transpiration.
Transpiration at the streak scales will be shown to be in-phase with the wall pressure when the drag is reduced, while transpiration at the roller scales is out-of-phase when spanwise rollers are generated and the drag increases.
\Cref{sec:comparisonPhaseDrag} will further show that the specific phase relations coincide with scale suppression or amplification, which drive the observed drag change.
\Cref{sec:summary} summarizes the results and relates them to the tailored surfaces.
Transpiration with streak scales will be shown to be dynamically equivalent to tailored surfaces that interact with the near-wall cycle, while transpiration with roller scales and positive phase shifts corresponds to the generation of spanwise rollers and drag increase.
Implications for higher Reynolds number flows are also discussed at the end of \cref{sec:summary}.

\section{Methodology}  \label{sec:methods}
This section describes the flow configuration and introduces the numerical methods.
\Cref{sec:governingEquations} summarizes the governing equations and \cref{sec:defFourAndPhase} defines the Fourier decomposition and notion of phase, which will be at the center of this study.
The details of the wall transpiration are introduced in \cref{sec:defTranspiration} and the DNS and pressure Poisson solver are summarized in \cref{sec:dnsCode,sec:pressurePoisson}, respectively.

\subsection{Flow configuration}  \label{sec:governingEquations}
We analyze an incompressible turbulent channel flow with periodic boundary conditions in the streamwise ($\dimensional{x}$) and spanwise direction ($\dimensional{z}$) and walls located at $\dimensional{y}_w = \pm \dimensional{h}$.
The walls impose a no-slip condition on the tangential velocity components ($\dimensional{u}(\dimensional{y}_w) = \dimensional{w}(\dimensional{y}_w) = 0$), as in the canonical configuration.
In contrast, the walls permit a transpiration in the wall-normal direction ($\dimensional{v}(\dimensional{y}_w) \neq 0$) with zero net mass flux.
This flow configuration can be mathematically described by the incompressible Navier-Stokes equations, which in nondimensional index notation are
\begin{equation}  \label{eq:nse}
\begin{aligned}
    \pd{u_j}{x_j} &= 0 \\
    \pd{u_i}{t} + u_j \pd{u_i}{x_j} &= -G_x \delta_{i1} - \pd{p}{x_i} + \frac{1}{\reBulk} \ppdSeparate{u_i}{x_j}{x_j}
\end{aligned}
\end{equation}
In the above expression, $t$ stands for time, $u_i = \{u, v, w\}$ are the velocity components in the directions $x_i = \{x, y, z\}$, and $p$ denotes pressure fluctuations about the time-dependent mean pressure gradient $G_x(t)$.
All quantities are nondimensionalized with respect to the channel half-height $\dimensional{h}$ and twice the bulk velocity $\dimensional{U}_b$.
This choice of reference scales introduces the bulk Reynolds number $\reBulk = 2 \dimensional{U}_b \dimensional{h} / \dimensional{\nu}$ as the governing problem parameter in \cref{eq:nse}.
The channel is driven by a constant mass flux, so that the bulk Reynolds number is held constant at $\reBulk = 5600$.
In contrast, the friction Reynolds number $\reTau = \dimensional{u}_{\tau} \dimensional{h} / \dimensional{\nu}$ changes as a function of the wall transpiration.
For a canonical channel without transpiration, the current $\mathrm{Re}_b$ corresponds to $(\mathrm{Re}_\tau)_0 = 180$.
The subscript $0$ will be used throughout to label a canonical channel configuration, which serves as reference for comparison.
In contrast, flows with wall transpiration will be denoted by a subscript $c$.

\subsection{Fourier decomposition and phase}  \label{sec:defFourAndPhase}
Given the periodicity in $x$ and $z$, a flow quantity $q$ can be expressed as a superposition of Fourier modes
\begin{equation}  \label{eq:fourierDecomposition}
        q(x,y,z,t) = \sum_l \sum_m \four{q}(l, m, y, t) \;
        e^{i \left(l \frac{2\pi}{L_x}x + m \frac{2\pi}{L_z}z \right)}
\end{equation}
Here, $\four{q}$ are the complex-valued Fourier coefficients, which depend on the wall-normal coordinate and time.
$L_x$ and $L_z$ denote the streamwise and spanwise periodicity of the channel domain, respectively, and $l, m$ are integer indices.
Each Fourier mode is characterized by its streamwise (${k_x = l \, 2 \pi / L_x}$) and spanwise wavenumber ($k_z = m \, 2\pi / L_z$), which together form the wavenumber vector $\bkappa = (k_x, k_z)^{\intercal}$ with magnitude $\kappa = \lVert \bkappa \rVert$.
Individual Fourier modes will be referred to as $\four{q}(l, m, y, t)$ or $\four{q}_{\bkappa}(y, t)$.

Most of the analysis in subsequent sections will focus on the complex-valued Fourier coefficients $\four{q}_{\bkappa}(y,t)$, which have an amplitude $\absVal{\four{q}_{\bkappa}(y,t)}$ and phase $\angle \four{q}_{\bkappa}(y,t)$.
In particular, we will represent the wall transpiration as a superposition of Fourier modes with coefficients $\four{v}_{\bkappa}(y_w, t)$.
Our aim is to understand how changes of the transpiration phase $\angle \four{v}_{\bkappa}(y_w, t)$ at specific length scales $\bkappa$ alter the wall pressure, the phase relation between $\four{p}_{\bkappa}$ and $\four{v}_{\bkappa}$, and the flow response.

\subsection{Wall transpiration}  \label{sec:defTranspiration}
The wall transpiration is set by the varying-phase opposition control scheme (\ref{eq:varyingPhaseOcDim}).
There is some ambiguity in the control law as to which shift-invariant coordinates are transformed to Fourier domain.
For the present study, we Fourier transform the streamwise and spanwise direction but retain the time dependency, as outlined in \cref{sec:defFourAndPhase}.
This is different from previous resolvent analyses, which also transformed the temporal coordinate \citep[see, e.g.][]{LuharSharmaMcKeon2014Oppositioncontrolresolvent}.
The appropriate form of the varying-phase opposition control law in nondimensional form then is then given by
\begin{equation}  \label{eq:varyingPhaseOc}
    \four{v}_{\bkappa}(y_w, t) = - \four{A}_d \, \four{v}_{\bkappa}(y_d, t)
\end{equation}
The complex-valued controller gain $\four{A}_d$ is constrained by Hermitian symmetry, but can otherwise be an arbitrary function of $\bkappa$.
A wavenumber-dependent gain enables optimization of the transpiration for each Fourier mode \citep[see e.g.][]{LuharSharmaMcKeon2014Oppositioncontrolresolvent}, but obfuscates the physical interpretation of $\angle \four{A}_d$.
This is because a multiplication of two wavenumber-dependent quantities implies a convolution in physical domain, which is challenging to interpret.
For the present study we limit the controller gain to the choice
\begin{equation}  \label{eq:controllerGainAllScales}
    \four{A}_d =
    \begin{cases}
        0, &  \text{if } k_x = k_z = 0 \\
        1, & \text{if } k_x=0, k_z \neq 0 \\
        e^{i \phaseShift}, & \text{if } k_x > 0, k_z=0 \\
        \frac{\min(\lvert \four{v}_{\bkappa}(t, y_d) \rvert, \lvert \four{v}_{\tilde{\bkappa}}(t, y_d) \rvert )}{\lvert \four{v}_{\bkappa}(t, y_d)\rvert } e^{i \phaseShift}, & \text{if } k_x > 0, k_z \neq 0
    \end{cases}
\end{equation}
where $\tilde{\bkappa} = (k_x, -k_z)^{\intercal}$ and controller gains for negative $k_{x}$ are omitted since they are determined by Hermitian symmetry.
This specific choice of $\four{A}_d$ enables a clear physical interpretation of $\phaseShift$, as will be shown next.

In Fourier domain, $\phaseShift$ alters the phase of the wall transpiration relative to the sensor signal
\begin{equation}  \label{eq:phsaeDifference}
    \angle \four{v}_{\bkappa}(y_w, t) = \angle \four{A}_{d} + \angle \four{v}_{\bkappa}(y_d, t) + \pi
\end{equation}
and will therefore be referred to as ``phase shift.''
The definition of $\four{A}_d$ according to (\ref{eq:controllerGainAllScales}) enables a physical interpretation of $\phaseShift$ as scale-dependent streamwise shifts.
The correspondence between phase shift and streamwise shift is clear for $k_{z} = 0$, but perhaps less evident for modes with streamwise and spanwise dependence (last line in \cref{eq:controllerGainAllScales}).
To illustrate the correspondence for those modes, consider the sensor signal at a specific $\bkappa$, which can be written as $\four{v}_{\bkappa}(y_{d},t) = A_{1} e^{i \phi_{1}}$, and the sensor signal at $\tilde{\bkappa}$, which is denoted by $\four{v}_{\tilde{\bkappa}}(y_{d},t) = A_{2} e^{i \phi_{2}}$.
The amplitudes ($A_1$ and $A_2$) and phases ($\phi_1$ and $\phi_2$) may be different instantaneously, and for the following discussion we assume $A_2 > A_1$ without loss of generality.
Applying the control law (\ref{eq:varyingPhaseOc}) with $\four{A}_d$ according to \cref{eq:controllerGainAllScales}, and evaluating \cref{eq:fourierDecomposition} at $\bkappa$, $\tilde{\bkappa}$, and their complex conjugates gives the physical flow structure of the wall transpiration at length scale $\blambda = (\lambda_x, \lambda_z)^{\intercal} = (2 \pi / k_x, 2 \pi / k_z)^{\intercal}$

\begin{equation} \label{eq:phaseShiftInterpretation}
    v_{\blambda}(x,y_{w},z,t) =
    - 4 A_{1} \cos \left( k_{x} \left(x + \frac{\angle \four{A}_{d}} {k_{x}} \right)
    + \frac{\phi_{1} + \phi_{2}}{2} \right) \,
    \cos \left( k_{z} z + \frac{\phi_{1} - \phi_{2}}{2} \right)
\end{equation}
The factor multiplying the phase shift in the last line of \cref{eq:controllerGainAllScales} is central to the derivation of \cref{eq:phaseShiftInterpretation} in two regards: first, it selects the smaller of $A_1$ and $A_2$ to generate the wall transpiration, and second, it enforces the same transpiration amplitude at $\bkappa$ and $\tilde{\bkappa}$.
The equal amplitude suppresses oblique structures, for which the phase shift has no clean physical interpretation, and thus establishes the correspondence between phase shift and streamwise shift.
Interested readers may refer to \citet{ToedtliLuharMcKeon2019Predictingresponseturbulent,Toedtli2021ControlWallBounded} for further details on the choice of the controller gain.
From here on, it is implied that $\four{A}_d$ of varying-phase opposition control is defined according to \cref{eq:controllerGainAllScales}.

\begin{figure}
    \centering
    \begin{subfigure}{0.45\linewidth}
        \caption{$\phaseShift = 0$}
        \includegraphics[width=\linewidth]{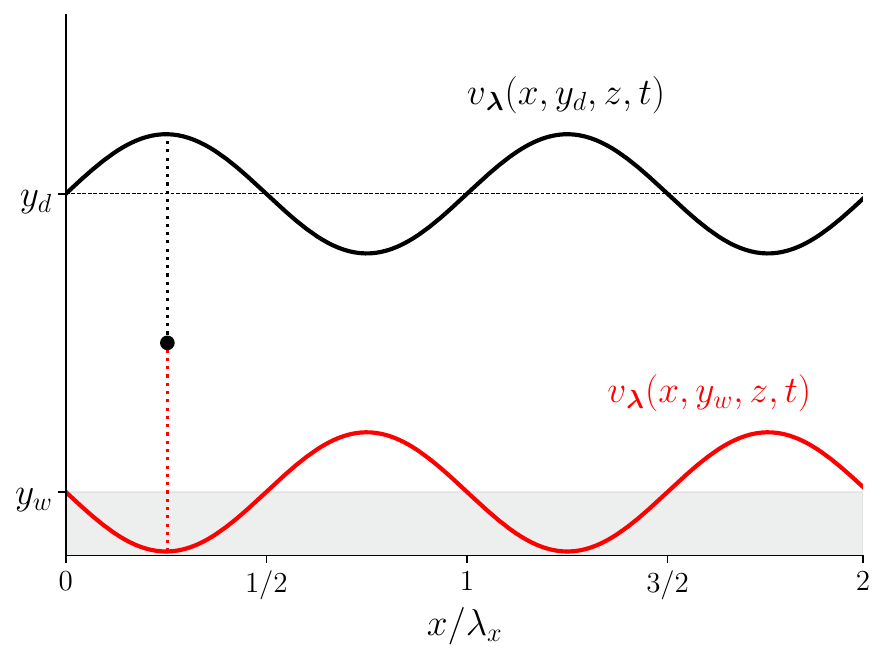}
        \label{fig:exampleOcNoShift}
    \end{subfigure}
    \hfill
    \begin{subfigure}{0.45\linewidth}
        \caption{$\phaseShift = -\pi / 2$}
        \includegraphics[width=\linewidth]{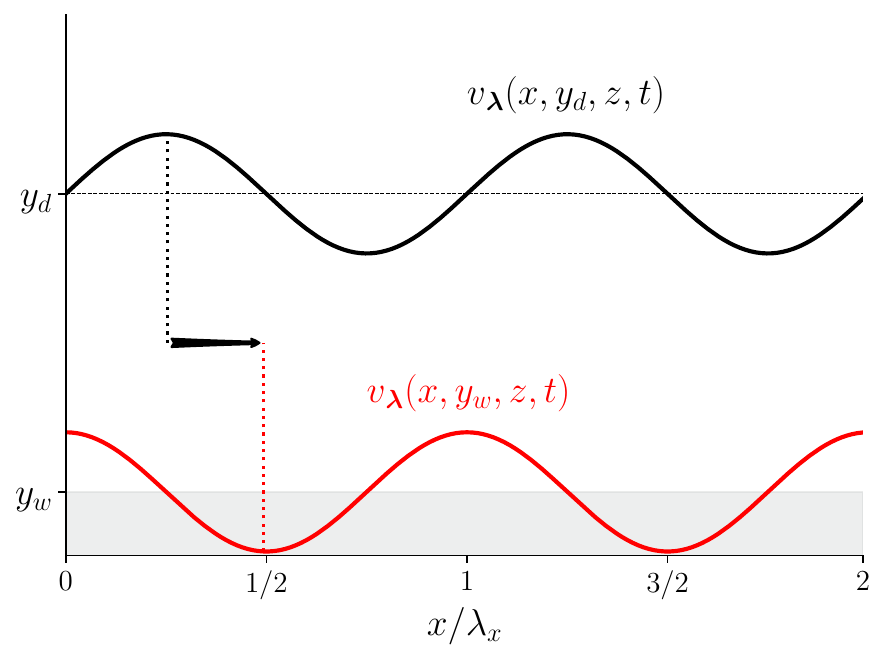}
        \label{fig:exampleOcNegShift}
    \end{subfigure}
    \caption{Interpretation of the phase shift as a streamwise shift in physical domain. The black curve represents an example sensor measurement, while the red curve shows the actuator response for two example phase shifts.}
    \label{fig:interpretationPhaseShift}
\end{figure}

\Cref{eq:phaseShiftInterpretation} demonstrates that each Fourier coefficient generates a building block of the wall transpiration, whose streamwise location can be altered through $\phaseShift$.{}
\Cref{fig:interpretationPhaseShift} illustrates the relative position of sensor measurement (black curve) and actuator response (red curve) for two different phase shifts.
Classical opposition control and varying-phase opposition control with $\angle \four{A}_{d} = 0$ generate an actuator response that is exactly out-of-phase with the sensor measurement, as shown in \cref{fig:exampleOcNoShift}.
In contrast, the sensor signal and actuator response not only differ by a sign, but are offset in the streamwise direction if $\phaseShift \neq 0$.
Negative phase shifts correspond to lead of the actuation relative to the sensor measurement, while positive phase shifts imply a lag of the actuation.
\Cref{fig:exampleOcNegShift} illustrates the relative arrangement for $\phaseShift = -\pi / 2$, which corresponds to a quarter wavelength lead of the actuation over the sensor measurement.
It is important to emphasize again that the interpretation of $\phaseShift$ according to \cref{eq:phaseShiftInterpretation} and \cref{fig:interpretationPhaseShift} is only valid if $\four{A}_d$ is defined according to \cref{eq:controllerGainAllScales}.

Under this restriction, the physical meaning of $\phaseShift$ suggests an alternative interpretation of varying-phase opposition control, which we adopt for this study:
the control scheme can be interpreted as a way to prescribe a wall-transpiration based on templates that occur naturally in the flow, namely $\four{v}_{\bkappa}(y_{d})$, and to change the relative streamwise phase between transpiration and background flow by altering $\phaseShift$.
In this spirit, we will use the varying-phase opposition control scheme to generate a wall transpiration and study how the wall pressure, the phase relation between $\four{p}_{\bkappa}$ and $\four{v}_{\bkappa}$ and the flow response change as a function of $\phaseShift$.

As mentioned in the introduction, the varying-phase opposition control scheme is related to the work of \citet{Lee2015Oppositioncontrolturbulent} and \citet{GusevaJimenez2022Linearinstabilityresonance}, who studied classical opposition control with streamwise shifts between sensor measurement and actuator response.
Despite conceptual similarities, these controllers are different from the current one in two important ways:
first, they apply a constant phase shift in physical space, which corresponds to a scale-dependent phase shift in Fourier domain.
And second, unlike the present controller, they do not suppress oblique waves in the actuation input.

\subsection{Direct numerical simulation}  \label{sec:dnsCode}
The flow response to varying-phase opposition control is studied by means of DNS.
The DNS solves the incompressible Navier-Stokes equations (\ref{eq:nse}) in velocity-vorticity form following the method of \citet{KimMoinMoser1987Turbulencestatisticsfully}.
The streamwise and spanwise coordinates are discretized by a Fourier pseudospectral method, while the wall-normal direction uses a compact finite difference scheme on a stretched sinusoidal mesh \citep[see][for details]{FloresJimenez2006Effectwallboundary}.
The stretching of the wall-normal grid is non-uniform in $y$ and controlled by a single parameter, analogous to the approach described in \citet{LeeMoser2015Directnumericalsimulation}.
A Runge-Kutta scheme with implicit viscous terms integrates the state variables in time.
The controller is implemented by replacing the no-through condition by the varying-phase opposition control scheme (\ref{eq:varyingPhaseOc}) at each time step.

\begin{figure}
    \centering
    \begin{subfigure}[b]{0.45\linewidth}
        \caption{DNS solver}
        \includegraphics[width=\linewidth]{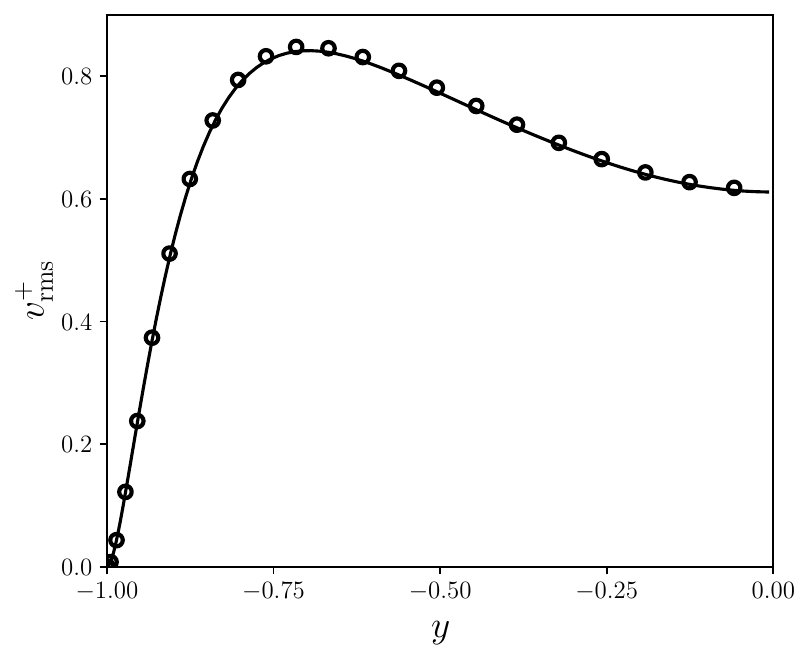}
        \label{fig:validationV}
    \end{subfigure}
    \hspace{0.1in}
    \begin{subfigure}[b]{0.45\linewidth}
        \caption{Pressure Poisson solver}
        \includegraphics[width=\linewidth]{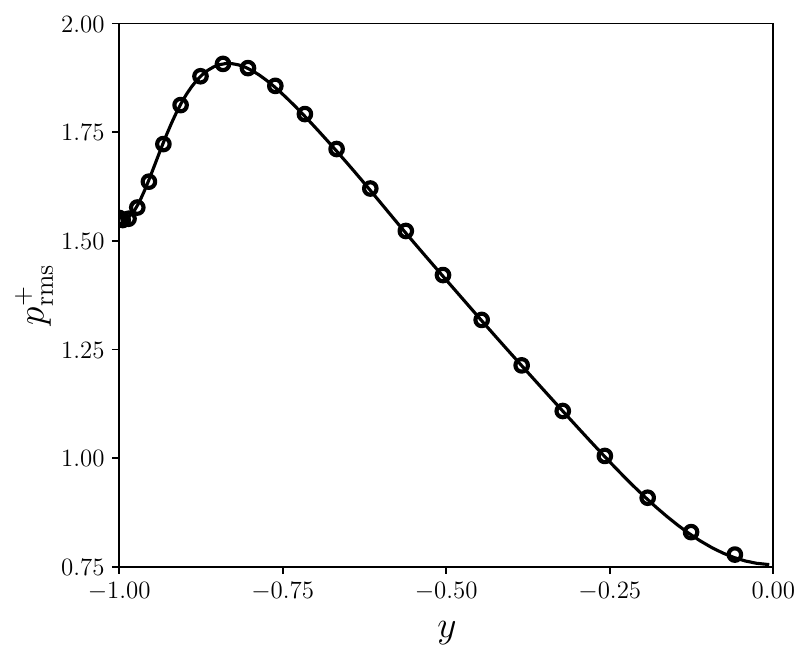}
        \label{fig:validationP}
    \end{subfigure}
    \caption{Validation of the present DNS and pressure Poisson equation solver (solid lines) against the data of \cite{LeeMoser2015Directnumericalsimulation} (open circles) for a canonical channel flow at $(\reTau)_0 = 180$.}
    \label{fig:validation}
\end{figure}

The channel is driven by a constant mass flow rate, which implies that $\mathrm{Re}_b = 5600$ is fixed.
The size of the computational domain in the streamwise and spanwise direction is $L_x = 4 \pi$ and $L_z = 2 \pi$, respectively, and $N_x = N_z = 256$ Fourier modes are used in these directions.
This corresponds to a resolution of $\Delta x^+ \approx 8.8$ and $\Delta z^+ \approx 4.4$ in terms of Fourier modes before dealiasing at the nominal $(\mathrm{Re}_\tau)_{0} = 180$ of the canonical flow.
$N_y = 172$ grid points are used in the wall-normal direction, which gives a resolution of $\Delta y^+_{\mathrm{min}} \approx 0.37$ at the wall and $\Delta y^+_{\mathrm{max}} \approx 3.09$ at the channel center.
All control runs are started from the same initial condition, and statistics are collected over at least $10$ eddy turnover times ($\dimensional{h} / \dimensional{u}_\tau)$ once a statistically steady state is reached.
\Cref{fig:validationV} shows a validation of the DNS solver against data of \citet{LeeMoser2015Directnumericalsimulation} for a canonical turbulent channel flow at $(\mathrm{Re}_{\tau})_0 = 180$.
The wall-normal velocity statistics are averaged over symmetric planes in the bottom and top half of the channel, and the results are shown for the lower channel half ($-1 \leq y \leq 0$) only.
The two curves collapse at all wall-normal locations and demonstrate that the solver settings are adequate.

\subsection{Pressure Poisson equation}  \label{sec:pressurePoisson}
The pressure field in a channel flow is a superposition of the mean pressure gradient ($G_{x}(t)$) and turbulent pressure fluctuations ($p(x,y,z,t)$).
Only the pressure fluctuations are relevant for the subsequent analysis and the term ``pressure'' will always refer to $p$ only.
The pressure fluctuations cancel out when the Navier-Stokes equations are written in the velocity-vorticity form that underlies the DNS.
We therefore recover the pressure fluctuations in a post-processing step by solving a Poisson equation of the form
\begin{equation}  \label{eq:pressurePoissonGeneral}
    \ppdSeparate{p}{x_i}{x_i} = - \pd{u_j}{x_k} \pd{u_k}{x_j}
\end{equation}
The pressure fluctuations are assumed to be periodic in $x$ and $z$ and the Neumann problem for the pressure field is solved based on the arguments presented by \citet{GreshoSani1987pressureboundaryconditions}.
The boundary conditions in $y$ are obtained from evaluating the wall-normal momentum equation at $y_{w}$, which results in
\begin{equation}  \label{eq:bcPressure}
    \pd{p}{y}(y_w)
    = \frac{1}{\reBulk} \ppdSeparate{v}{x_j}{x_j}(y_w) - \pd{v}{t}(y_w) \equiv \begin{cases}
        P_b & \text{if } y_w = -1 \\
        P_t & \text{if } y_w = +1
    \end{cases}
\end{equation}
for a no-slip wall with transpiration.
In the above expression, $P_b$ and $P_t$ are shorthand for the boundary conditions at the bottom and top wall, respectively.
Further note that the transpiration adds inertial and viscous terms which are not present in canonical flows.

The Poisson equation (\ref{eq:pressurePoissonGeneral}) is linear in $p$ and the superposition principle applies.
The pressure can be split into three components by considering the homogeneous solution in isolation and separating the inhomogeneous solution into a part with linear and nonlinear source term (the notion of linearity is in terms of the velocity fluctuations $\fluct{u}_i$ about a spatio-temporal mean $\mean{u}$).
The contributions are usually referred to as fast ($p_{f}$, linear source term), slow ($p_{s}$, nonlinear source term), and Stokes pressure ($p_{St}$, homogeneous term), respectively, and are defined as \citep[see e.g.][]{Kim1989structurepressurefluctuations}
\begin{equation}  \label{eq:pressurePoissonSplit}
\begin{aligned}
    \ppdSeparate{p_f}{x_i}{x_i} &= - 2 \td{\mean{u}}{y} \pd{\fluct{v}}{x},  && \pd{p_f}{y}(y_w) = 0 \\
    \ppdSeparate{p_s}{x_i}{x_i} &= - \pd{\fluct{u}_j}{x_k} \pd{\fluct{u}_k}{x_j}, &&  \pd{p_s}{y}(y_w) = 0 \\
    \ppdSeparate{p_{St}}{x_i}{x_i} &= 0, && \pd{p_{St}}{y} (y_w) = \frac{1}{\reBulk} \ppdSeparate{v}{x_j}{x_j}(y_w) - \pd{v}{t}(y_w)
\end{aligned}
\end{equation}
Note that $\mean{v} = \mean{w} = 0$, so that $v^{\prime} = v$ and $w^{\prime} = w$ above.
Furthermore, $p = p_{f} + p_{s} + p_{St}$, and we will sometimes refer to $p$ as the ``total pressure'' in order to distinguish it from its components.
The decomposition of \cref{eq:pressurePoissonSplit} is by no means the only possible one, but it will turn out to be useful in the subsequent analysis.

The pressure Poisson equation is solved using Fourier transforms in $x$ and $z$, which reduces each equation in (\ref{eq:pressurePoissonSplit}) to ordinary differential equations in $y$ with known analytical solutions.
The Stokes pressure (homogeneous solution) at $\kappa \neq 0$ is given by
\begin{equation}  \label{eq:solStokesPressure}
    \hat{p}_{\bkappa, St}(y) = \frac{1}{\kappa \sinh(2 \kappa)} \left(\four{P}_t \cosh \left[ \kappa(y+1) \right] - \four{P}_b \cosh \left[ \kappa(y-1) \right] \right)
\end{equation}
where $\four{P}_{b}$ and $\four{P}_{t}$ are the Fourier transforms of \cref{eq:bcPressure}.
The solution for the inhomogeneous fast and slow pressure equations at $\kappa \neq 0$ can be written in terms of a Green's function
\begin{equation}  \label{eq:solFastSlowPressure}
    \four{p}_{\bkappa, \{f/s \}}(y) = \int_{-1}^{1} G(y, a) \four{f}(a) \, \mathrm{d}a
\end{equation}
where $\four{f}$ is the Fourier transform of the respective inhomogeneous source term in \cref{eq:pressurePoissonSplit} and $G(y,a)$ is the Green's function kernel \citep[see e.g.][for futher details]{Kim1989structurepressurefluctuations}
\begin{equation}  \label{eq:GreensFuncPressure}
    G(y, a) = \begin{cases}
        - \frac{\cosh \left[ \kappa (a - 1) \right] \cosh \left[ \kappa (y + 1) \right]}{\kappa \sinh ( 2 \kappa)} & \text{for } y < a\\
        - \frac{\cosh \left[ \kappa (a + 1) \right] \cosh \left[ \kappa (y - 1) \right]}{\kappa \sinh ( 2 \kappa)} & \text{for } y > a
    \end{cases}
\end{equation}
A few aspects of the limiting behavior are worthwhile noting:
The difference $\four{P}_t - \four{P}_b$, and therefore also $\partial_t (\four{v}_{\bkappa}(y=-1) - \four{v}_{\bkappa}(y=1))$, must scale like $\mathcal{O}(\kappa^2)$ or smaller for the Stokes pressure to be bounded as $\kappa \rightarrow 0$.
The Stokes and fast pressure vanish in the limit $\kappa = 0$, and the wall-parallel mean of the slow pressure can be obtained by integrating \cref{eq:pressurePoissonSplit} twice in the wall-normal direction
\begin{equation}
    \four{p}_{\bm{0},s} (y) = P_c - \avgxz{v^{\prime} v^{\prime}}(y)
\end{equation}
where $P_c$ is an undefined constant and $\avgxz{\cdot}$ denotes the wall-parallel average.
The same result is obtained if the wall-normal momentum equation is averaged in the streamwise and spanwise direction and integrated in $y$ \citep[see e.g.][]{TennekesLumley1972firstcourseturbulence}.

We further note that the numerator and denominator in \cref{eq:solStokesPressure,eq:GreensFuncPressure} can lead to floating point overflow for sufficiently large $\kappa$, but this issue can be remedied by rewriting the expressions in terms of exponentials.
The Stokes pressure can then be calculated directly from the analytical solution \cref{eq:solStokesPressure}, while the fast and slow pressure are obtained by numerically integrating \cref{eq:solFastSlowPressure} using the trapezoidal rule on the DNS wall-normal grid.
\Cref{fig:validationP} shows a validation of the pressure solver against the data of \citet{LeeMoser2015Directnumericalsimulation} for a canonical turbulent channel flow at $(\mathrm{Re}_{\tau})_0 = 180$.
The two curves collapse at all wall-normal locations, which confirms the adequacy of the solver and settings.

\section{Summary of previous works: drag reduction and transpiration structure}  \label{sec:dr}
The present study builds on earlier works by the same authors \citep{ToedtliLuharMcKeon2019Predictingresponseturbulent,ToedtliYuMcKeon2020origindragincrease,Toedtli2021ControlWallBounded}.
This section provides a brief summary of the main findings of these previous studies, which are a prerequisite for the analyses in \cref{sec:wallPressure,sec:pvPhaseWall,sec:comparisonPhaseDrag}.
\Cref{sec:drOc} describes the drag reduction for varying-phase opposition control with various $\phaseShift$ and sensors located at $y_d^+ = 15$.
The transpiration structure of three example controlled flows is discussed in \cref{sec:transpirationStructure} and reveals the imprint of two families of spatial scales.
\Cref{sec:scaleFamilies}, together with \cref{sec:scaleRestrictedControllers}, shows how the flow response to varying-phase opposition control can be understood as a superposition of the two scale families.

\subsection{Drag reduction under varying-phase opposition control}  \label{sec:drOc}
The drag change $\Delta D$ is defined as the change in friction coefficient $c_{f}$ under control
\begin{equation}  \label{eq:defDr}
    \Delta D = 1 - \underbrace{\frac{(c_{f})_{c}}{(c_{f})_{0}}}_{\equiv \xi}; \qquad c_{f} = \frac{\dimensional{\tau}_w}{\frac{1}{2} \dimensional{\rho} (\dimensional{U}_b)^2}
\end{equation}
Positive values of $\Delta D$ indicate drag reduction (DR for short), while negative values represent drag increase (DI).
In order to characterize the drag change, it is sufficient to keep track of the friction coefficient ratio $\xi$, which is interpreted as follows
\begin{equation}
    \xi \in \begin{cases}
        [0, 1) & \text{drag reduction (smaller is better)} \\
        (1, \infty) & \text{drag increase (larger is worse)}
    \end{cases}
\end{equation}
and the drag is unchanged if $\xi = 1$.
We will use  $\Delta D$ and $\xi$ interchangeably throughout this manuscript, with an emphasis on $\xi$ for qualitative visualizations and $\Delta D$ for quantitative statements.
It is important to point out that $\Delta D$ does not account for the work done by the actuation.
\Cref{eq:defDr} does therefore not quantify energetic savings, but rather categorizes the flow response into two broad classes, namely a drag-reducing ($\xi < 1)$ and a drag-increasing ($\xi>1$) one.

\begin{figure}
    \centering
    \begin{subfigure}[b]{0.51\linewidth}
        \caption{Full range of $\phaseShift$}
        \includegraphics[width=\linewidth]{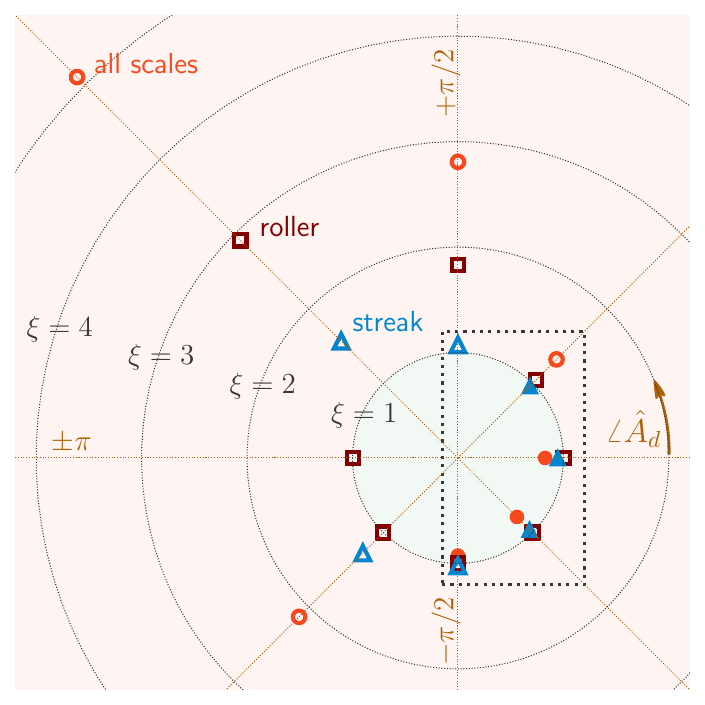}
        \label{fig:drOverall}
    \end{subfigure}
    \hspace{0.3in}
    \begin{subfigure}[b]{0.3\linewidth}
        \caption{Detail view}
        \includegraphics[width=\linewidth]{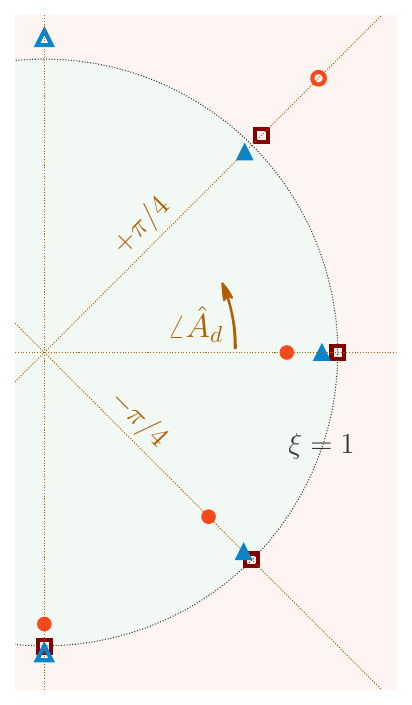}
        \label{fig:drDetail}
    \end{subfigure}
    \caption{Drag reduction of varying-phase opposition control with sensors located at $y_d^+ = 15$. Filled symbols inside the green shaded area indicate drag reduction ($\xi < 1$), while open symbols in the red region indicate unchanged or increasing drag ($\xi \geq 1$). The area inside the black square in \cref{fig:drOverall} is magnified in \cref{fig:drDetail}.}
    \label{fig:dragReduction}
\end{figure}

Varying-phase opposition control with $\phaseShift$ according to \cref{eq:controllerGainAllScales} has two parameters, the sensor location ($y_d$) and phase shift ($\phaseShift$).
The present study focuses on the role of the phase shift.
We therefore fix $y_d^+ = 15$, which corresponds to a sensor location with a strong flow response \citep[see][for details]{ToedtliLuharMcKeon2019Predictingresponseturbulent}, and vary the phase shift over its entire range $\phaseShift \in [-\pi, +\pi]$.
The associated drag change in terms of $\xi$ is shown by the orange circles in \cref{fig:drOverall} (entire parameter range) and \cref{fig:drDetail} (detail view of the black square in \cref{fig:drOverall}).
Filled symbols in the green-shaded area indicate drag reduction ($\xi < 1$), while open symbols in the red-shaded region denote unchanged or increasing drag ($\xi \geq 1$).
The data clearly show that the drag change strongly depends on the phase shift of the wall transpiration.
Small negative phase shifts lead to drag reduction, while positive or large negative phase shifts result in drag increase.
In particular, the drag increase at $\phaseShift = \pm \pi$ (which corresponds to positive control feedback) was so pronounced that the simulations diverged at the current resolution and no data point is therefore plotted at this phase shift.

In the following sections, we will analyze the three example controllers listed in \cref{tab:exampleControllers} in detail.
An example configuration will either be referred to by its label (e.g. N25) or its phase shift (e.g. $\phaseShift = -\pi / 4$).
Configuration N25 leads to 21\% DR, which is the largest drag reduction over the $\{y_d, \phaseShift\}$ parameter range considered by \cite{ToedtliLuharMcKeon2019Predictingresponseturbulent}.
The remaining two example configurations lead to a drag increase of 113\% (N75) and 180\% (P50), respectively.
Two drag-increasing example controllers are chosen to investigate if the physical mechanisms that drive the flow response are different for positive and negative $\phaseShift$.

\begin{table}
    \centering
    \begin{tabular}{*{3}{c}}
        Label & $\phaseShift$ & $\Delta D$\\
        \hline
        N75 & $- 3 \pi / 4$ & -113\% (DI) \\
        N25 & $- \pi / 4$ & +21\% (DR) \\
        P50 & $+ \pi / 2$ & -180\% (DI)
    \end{tabular}
    \caption{Parameters for example controllers. Positive values of $\Delta D$ indicate drag reduction (DR), while negative values represent drag increase (DI). In all cases, the sensors are located at $y_d^+ = 15$.}
    \label{tab:exampleControllers}
\end{table}

\subsection{Wall transpiration structure}  \label{sec:transpirationStructure}
\begin{figure}
    \centering
    \begin{subfigure}[b]{0.57\linewidth}
        \caption{$\phaseShift = - 3 \pi / 4$}
        \includegraphics[width=\linewidth]{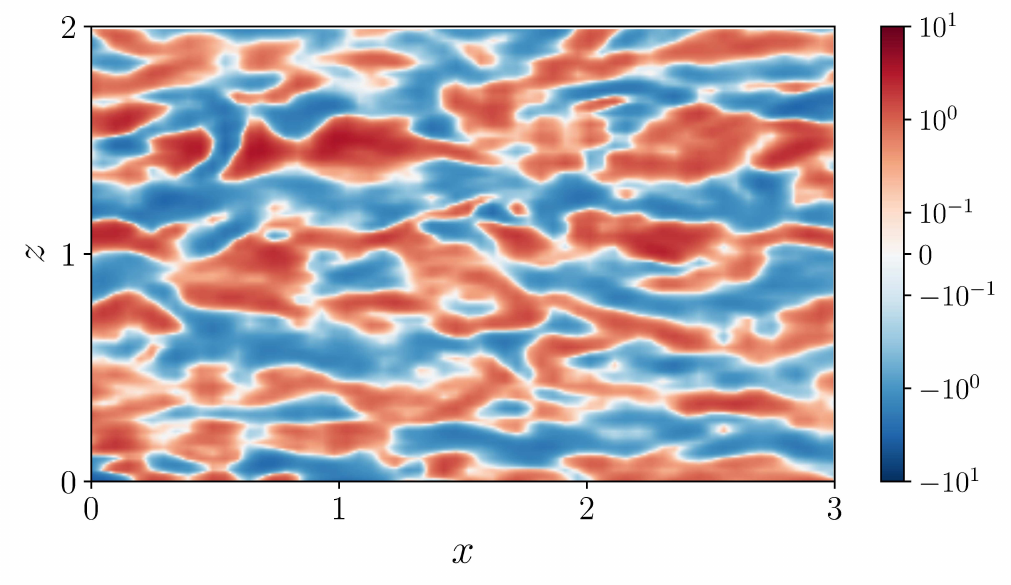}
        \label{fig:vSpatialN075pi}
    \end{subfigure}
    \hfill
    \begin{subfigure}[b]{0.41\linewidth}
        \caption{$\phaseShift = - 3 \pi / 4$}
        \includegraphics[width=\linewidth]{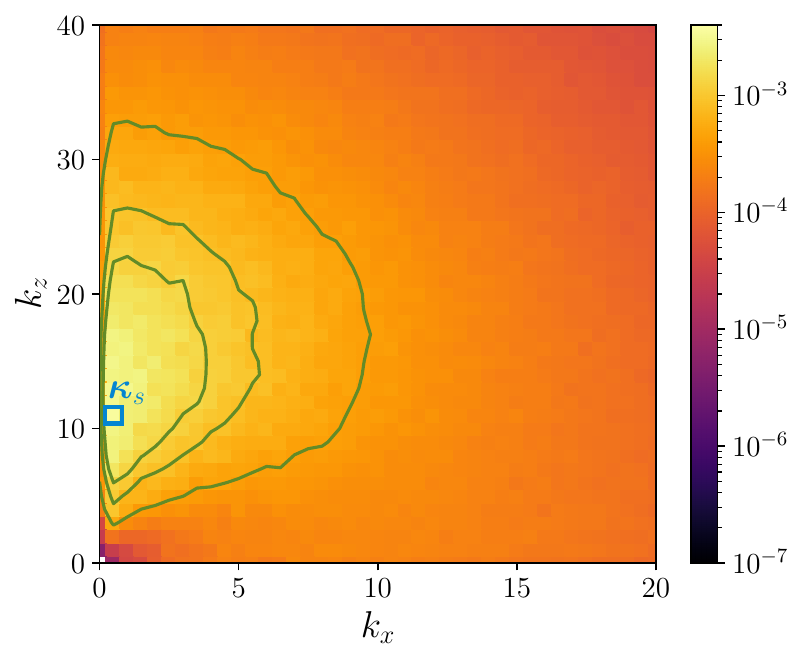}
        \label{fig:actSpectrumN075pi}
    \end{subfigure}
    \begin{subfigure}[b]{0.57\linewidth}
        \caption{$\phaseShift = - \pi / 4$}
        \includegraphics[width=\linewidth]{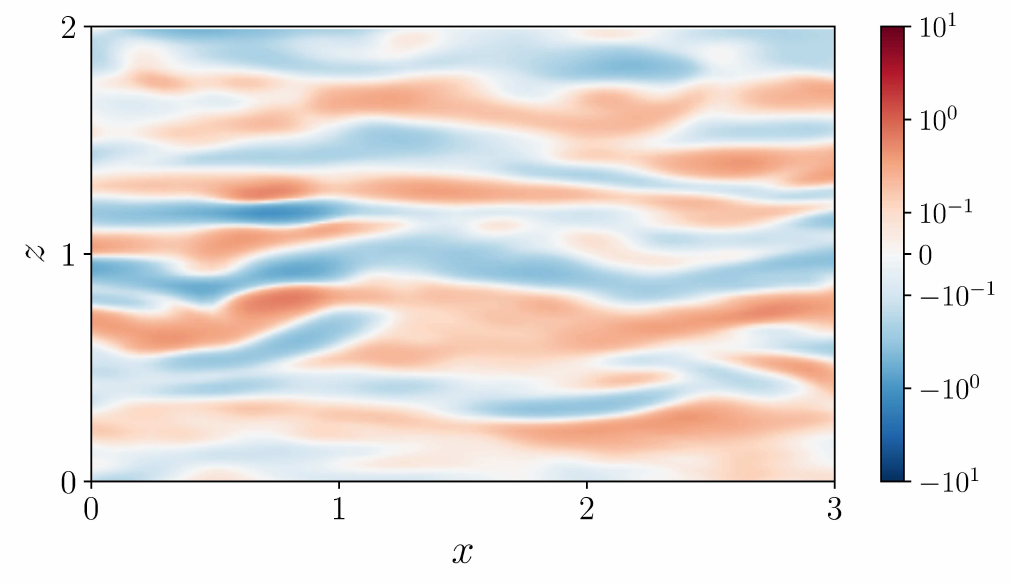}
        \label{fig:vSpatialN025pi}
    \end{subfigure}
    \hfill
    \begin{subfigure}[b]{0.41\linewidth}
        \caption{$\phaseShift = - \pi / 4$}
        \includegraphics[width=\linewidth]{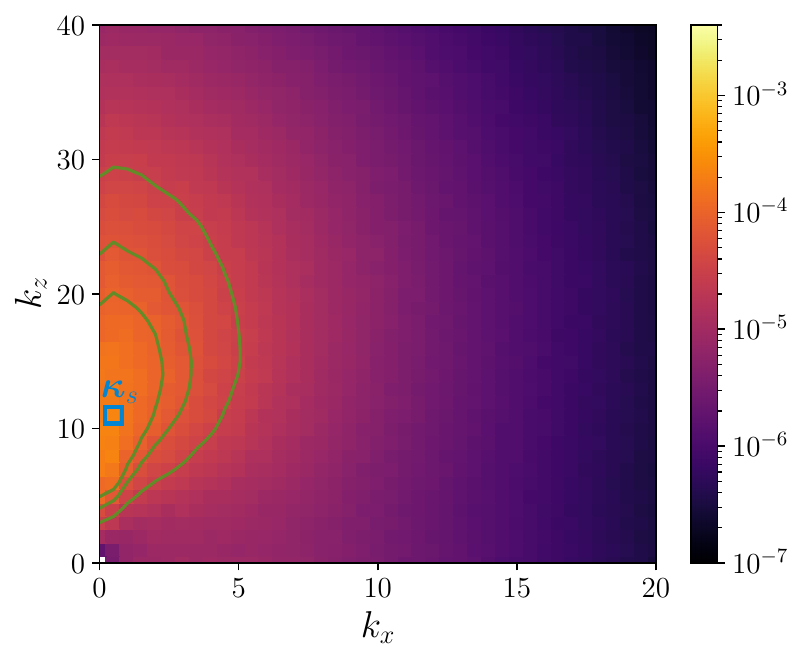}
        \label{fig:actSpectrumN025pi}
    \end{subfigure}
    \begin{subfigure}[c]{0.57\linewidth}
        \caption{$\phaseShift = + \pi / 2$}
        \includegraphics[width=\linewidth]{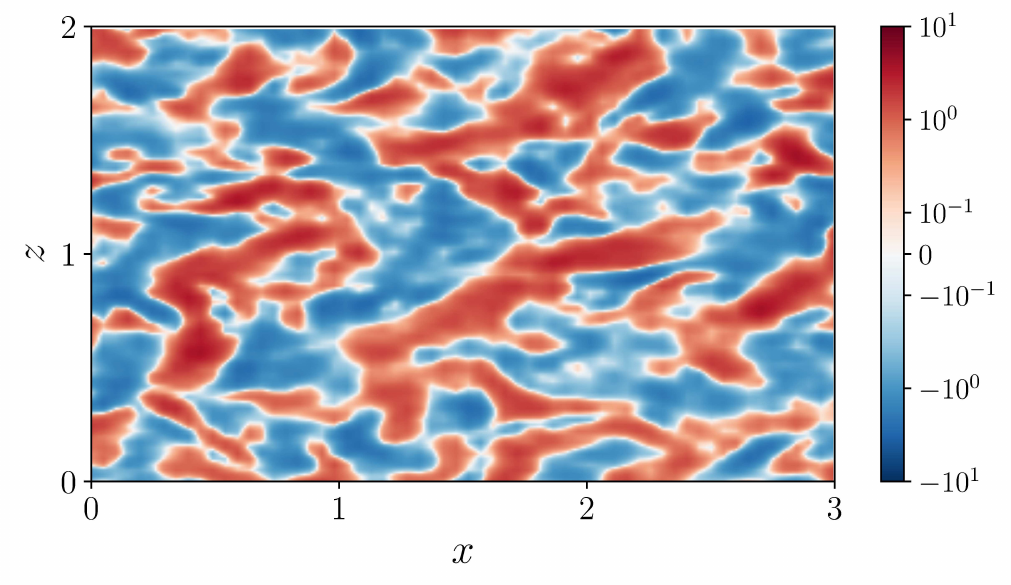}
        \label{fig:vSpatialP05pi}
    \end{subfigure}
    \hfill
    \begin{subfigure}[c]{0.42\linewidth}
        \caption{$\phaseShift = + \pi / 2$}
        \includegraphics[width=\linewidth]{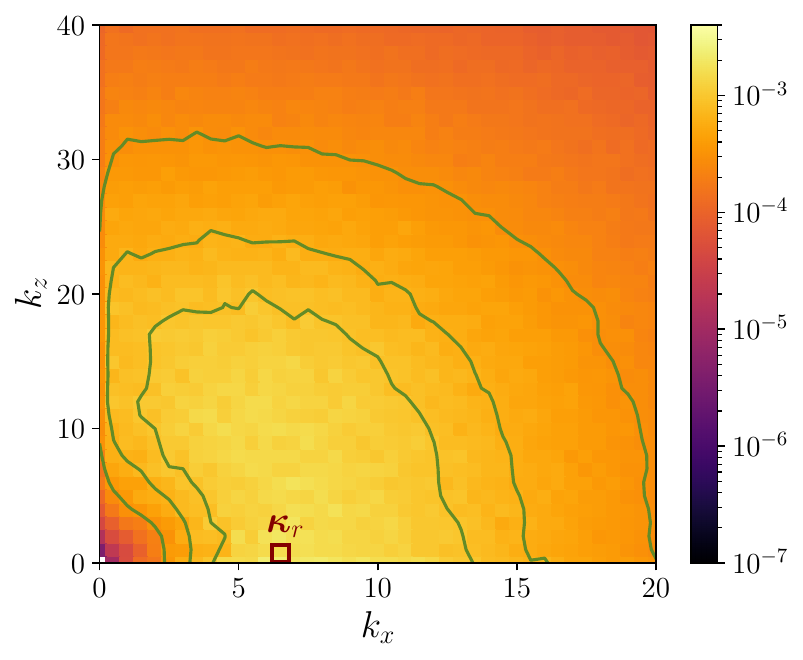}
        \label{fig:actSpectrumP05pi}
    \end{subfigure}
    \caption{Left column: instantaneous spatial structure of the transpiration $v^+(y_w)$ at the wall. Right column: time-averaged actuation spectrum $\Phi_{vv}^+(y_w)$. The green contour lines denote $\Phi_{vv}^+ / \max(\Phi_{vv}^+) = (0.15, 0.3, 0.45)$ and the blue and red square indicate the example spatial scales $\bkappa_s$ and $\bkappa_r$, respectively.}
    \label{fig:vSpatialSpectraAct}
\end{figure}

The left column of \cref{fig:vSpatialSpectraAct} shows representative instantaneous snapshots of the wall transpiration for each example controller.
These qualitative flow visualizations highlight two important aspects.
First, the magnitude of the transpiration varies significantly with $\phaseShift$.
For example, the control input of the drag-reducing configuration N25 is an order of magnitude smaller compared to the drag-increasing controllers N75 and P50 (note the logarithmic color scale).
A larger wall transpiration is therefore not necessarily beneficial in terms of drag reduction.
The second observation is the different spatial structure of the transpiration for positive and negative phase shifts.
The control input of controllers N25 and N75 consists of streamwise-elongated streaky structures and we observe larger, more coherent structures in case N25.
In contrast, the transpiration of controller P50 exhibits little coherence in the streamwise direction and coherent patches instead occur at an angle to the mean flow or even oriented along the span.

The wall transpiration varies over time, and a statistical characterization of the transpiration structure is given by the time-averaged power spectra in the right column of \cref{fig:vSpatialSpectraAct}.
The spectra confirm the observations from the instantaneous flow fields.
The transpiration magnitude of the drag-reducing configuration is about an order of magnitude smaller compared to the drag-increasing controllers.
And the structural difference between transpiration with positive and negative phase shifts is apparent as well.
The most energetic scales in the spectrum of N25 (\cref{fig:actSpectrumN025pi}) are streamwise-elongated ($k_x$ small and $k_x < k_z$) and the peak occurs at $\bkappa_s = (k_x = 0.5, k_z = 11)^{\intercal}$, which is indicated by the blue square.
Scales centered around $\bkappa_s$ are also the most active in configuration N75 (\cref{fig:actSpectrumN075pi}) but the energetic region extends to larger $k_x$ and $k_z$, which confirms that this transpiration is more multi-scale than N25.
In contrast, the most active scales in the spectrum of P50 (\cref{fig:actSpectrumP05pi}) are short in the streamwise direction and wide in the span.
The peak occurs at $\bkappa_r = (k_x = 6.5, k_z = 0)^{\intercal}$, which is marked by the red square.
The streamwise-elongated structures observed in cases N25 and N75 are not very pronounced in \cref{fig:actSpectrumP05pi}, which underscores that the transpiration structure at positive and negative phase shifts is markedly different.

Additional visualizations and a more in-depth discussion of the controlled flow structure can be found in \citet{ToedtliChristineMcKeon2019StructuralSpectralAnalysis,Toedtli2021ControlWallBounded}.
One point to highlight from these references is that the flow structure of the three example controllers not only differs close to the wall, but throughout the channel.
Perhaps unsurprisingly, drag-reduced flows exhibit less vortical activity and turbulence intensity compared to drag-increased flows at all $y$.

\subsection{Scale families}  \label{sec:scaleFamilies}
The analysis of the controlled flows would be much simplified if the overall flow response and drag change were mainly due to transpiration at a limited number of scales.
The transpiration spectra in \cref{fig:vSpatialSpectraAct} suggest that two scale families are energetic at different $\phaseShift$:
Streamwise-elongated streaky scales centered around $\bkappa_s$ dominate the control input for negative phase shifts and will be referred to as ``streak scales'' hereafter.
On the other hand, scales with short $x$ and wide $z$ extents clustered around $\bkappa_r$ are the most active scales for $\phaseShift > 0$ and will be termed ``roller scales'' (the reason behind this nomenclature will become clear shortly).
However, just because a scale $\bkappa$ is energetic in the transpiration spectra does not imply that this scale is relevant for drag reduction or that its behavior is dictated by $\phaseShift$, because the governing equations are nonlinear.

In order to tie the drag behavior to individual scales and the change in transpiration to $\phaseShift$ at that $\bkappa$, possible nonlinear control effects have to be minimized.
This can be achieved by restricting control to a single wavenumber, or a small set of wavenumbers that are not triadically consistent.
We refer to these controllers as ``scale-restricted'' controllers, and have analyzed the flow response to two scale-restricted controllers:
the first controller generates a transpiration at a small number of triadically-inconsistent streak scales and the second one generates a transpiration with a single roller scale.
The details of the scale-restricted controllers are summarized in \cref{sec:scaleRestrictedControllers} and an in-depth discussion can be found in \citet{Toedtli2021ControlWallBounded}.

The drag reduction of the scale-restricted controllers is shown by the blue triangles (streak scales) and red squares (roller scales) in \cref{fig:dragReduction}.
Varying-phase opposition control with streak scales can reduce ($\lvert \phaseShift \rvert \lesssim \pi / 4$) or increase ($\lvert \phaseShift \rvert \gtrsim \pi / 2$) the drag and the controller with $\phaseShift = \pm \pi$ also diverged in this case.
At the present $\reTau$, the streak scales are associated with the near-wall cycle and the observed drag changes suggest that varying-phase opposition control can suppress or amplify the near-wall cycle, depending on $\phaseShift$.
This hypothesis will be confirmed by the quantitative scale suppression analysis in \cref{sec:comparisonPhaseDrag}.
On the other hand, control with the roller scales leaves the drag unchanged if the phase shift is negative ($\phaseShift \in [-\pi , 0]$) and increases the drag significantly for positive phase shifts ($\phaseShift \in [+\pi / 4, +3 \pi / 4]$).
These roller scales are not very energetic in a canonical turbulent channel flow, but are energized by control with positive phase shifts.
Conditional averaging shows that control with the roller scales and positive phase shifts induces spanwise rollers (thus the name) and interested readers can find an example visualization in \citet{ToedtliYuMcKeon2020origindragincrease}.
In addition, the onset of the drag increase and appearance of the spanwise rollers coincides with the presence of an amplified eigenvalue in the linearized Navier-Stokes equations \citep{ToedtliYuMcKeon2020origindragincrease}.

It is important to emphasize again that nonlinear control effects are reduced as much as possible in the scale-restricted controllers.
The change in transpiration at a specific scale can thus be tied to $\phaseShift$ at that $\bkappa$ and the overall drag change can be attributed to the transpiration at the few active scales.
A comparison between the drag change of the scale-restricted and the full controller therefore suggests that the flow response to varying-phase opposition control can be understood from a superposition of the two scale families.
The roller scales are inactive for $\phaseShift \in [-\pi, +\pi / 4)$ and the flow response is fully determined by the streak scales, which reduce (small phase shifts) or increase (large positive or negative $\phaseShift$) the drag.
Both scale families contribute to the drag increase at $\phaseShift \in [+\pi / 4, 3 \pi / 4]$ but the roller scales dominate due to the stronger flow response (compare the drag increase due to the two-scale restricted controllers in \cref{fig:dragReduction}).
The dominance of the roller scales for positive phase shifts is further consistent with the actuation spectra of \cref{fig:vSpatialSpectraAct}.

\section{Wall pressure in the presence of transpiration}  \label{sec:wallPressure}
We next consider the properties of the wall pressure in the presence of transpiration.
\Cref{sec:wallPressureStructure} explores the instantaneous and time-averaged structure of the wall pressure field and compares it to the transpiration structure.
The pressure field is then decomposed into its fast, slow and Stokes components and the relative magnitude of the three components is studied in \cref{sec:relImportancePressureComp}.
Following the approach from the previous section, the discussion focuses on the three example controllers of \cref{tab:exampleControllers}.

\subsection{Wall pressure structure}  \label{sec:wallPressureStructure}
\begin{figure}
    \centering
    \begin{subfigure}[b]{0.57\linewidth}
        \caption{$\phaseShift = - 3 \pi / 4$}
        \includegraphics[width=\linewidth]{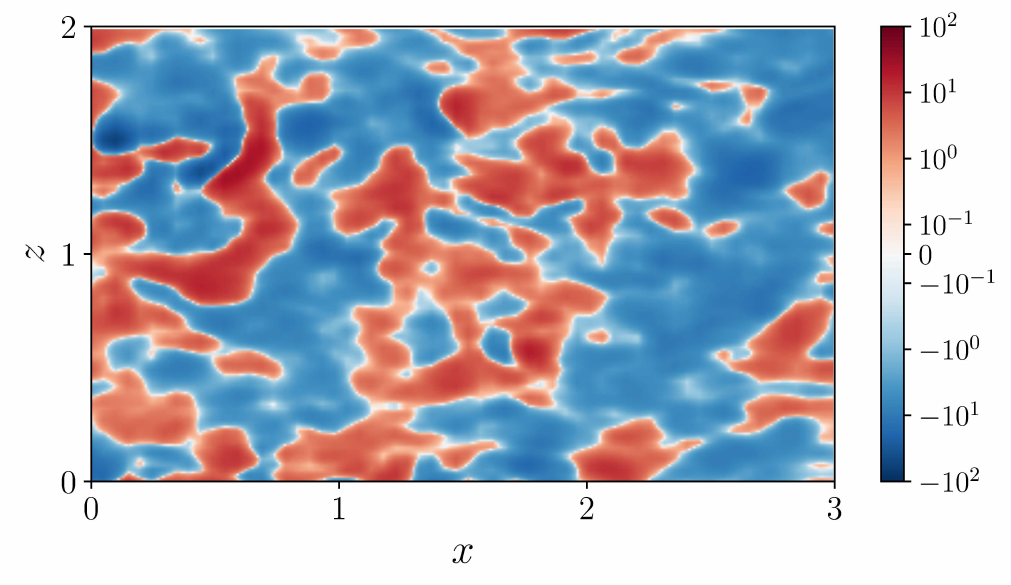}
        \label{fig:pSpatialN075pi}
    \end{subfigure}
    \begin{subfigure}[b]{0.41\linewidth}
        \caption{$\phaseShift = -3 \pi / 4$}
        \includegraphics[width=\linewidth]{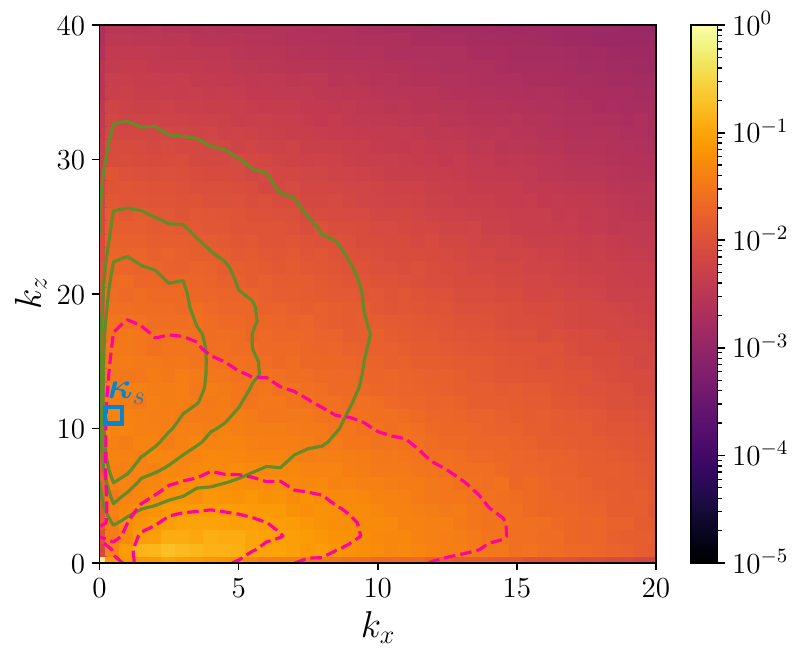}
        \label{fig:wallPressureN075pi}
    \end{subfigure}
    \begin{subfigure}[b]{0.57\linewidth}
        \caption{$\phaseShift = - \pi / 4$}
        \includegraphics[width=\linewidth]{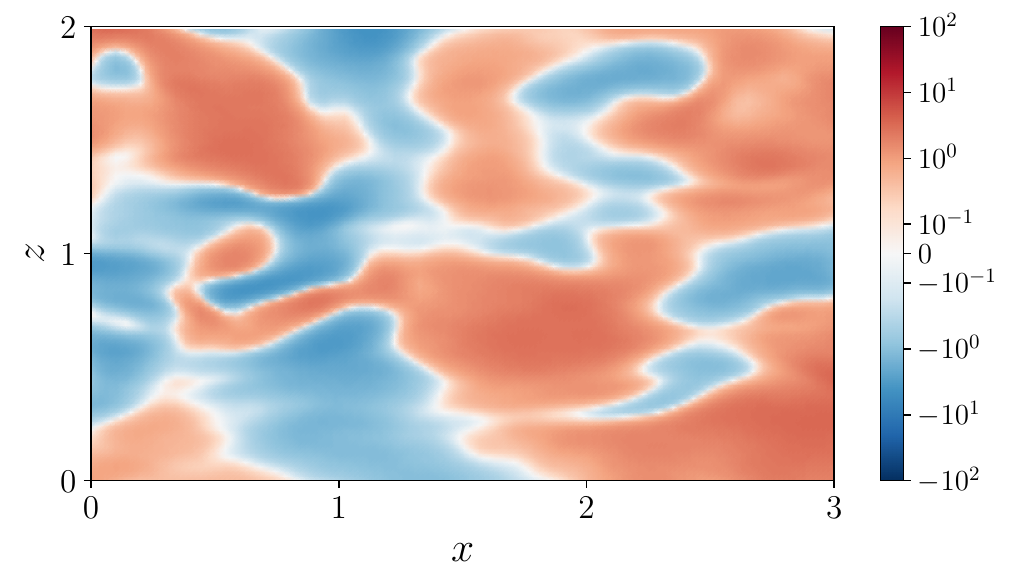}
        \label{fig:pSpatialN025pi}
    \end{subfigure}
    \begin{subfigure}[b]{0.41\linewidth}
        \caption{$\phaseShift = -\pi / 4$}
        \includegraphics[width=\linewidth]{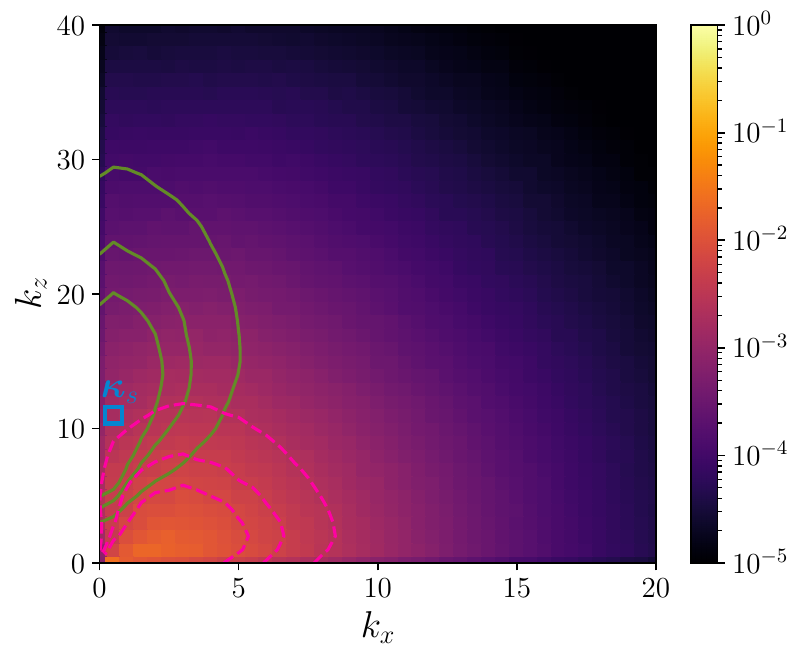}
        \label{fig:wallPressureN025pi}
    \end{subfigure}
    \begin{subfigure}[b]{0.57\linewidth}
        \caption{$\phaseShift = + \pi / 2$}
        \includegraphics[width=\linewidth]{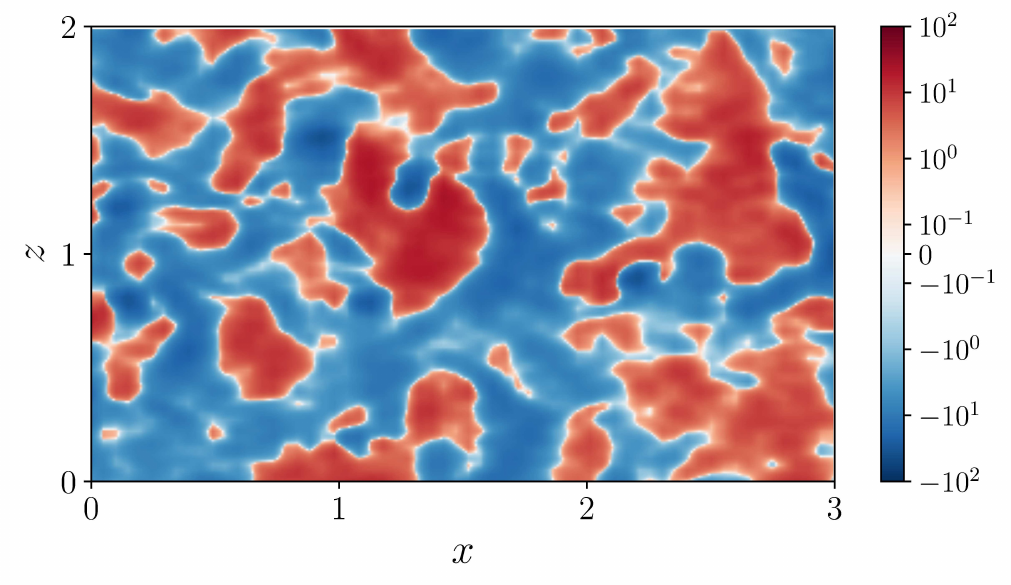}
        \label{fig:pSpatialP05pi}
    \end{subfigure}
    \begin{subfigure}[b]{0.41\linewidth}
        \caption{$\phaseShift = +\pi / 2$}
        \includegraphics[width=\linewidth]{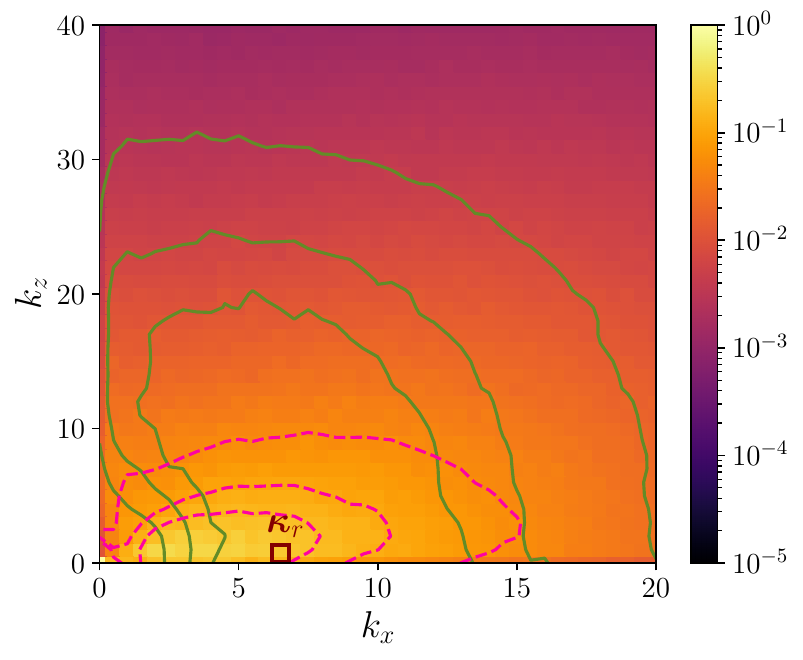}
        \label{fig:wallPressureP05pi}
    \end{subfigure}
    \caption{Left column: Instantaneous spatial structure of $p^+(y_w)$, right column: time averaged wall pressure fluctuation spectrum $\Phi_{pp}^+$. The lines outline the levels $\Phi_{qq}^+ / \max(\Phi_{qq}^+) = (0.15, 0.3, 0.45)$ of $v$ (solid green lines, replotted from \cref{fig:vSpatialSpectraAct}) and $p$ (dashed pink lines).}
    \label{fig:pSpatialSpectra}
\end{figure}

The spatial structure of the wall pressure for the three example controllers is depicted in the left column of \cref{fig:pSpatialSpectra}.
The time instant and location are identical to the earlier visualizations of the transpiration (\cref{fig:vSpatialN075pi,fig:vSpatialN025pi,fig:vSpatialP05pi}).
Similar to the earlier discussion of the transpiration, we observe differences in both magnitude and structure across the three example wall pressure fields.
The wall pressure of the drag-reducing controller N25 (\cref{fig:pSpatialN025pi}) features large-scale patches of positive and negative fluctuations of order one.
The drag-increasing controllers (\cref{fig:pSpatialN075pi,fig:pSpatialP05pi}) on the other hand present a less coherent wall pressure field with variations of order ten (note the logarithmic color scale).
The pressure field of both drag-increasing controllers are similar and feature structures with shorter $x$ and larger $z$ extent.

\Cref{fig:wallPressureN075pi,fig:wallPressureN025pi,fig:wallPressureP05pi} show the time-averaged power spectra of the wall pressure, with distinct levels outlined by the dashed pink lines.
In addition, the green contour lines represent the spectral content of the actuator input (see \cref{fig:vSpatialSpectraAct}) and the blue and red squares indicate the representative streak ($\bkappa_s$) and roller scales ($\bkappa_r$), respectively.
The overall structure of the three wall pressure spectra is similar, with the most energetic region located at small streamwise and spanwise wavenumbers.
The magnitude of the scale-by-scale contribution and the extent of the energetic scales depends strongly on $\phaseShift$, with drag-increasing configurations leading to more energetic wall pressure fluctuations (note the logarithmic color scale) over a wider range of scales.
This is, on average, the spectral analogue of the more intense and fragmented instantaneous wall pressure structures observed in \cref{fig:pSpatialN075pi,fig:pSpatialP05pi}.

\begin{figure}
    \centering
    \includegraphics[width=0.52\textwidth]{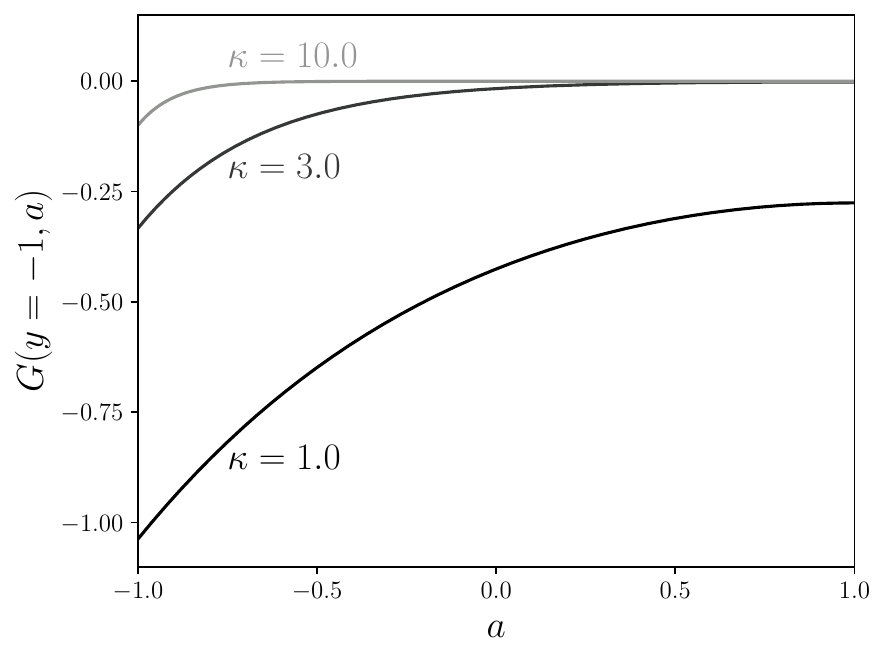}
    \caption{Example Green's function kernels for the wall pressure ($y = -1$) at various values of $\kappa$. The Green's function is plotted versus the integration variable $a$.}
    \label{fig:greensFunctionKernels}
\end{figure}

The Greens function solution, \cref{eq:solStokesPressure,eq:solFastSlowPressure}, provides some insight into the structure of the wall pressure spectra and their relation to the transpiration.
The Green's function kernel $G(y,a)$ is real-valued and only depends on the magnitude of the wavenumber vector, $\kappa$.
Different combinations of $k_x$ and $k_z$ with the same $\kappa$ thus have an identical $G(y,a)$ and example kernels for the wall pressure at various $\kappa$ are shown in \cref{fig:greensFunctionKernels}.
For increasing $\kappa$, the wall-normal support of the Green's function decreases and the maximum amplitude decreases like $1/\kappa$.{}
It is important to reiterate the observation of \cite{Kim1989structurepressurefluctuations} that the Green's function kernels for small $\kappa$ decrease slowly with $y$ and can be nonzero throughout the channel (see e.g. the profile $\kappa = 1$ in \cref{fig:greensFunctionKernels}).
The slow decay implies a large domain of dependence in $y$ and can even introduce a mutual influence of the two walls.
Similar observations apply to the Stokes pressure as well, but the details are omitted for the sake of brevity.

The wall pressure spectra peak at low $k_{x}$ and $k_{z}$, where the Green's function kernels have the largest $y$ support.
Regions far from the wall can thus potentially contribute to these maxima.
Interestingly, the wall pressure spectra are a function of $k_x$ and $k_z$, which is different from the $\kappa$ dependence of the Green's function kernel.
This dependence on $(k_x, k_z)$ must arise from the source terms (slow and fast pressure) and from the boundary conditions (Stokes pressure).
One aspect that certainly contributes to the observed asymmetry is the fast pressure, whose source term contains an $x$-derivative that amplifies smaller streamwise scales (large $k_x$) and damps larger ones (small $k_x$).
The edge case is $k_{x} = 0$, where the fast pressure disappears all together.
Consistent with this argument, the wall pressure spectra become less energetic as $k_x \rightarrow 0$.

Likely for the same reason, actuation with streak scales does alter the wall pressure at low $k_x$ significantly (\cref{fig:wallPressureN075pi,fig:wallPressureN025pi}) and preserves the overall wall pressure structure.
In contrast, actuation with roller scales goes hand-in-hand with an increase in pressure signature at small $k_z$, where the fast pressure source term is not damped.
It is also interesting to note that the drag-increasing controllers increases the wall pressure in spectral regions beyond the actuator input.
This hints at the importance of nonlinear interactions, either through the slow pressure or energy transfer across velocity scales.

\subsection{Relative importance of the pressure components}  \label{sec:relImportancePressureComp}
\begin{figure}
    \centering
    \begin{subfigure}[t]{0.32\textwidth}
        \caption{$\phaseShift = - 3 \pi / 4$}
        \includegraphics[width=\textwidth]{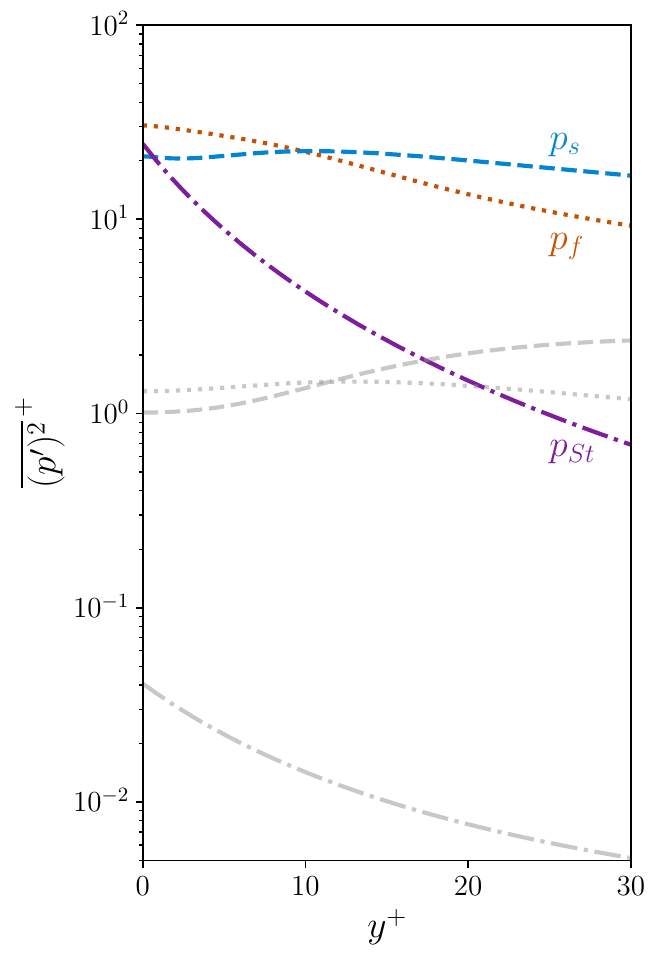}
        \label{fig:pvVarN075pi}
    \end{subfigure}
    \begin{subfigure}[t]{0.32\textwidth}
        \caption{$\phaseShift = - \pi / 4$}
        \includegraphics[width=\textwidth]{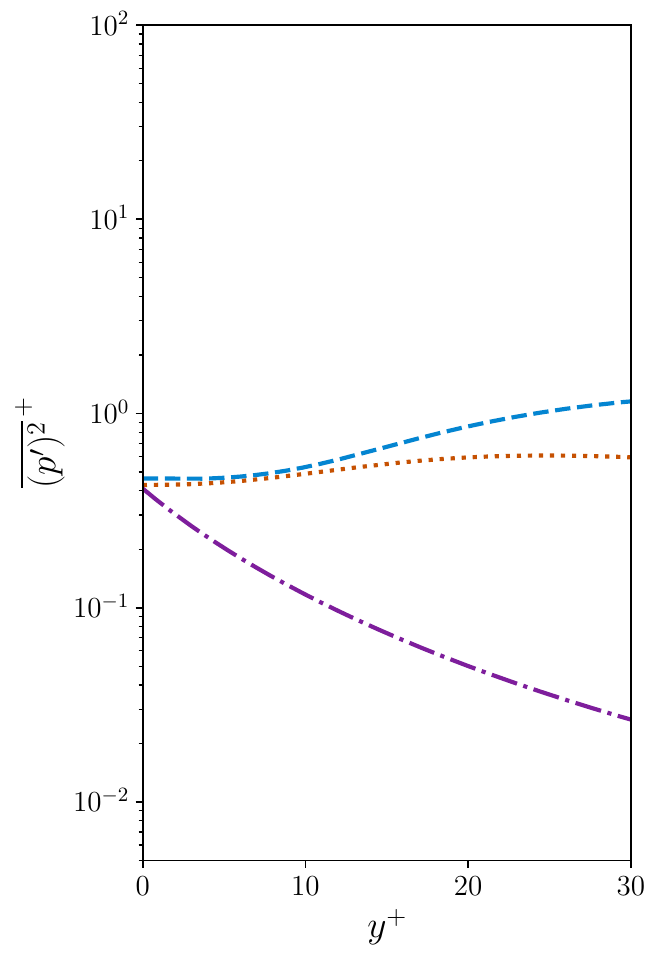}
        \label{fig:pvVarN025pi}
    \end{subfigure}
    \begin{subfigure}[t]{0.32\textwidth}
        \caption{$\phaseShift = + \pi / 2$}
        \includegraphics[width=\textwidth]{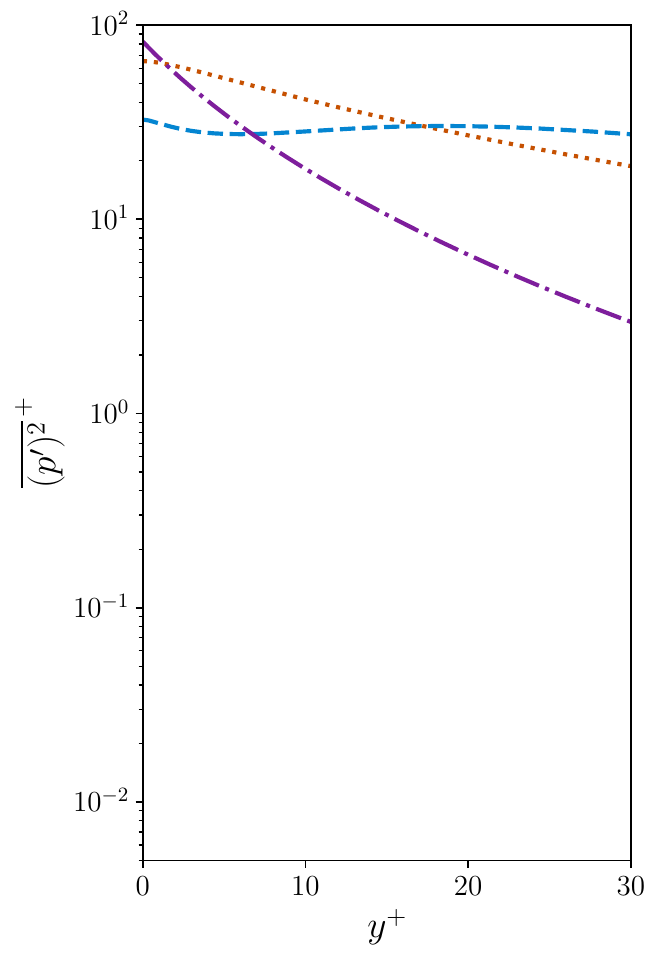}
        \label{fig:pvVarP05pi}
    \end{subfigure}
    \begin{subfigure}[t]{0.32\textwidth}
        \captionsetup{width=\extraCaptionWidth}
        \caption{$\phaseShift = - 3 \pi / 4, \bkappa_s$}
        \includegraphics[width=\textwidth]{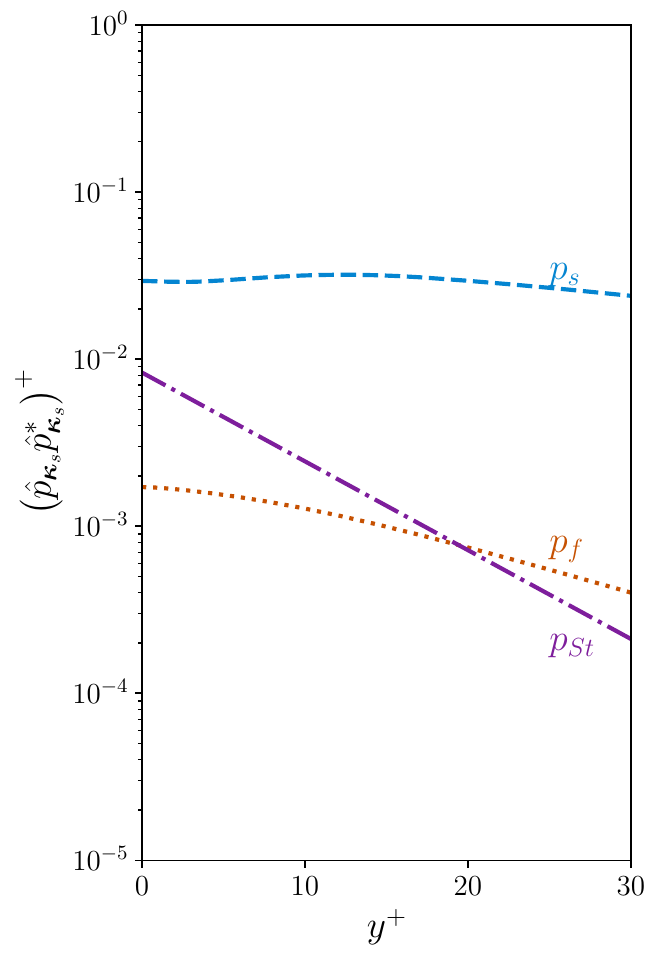}
        \label{fig:pvStatsN075pi}
    \end{subfigure}
    \begin{subfigure}[t]{0.32\textwidth}
        \captionsetup{width=\extraCaptionWidth}
        \caption{$\phaseShift = - \pi / 4, \bkappa_s$}
        \includegraphics[width=\textwidth]{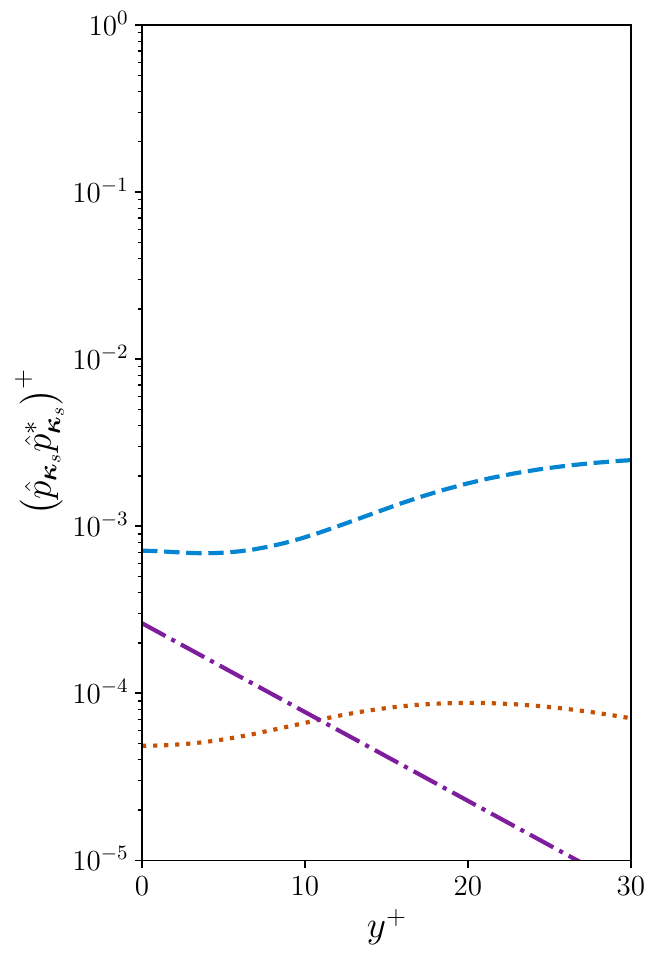}
        \label{fig:pvStatsN025pi}
    \end{subfigure}
    \begin{subfigure}[t]{0.32\textwidth}
        \captionsetup{width=\extraCaptionWidth}
        \caption{$\phaseShift = + \pi / 2, \bkappa_r$}
        \includegraphics[width=\textwidth]{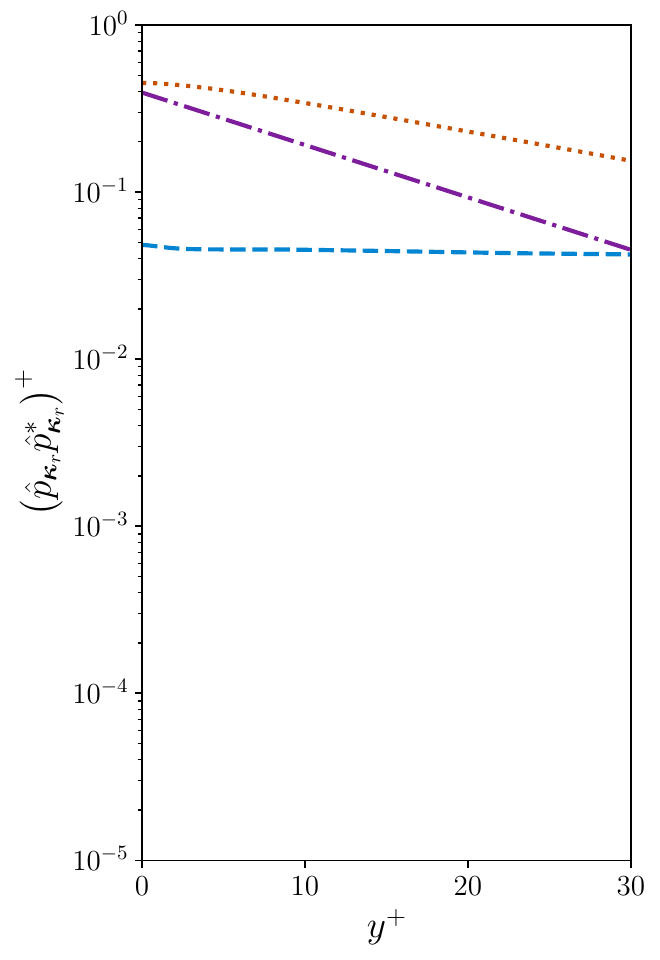}
        \label{fig:pvStatsP05pi}
    \end{subfigure}
    \caption{Total variance (top row) and example scale contribution (bottom row) to the fast (dotted orange line), slow (dashed blue line), and Stokes pressure (dash-dotted purple line). The gray lines in \cref{fig:pvVarN075pi} outline the reference profiles for a canonical turbulent channel flow at $(\reTau)_0 = 180$.}
    \label{fig:pvStats}
\end{figure}

The discussion thus far considered the total wall pressure field.
Subsequent analyses will consider the pressure components (fast, slow, and Stokes pressure) individually.
Interpretation of these more finely partitioned data requires quantification of the relative importance of the pressure components, which is shown in \cref{fig:pvStats}.
The colored lines in \cref{fig:pvVarN075pi,fig:pvVarN025pi,fig:pvVarP05pi} show the total pressure component variance of the controlled flows, while the gray lines in \cref{fig:pvVarN075pi} outline the reference profiles for a canonical turbulent channel flow without transpiration.
The sum of the component variances at $y_w$ represents a subset of the terms in the Parseval identity and is thus related to the wall pressure spectra of \cref{fig:pSpatialSpectra}.
The relative magnitude of the pressure component fluctuations underscore again the earlier observations that drag-increasing controllers lead to larger pressure fluctuations at the wall compared to the drag-reducing configuration.

It is also apparent that wall transpiration changes the relative importance of the pressure components compared to a canonical turbulent channel flow.
In the absence of transpiration, the slow pressure is larger than the fast one, except very near the wall, and the Stokes pressure is the smallest of the three components.
The Stokes pressure further scales as $1/\reBulk$ and is therefore assumed negligible at high Reynolds numbers (see \cref{eq:solStokesPressure} and discussion in \cite{Kim1989structurepressurefluctuations} for details).
In contrast, the magnitude of the Stokes pressure becomes comparable to the other components close to the wall when the no-throughflow condition is relaxed, regardless of whether the drag is reduced or increased.
This increase in magnitude occurs because the transpiration adds a leading-order inertial term (see \cref{eq:solStokesPressure}).
The Stokes pressure in transpiration-based controlled flows can thus be leading-order, possibly even at high Reynolds numbers.
Moreover, the relative magnitude of the slow and fast pressure can change with transpiration.
For example, the relative magnitude of the fast pressure decreases when the drag is reduced (\cref{fig:pvVarN025pi}) and increases when the drag is increased, especially near the wall.
This is likely a consequence of control reducing or amplifying the wall-normal velocity (see spectra in \cref{fig:vSpatialSpectraAct}), which enters the source term of the fast pressure.

The spatially averaged variances discussed thus far represent a sum over all spatial scales, but the relative importance of the pressure components can in principle vary from scale to scale.
This aspect is explored in \cref{fig:pvStatsN075pi,fig:pvStatsN025pi,fig:pvStatsP05pi}, which show the variance contributions of the relevant example scale (streak scale $\bkappa_s$ for negative $\phaseShift$ and roller scale $\bkappa_r$ for positive $\phaseShift$) for each controller.
A comparison between the top and bottom row demonstrates that the relative magnitude is indeed scale-dependent.
For example, the fast pressure contributes most to the wall pressure when summed over all scales at $\phaseShift = - 3\pi / 4$, but it is the smallest component at scale $\bkappa_s$ at that phase shift.

The scale-by-scale analysis further shows that the dominant pressure components change with $\phaseShift$ and $\bkappa$.
For negative $\phaseShift$, the slow and Stokes pressure at $\bkappa_s$ are almost an order of magnitude larger at the wall compared to the fast pressure.
It can therefore be expected that they determine the overall wall pressure signature at this scale.
On the other hand, for positive $\phaseShift$, the fast and the Stokes pressure are dominant at the scale $\bkappa_r$.
\Cref{fig:pvStatsN075pi,fig:pvStatsP05pi} further show that the different relative importance occurs even when the transpiration magnitude is comparable.
This suggests that the different drag increase mechanisms, amplification of the near-wall cycle for $\bkappa_s$ and generation of spanwise rollers for $\bkappa_r$, lead to fundamentally different source terms for the pressure components and therefore to different relative magnitudes.
Some aspects of this will be further explored in \cref{sec:pvPhaseDiffStrucExample} and \cref{sec:comparisonPhaseDrag}.

\section{Relative phase between wall pressure and transpiration}  \label{sec:pvPhaseWall}
\begin{figure}
    \centering
    \begin{subfigure}[b]{0.49\textwidth}
        \caption{$\phaseShift = - \pi / 4$}
        \includegraphics[width=\textwidth]{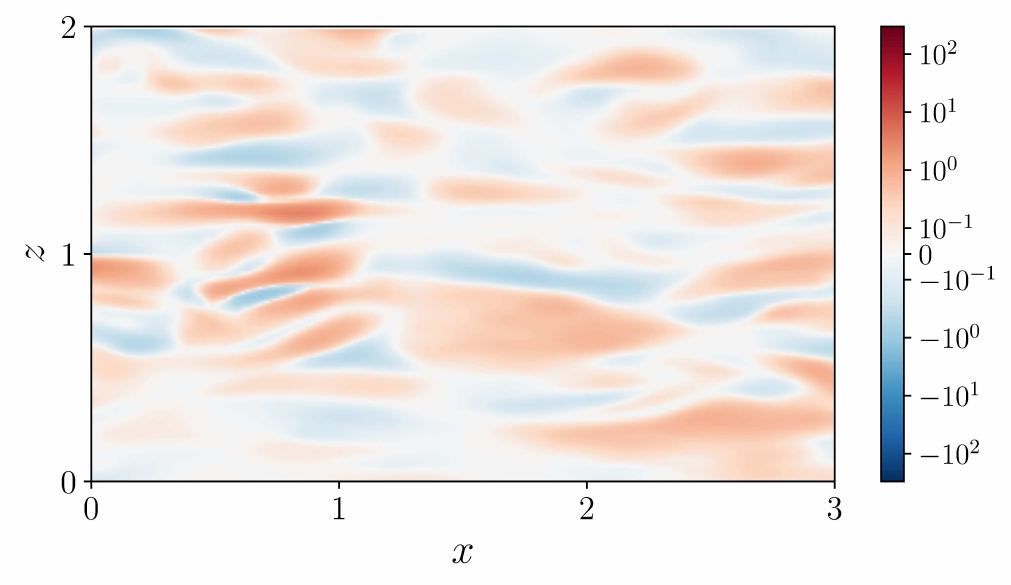}
        \label{fig:pvSpatialN25pi}
    \end{subfigure}
    \hfill
    \begin{subfigure}[b]{0.49\textwidth}
        \caption{$\phaseShift = + \pi / 2$}
        \includegraphics[width=\textwidth]{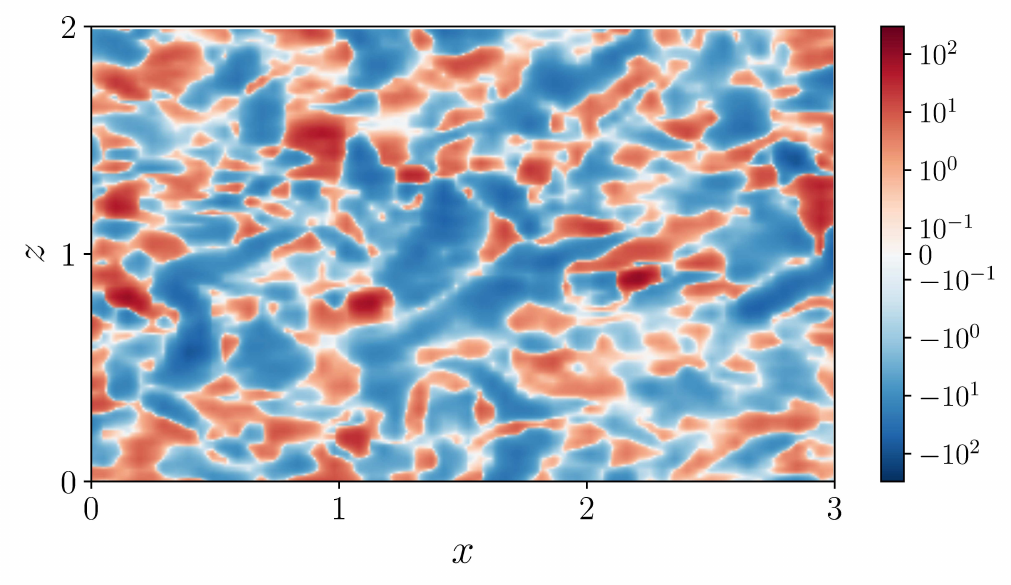}
        \label{fig:pvSpatialP05pi}
    \end{subfigure}
    \caption{Instantaneous spatial structure of $p^+ v^+$ at the wall.}
    \label{fig:pvSpatial}
\end{figure}

We next discuss the relative spatial arrangement between wall transpiration and pressure.
\Cref{fig:pvSpatial} shows the instantaneous point-wise product of the wall transpiration (left column of \cref{fig:vSpatialSpectraAct}) and pressure (left column of \cref{fig:pSpatialSpectra}) for configurations N25 (left) and P50 (right).
A detailed analysis of the magnitude and structure of $vp$ is beyond the scope of this section and omitted.
Our focus is instead on the sign of the product, which results from the relative spatial arrangement of the transpiration and the pressure.

Unlike the transpiration, which induces no net mass flux and therefore has a zero wall-parallel mean, the product $vp$ can have a nonzero $xz$-mean due to the modulation by the pressure.
For example, the dominance of red patches in \cref{fig:pvSpatialN25pi} suggests that $vp$ is on average positive for controller N25 (the figure only shows a subset of the $xz$-domain, but the observation is true for the entire wall as well).
This indicates that $v$ and $p$ have locally the same sign at the wall, a configuration that will be referred to as ``in-phase'' subsequently.
In contrast, the product is mostly negative for configuration P50 (\cref{fig:pvSpatialP05pi}), which implies that $v$ and $p$ have opposite signs (``out-of-phase'').

The following analysis explores the relative position of signed transpiration and pressure fluctuations for the three example control configurations in detail.
We again take a Fourier perspective and quantify the instantaneous relative spatial arrangement in terms of the phase difference between wall pressure and transpiration at length scale $\bkappa$
\begin{equation}
    \Delta \theta_{\bkappa}(t) = \angle \four{p}_{\bkappa}(y_w, t) - \angle \four{v}_{\bkappa}(y_w, t)
\end{equation}
The co-spectrum of $p$ and $v$ is not suitable to infer statistics about $\Delta \theta_{\bkappa}$ itself, because it weights each phase difference by $\absVal{\four{p}_{\bkappa}(y_w,t)}$ and $\absVal{\four{v}_{\bkappa}(y_w,t)}$ in the time-average.
Appropriate tools to statistically characterize the phase difference are instead introduced in \cref{sec:circularStats}.
\Cref{sec:pvPhaseDiffObsExample} reports the mean phase difference between the transpiration and wall pressure components at the example spatial scales $\bkappa_s$ and $\bkappa_r$.
The observations are explained in \cref{sec:domainDependencePwPhase,sec:pvPhaseDiffStrucExample} in terms of the Green's function structure and the properties of the example scales.
\Cref{sec:pvPhaseDiffAllScales} discusses the phase difference at all spatial scales and \cref{sec:connectionKinEn} connects the phase difference to the kinetic energy of the flow.

\subsection{Circular statistics}  \label{sec:circularStats}
\begin{figure}
    \centering
    \includegraphics[width=0.5\textwidth]{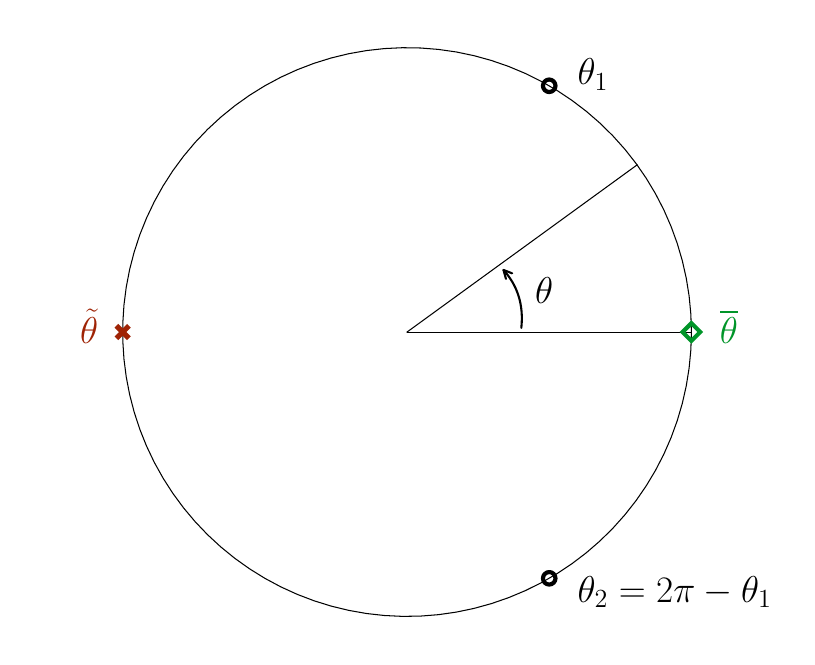}
    \caption{Circular data ($\theta_1$, $\theta_2$) and their linear ($\tilde{\theta}$, red cross) and circular mean ($\overline{\theta}$, green square).}
    \label{fig:exampleCircularStats}
\end{figure}

Phase differences or, more generally, phase angles in the complex plane are periodic quantities and take on values in the interval $[0, 2\pi)$ or any multiple of it.
They are an instance of so-called directional data and have a number of properties that complicate their statistical characterization.
For example, the numerical representation of directional data is non-unique, since the angular value depends on the choice of zero direction and sense of rotation.
In addition, the periodicity violates the Euclidean notion of distance: for example, a value of $2\pi - \epsilon$ is closer to $0$ than to say $\pi$ for sufficiently small $\epsilon$.
As a consequence, methods from linear statistical analysis are inappropriate to compute statistics of directional data.
This is best illustrated by the example dataset shown in \cref{fig:exampleCircularStats}, which consists of two angles ($\theta_1, \theta_2$).
East is taken as the zero direction, and positive angles are measured in counter-clockwise direction, so that $\theta_2 = 2\pi - \theta_1$.
The standard linear mean $\tilde{\theta}$
\begin{equation}
    \tilde{\theta} = \frac{1}{N} \sum_{i=1}^N \theta_i = \pi
\end{equation}
is represented as red cross in \cref{fig:exampleCircularStats} and is clearly not representative of the average direction of the data.

An appropriate mean value can instead be obtained from the circular mean, which interprets the directional data as points on the unit circle in the complex plane.
Starting from the resultant $\bm{R}$ of the associated unit vectors
\begin{equation}
    \bm{R} = \left( \sum_i \cos(\theta_i), \sum_i \sin(\theta_i) \right)^{\intercal} = \left(C, S\right)^{\intercal}, \qquad R = \sqrt{C^2 + S^2} \\
\end{equation}
the circular mean $\mean{\theta}$ is defined as the direction of $\bm{R}$ in the complex plane
\begin{equation}  \label{eq:defCircularMean}
    \mean{\theta} = \angle \left( C + i S \right)
\end{equation}
In the example of \cref{fig:exampleCircularStats}, the resultant of the two angles is
\begin{equation}
    \bm{R} = \left(2 \cos (\theta_1), 0 \right)^{\intercal}
\end{equation}
with circular mean $\mean{\theta} = 0$, indicated by the green square in the figure.
This circular mean provides an appropriate measure for the mean direction, and an overbar over an angular variable will refer to the circular mean from here on.
We note that higher-order statistical measures can be defined as well, but are not further explored here.
Interested readers may refer to \cite{JammalamadakaSenGupta2001TopicsCircularStatistics} for an in-depth discussion of the topic.

\subsection{Phase difference at example scales: observations}  \label{sec:pvPhaseDiffObsExample}

\begin{figure}
    \centering
    \begin{subfigure}[b]{0.32\textwidth}
        \captionsetup{width=\extraCaptionWidth}
        \caption{$\phaseShift = - 3 \pi / 4, \bkappa_s$}
        \includegraphics[width=\textwidth]{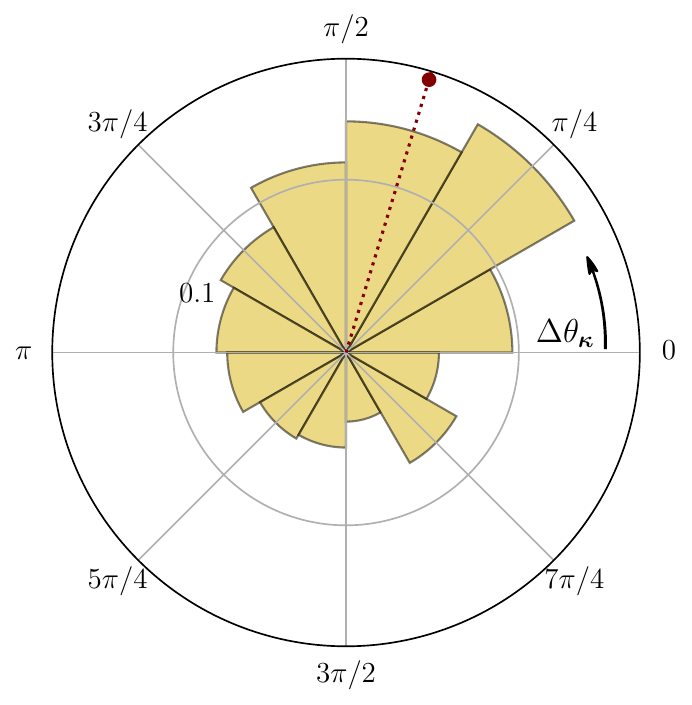}
        \label{fig:phaseHistogramsN075pi}
    \end{subfigure}
    \begin{subfigure}[b]{0.32\textwidth}
        \captionsetup{width=\extraCaptionWidth}
        \caption{$\phaseShift = - \pi / 4, \bkappa_s$}
        \includegraphics[width=\textwidth]{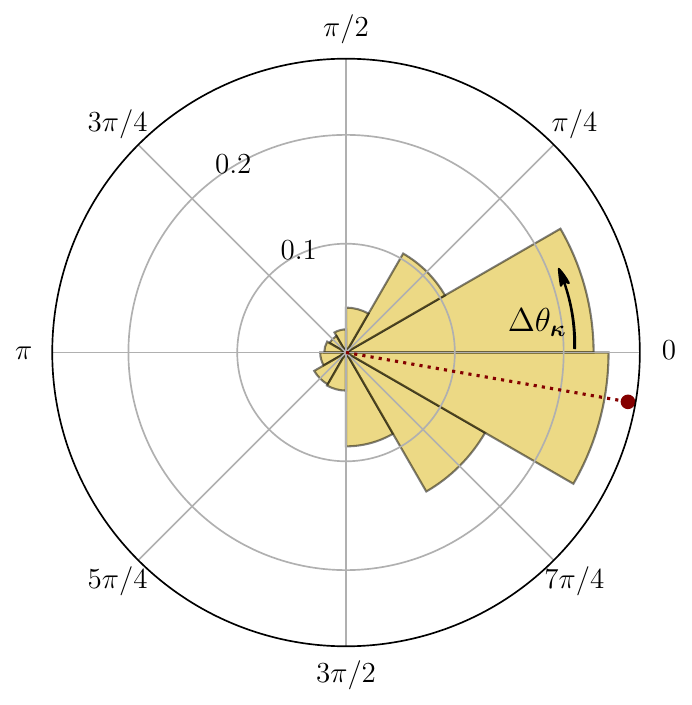}
        \label{fig:phaseHistogramsN025pi}
    \end{subfigure}
    \begin{subfigure}[b]{0.32\textwidth}
        \captionsetup{width=\extraCaptionWidth}
        \caption{$\phaseShift = + \pi / 2, \bkappa_r$}
        \includegraphics[width=\textwidth]{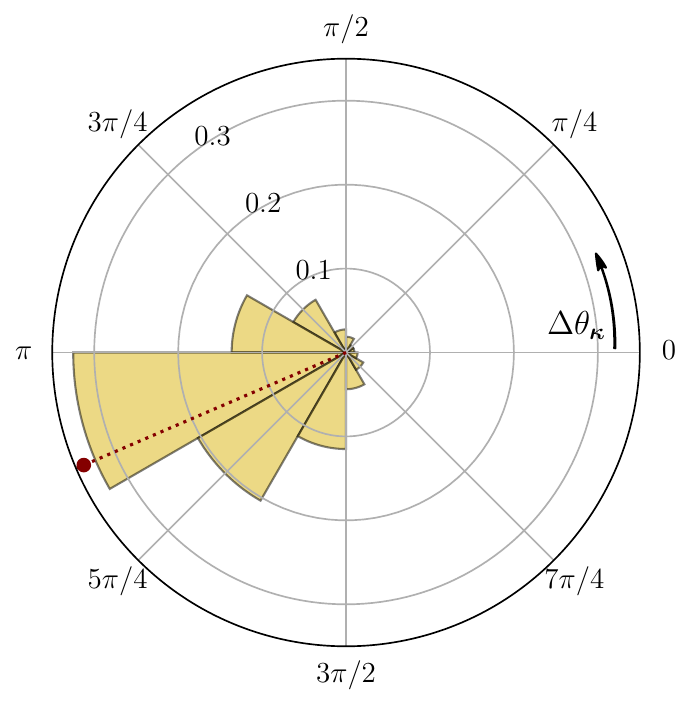}
        \label{fig:phaseHistogramsP05pi}
    \end{subfigure}
    \begin{subfigure}[b]{0.32\textwidth}
        \includegraphics[width=\textwidth]{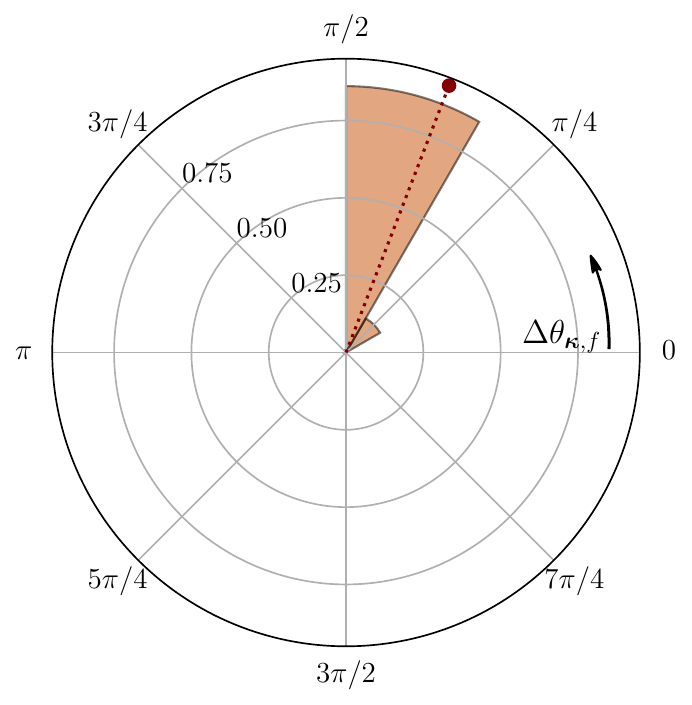}
    \end{subfigure}
    \begin{subfigure}[b]{0.32\textwidth}
        \includegraphics[width=\textwidth]{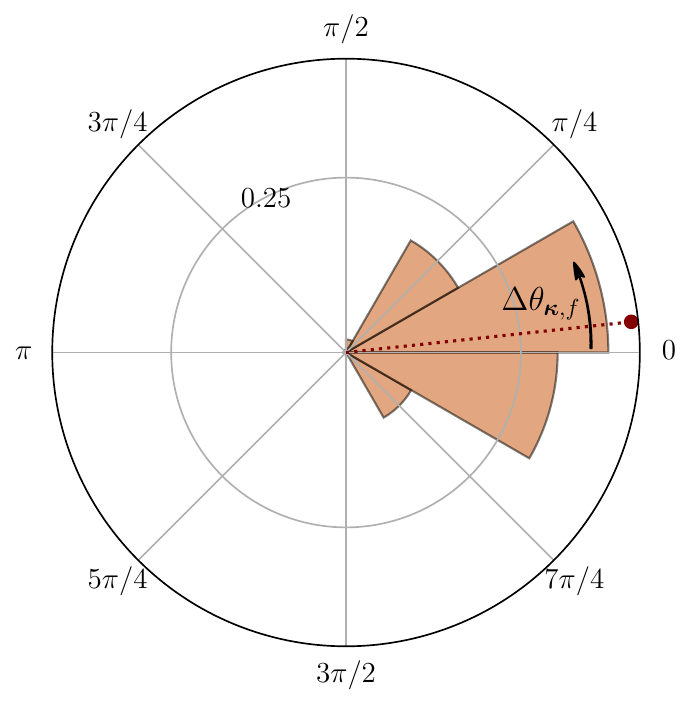}
    \end{subfigure}
    \begin{subfigure}[b]{0.32\textwidth}
        \includegraphics[width=\textwidth]{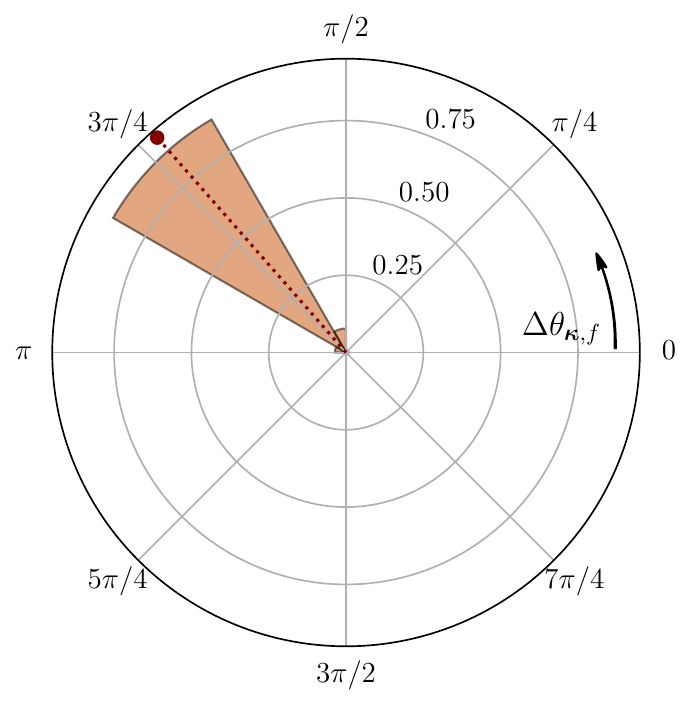}
    \end{subfigure}
    \begin{subfigure}[b]{0.32\textwidth}
        \includegraphics[width=\textwidth]{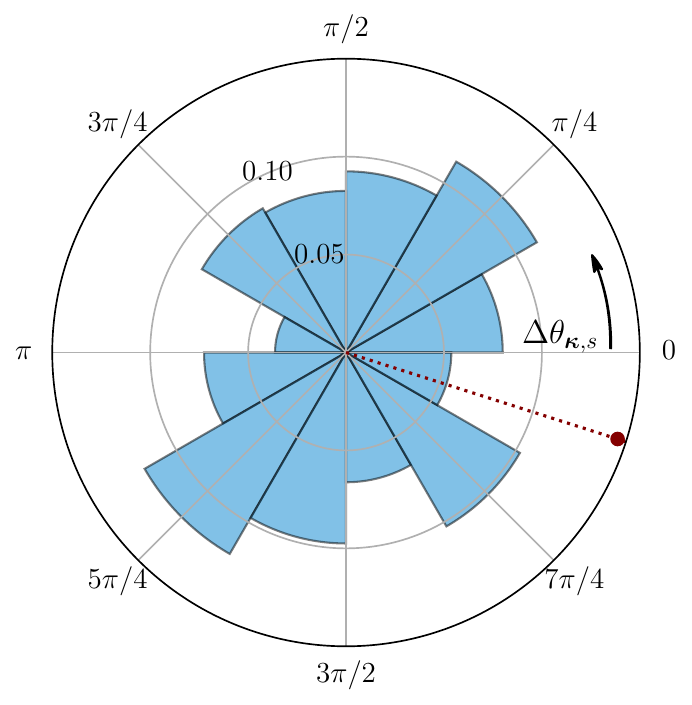}
    \end{subfigure}
    \begin{subfigure}[b]{0.32\textwidth}
        \includegraphics[width=\textwidth]{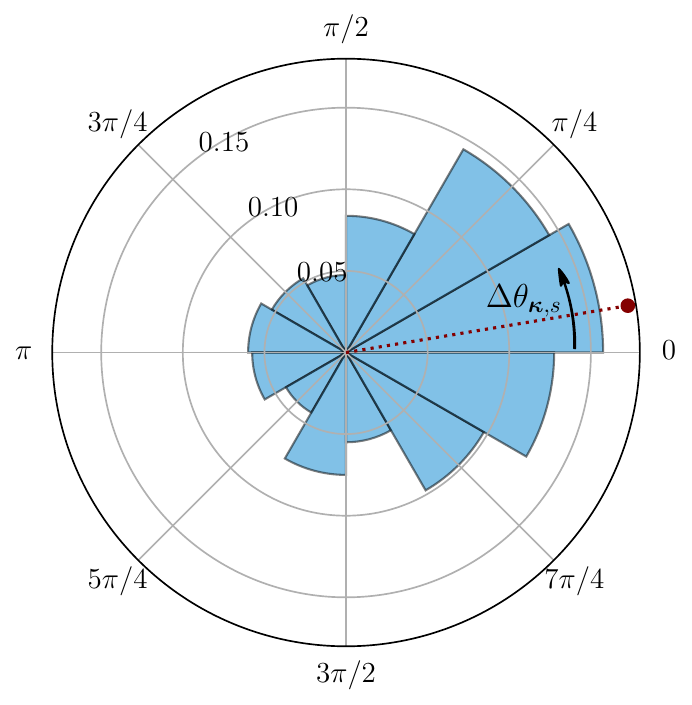}
    \end{subfigure}
    \begin{subfigure}[b]{0.32\textwidth}
        \includegraphics[width=\textwidth]{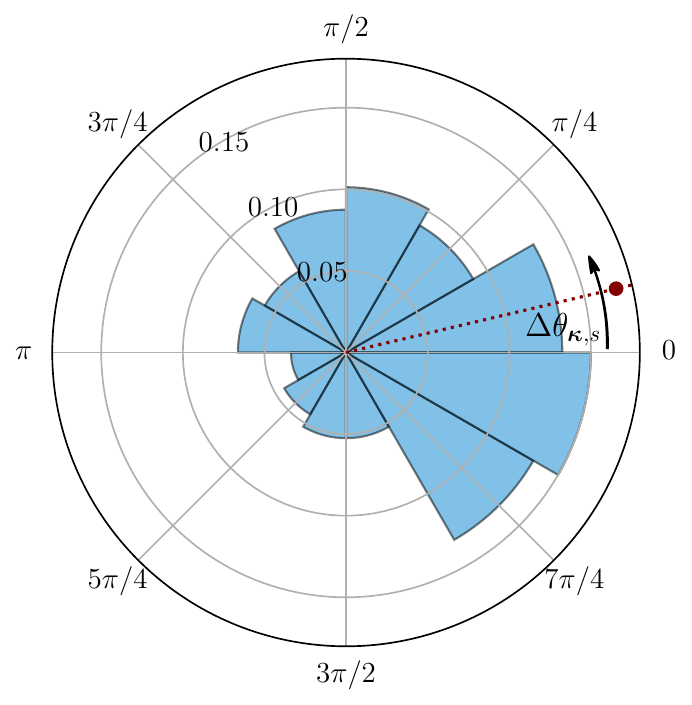}
    \end{subfigure}
    \begin{subfigure}[b]{0.32\textwidth}
        \includegraphics[width=\textwidth]{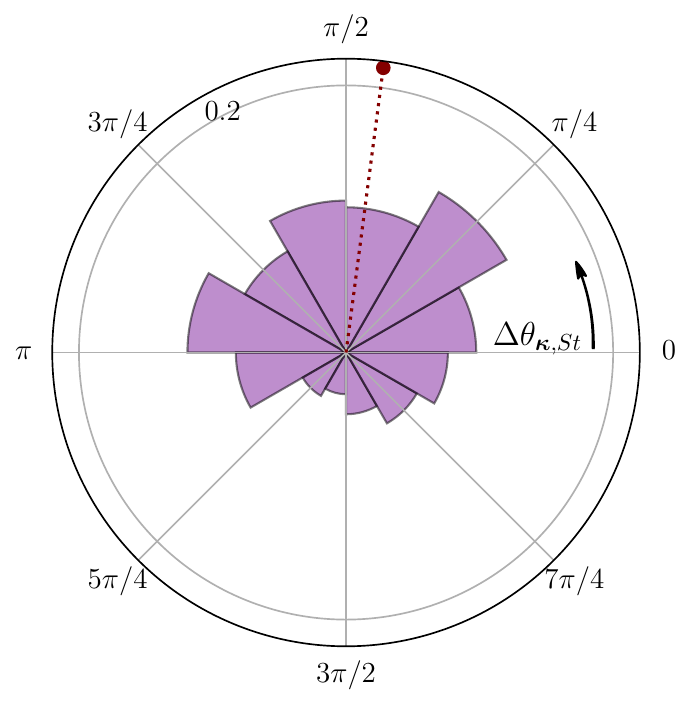}
    \end{subfigure}
    \begin{subfigure}[b]{0.32\textwidth}
        \includegraphics[width=\textwidth]{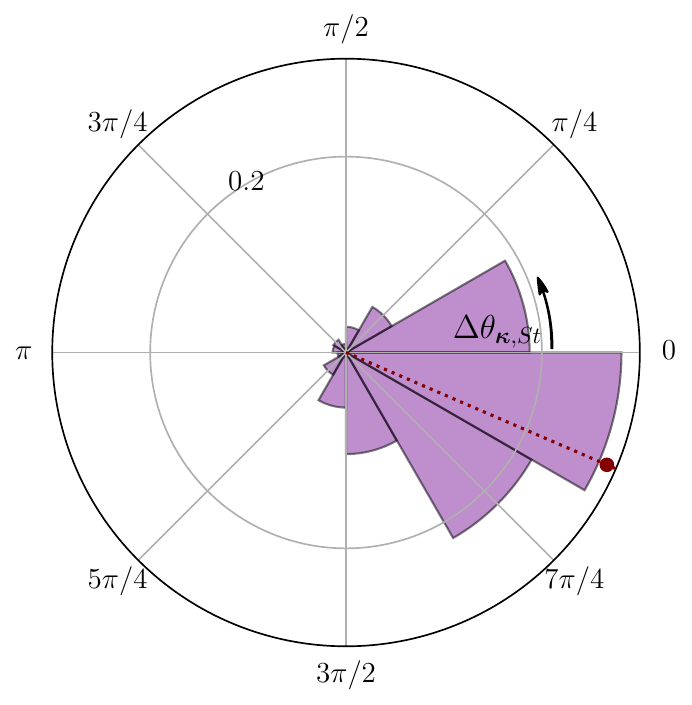}
    \end{subfigure}
    \begin{subfigure}[b]{0.32\textwidth}
        \includegraphics[width=\textwidth]{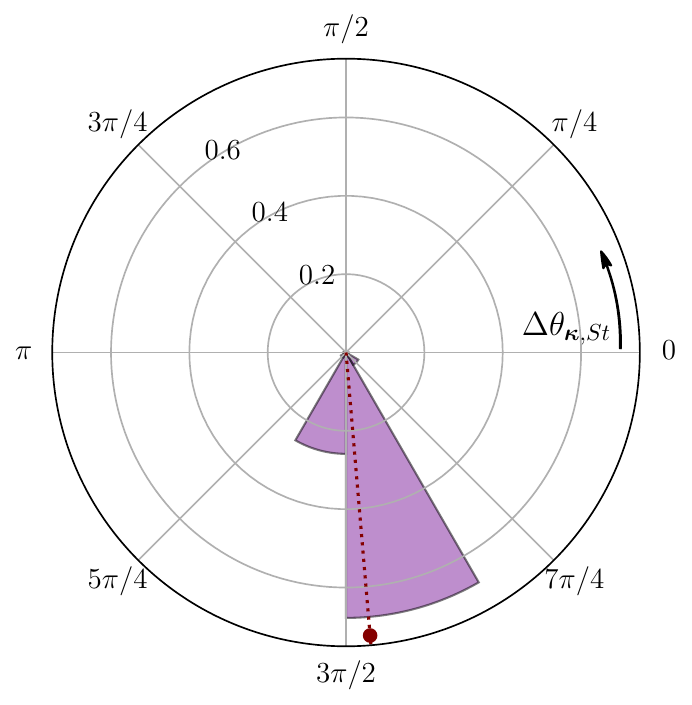}
    \end{subfigure}
    \caption{Histograms of the phase difference between wall pressure and transpiration. Each column represents a different phase shift and scale. The rows show from top to bottom: total, fast, slow, and Stokes pressure. The red symbol indicates the circular mean of each histogram.}
    \label{fig:phaseHistrograms}
\end{figure}

This section focuses on spatial scales $\bkappa_s$ and $\bkappa_r$ in the regime where they are dynamically relevant.
Select histograms of the phase difference $\Delta \theta_{\bkappa}$ are shown in \cref{fig:phaseHistrograms}.
Each column represents a different control configuration and scale, while each row shows a different pressure component.
The bin count of the phase difference is represented by the segment radius and the circular mean is indicated by the red circle.
The histograms further give a sense for the spread about the mean value, even if the circular variance is not explicitly quantified.

The total pressure is the measurable quantity in experiments and real-world applications, which makes the associated phase difference $\Delta \theta_{\bkappa}$ the most relevant one.
This phase difference is shown in the first row (yellow segments) of \cref{fig:phaseHistrograms} and varies significantly across control configurations and spatial scales.
For example, the mean phase difference at scale $\bkappa_s$ changes from a value of approximately $+\pi / 2$ for configuration N75 (left column) to $0$ (i.e. in-phase) for controller N25 (center column).
On the other hand, $v$ and $p$ in configuration P50 are out-of-phase at scale $\bkappa_r$ (right column).
The spread about the circular mean is also different in each case.
The spread is small for N25 and P50, which makes conclusions for these configurations statistically strong.
On the other hand, the spread for N75 is much larger, which renders the interpretation of this controller more ambiguous.

To understand the differences in mean values and spread across configurations, it is helpful to decompose the pressure into its components and study the associated phase relation separately.
The second row of \cref{fig:phaseHistrograms} (orange segments) shows the phase difference between the fast pressure and the wall transpiration $\Delta \theta_{\bkappa,f}$.
The mean phase difference varies in each case and appears to shift in tandem with $\phaseShift$.
For example, $\mean{\Delta \theta}_{\bkappa,f} \approx 0$ at $\phaseShift = - \pi / 4$ and increases to $\mean{\Delta \theta}_{\bkappa,f} \approx 3\pi / 8$ at $\phaseShift = -3 \pi / 4$.
It is also evident that all values in the time series are clustered around the circular mean, with the greatest spread observed for configuration N25.
In contrast, the phase difference between slow pressure and transpiration $\Delta \theta_{\bkappa, s}$ (blue segments in the third row) shows significant variation in all cases.
The three circular means have similar values, but their significance is likely reduced due to the large variance.
The last row of \cref{fig:phaseHistrograms} (purple segments) completes the picture and shows the phase difference between the Stokes pressure and the transpiration $\Delta \theta_{\bkappa,St}$.
The phase difference for negative $\phaseShift$ shows considerable spread around the mean, whereas all values are tightly clustered near the mean for controller P50.

\Cref{fig:pvStats,fig:phaseHistrograms} together offer insights into how the phase difference between total pressure and transpiration ($\Delta \theta_{\bkappa}$) is established.
Consider, for example, scale $\bkappa_r$ in control configuration P50.
The dominant Stokes and fast pressure (see \cref{fig:pvStatsP05pi}) add to a resultant vector in the complex plane that is close to out-of-phase with the transpiration.
The slow pressure adds a stochastic element to the resultant, but its magnitude is too small to significantly change the orientation of the pressure vector.
The fast and Stokes pressure thus establish the out-of-phase relationship observed for the total pressure in this case.
In contrast, the slow and Stokes pressure are dominant at $\bkappa_s$ and negative phase shifts and imprint the overall phase relation between transpiration and wall pressure.
Given the significant variation in the direction of the slow pressure, it remains somewhat unclear how controller N25 establishes a robust overall phase relationship.

\subsection{Phase of wall pressure components: domain of dependence}  \label{sec:domainDependencePwPhase}
The relative phase between wall pressure and transpiration for a given spatial scale and control configuration can vary significantly across pressure components.
This is most evident in the right column of \cref{fig:phaseHistrograms}, which shows scale $\bkappa_r$ at $\phaseShift = +\pi / 2$.
The present section analyzes the structure of the Greens function kernel and the analytical Stokes pressure solution, with a focus on the domain of dependence in the wall-normal direction and in spectral space.
The discussion concentrates on the wall pressure, particularly on how its phase is established.
These observations apply to turbulent channel flows in general and will be applied to the controlled flows in \cref{sec:pvPhaseDiffStrucExample} to explain the observations from the previous section.

The phase of the fast and slow wall pressure results from a convolution of the Green's function kernel with the associated source term
\begin{equation}  \label{eq:phaseFastSlowPressure}
\begin{aligned}
    \angle \four{p}_{\bkappa, \{f/s\}}(y_w, t)
    &= \angle \left \lbrace \int_{-1}^{1} G(y_w, a) \hat{f}_{\{f/s\}}(a, t) \, \mathrm{d} a \right \rbrace \\
    &= \angle \left \lbrace \int_{-1}^{1} \lvert G(y_w, a) \hat{f}_{\{f/s\}}(a, t) \rvert e^{i \left( \angle \hat{f}_{\{f/s\}}(a,t) + \pi \right) }  \mathrm{d} a \right \rbrace
\end{aligned}
\end{equation}
Recall from \cref{eq:GreensFuncPressure} that the Green's function kernel $G(y,a)$ is real-valued and negative throughout the channel, which is reflected in the additional $\pi$ phase above.
The integral expression in \cref{eq:phaseFastSlowPressure} shows that the phases of the fast and slow wall pressure are weighted wall-normal averages of $\angle \hat{f}_{\{f/s\}}(y)$.
The weight function is given by $\lvert G(y_w, a) \hat{f}_{\{f/s\}}(a) \rvert$, and differs for the fast and slow pressure due to the different source terms (see \cref{eq:pressurePoissonSplit}).
Both weight functions have compact support in $y$ for sufficiently large $\kappa$, since $G(y_w, a)$ is localized (see \cref{fig:greensFunctionKernels}).
The wall-normal support of the weight function for the fast pressure may be further restricted by the mean shear $S$, which is part of the source term.
This is especially relevant at low $\kappa$ when the magnitude of $G(y_w, a)$ decays slowly (see e.g. the profile for $\kappa = 1$ in \cref{fig:greensFunctionKernels}).
Despite their similarities in the wall-normal domain of dependence, the fast and slow pressure have very different spectral dependencies.
The source term of the fast pressure is linear, which implies that $\four{p}_{\bkappa, f}(y_w)$ and its phase only depend on the flow state at that same $\bkappa$.
Conversely, the source term of the slow pressure is nonlinear, so that its value and phase at the wall do not depend on the flow state at $\bkappa$ itself, but on all wavenumbers that are triadically consistent with $\bkappa$.
Robust phase relations between wall transpiration and fast or slow pressure can result if the pressure source term correlates with $\four{v}_{\bkappa}(y_w)$ over the wall-normal layer where the weight function is non-zero.

The Stokes pressure on the other hand is forced by the boundary condition.
The relation is linear, which implies that $\four{p}_{St,\bkappa}$ only depends on the flow state at that same wave number.
For sufficiently large $\kappa$, when the influence of one wall is negligible at the other, the Stokes pressure further depends on local wall information only.
In general, the Stokes pressure is a complex function of the wall transpiration gradients (see \cref{eq:pressurePoissonSplit,eq:solStokesPressure}), which makes its analysis challenging.
Substantial simplifications are possible if $\kappa$ is large enough, so that the two walls are approximately independent, and if the rate of change of the transpiration is large relative to the viscous terms.
These assumptions are reasonable for a realistic flow and enable approximating the Stokes pressure at the lower wall as
\begin{equation}
    \four{p}_{\bkappa,St}(y=-1, t) \approx - \frac{\hat{P}_{b}(t)}{\kappa \tanh(2 \kappa)} \approx \frac{1}{\kappa \tanh (2 \kappa)} \pd{\hat{v}_{\bkappa}}{t}(y=-1,t)
\end{equation}
and a similar expression can be derived for the upper wall as well.
The phase of the Stokes pressure now depends on the temporal frequency content of $\partial \hat{v}_{\bkappa}/ \partial t$
\begin{equation}  \label{eq:phaseStokesPressure}
    \angle \four{p}_{\bkappa, St}(y_w, t) = \angle \left \lbrace \int_{-\infty}^{\infty} -i \omega \four{v}_{\bkappa}(y_w, \omega) \, e^{-i \omega t} \, \mathrm{d} \omega \right \rbrace
\end{equation}
where the the sign of the complex exponential in \cref{eq:phaseStokesPressure} is inverted relative to the Fourier series in \cref{eq:fourierDecomposition}, as is common practice in the stability literature.
Robust phase relations between wall transpiration and Stokes pressure can result if the temporal frequency content in \cref{eq:phaseStokesPressure} is approximately sparse.

\subsection{Phase difference at example scales: underlying structure}  \label{sec:pvPhaseDiffStrucExample}
This section combines pressure source term data with the analytical insights from \cref{sec:domainDependencePwPhase} to explain the phase relations observed in \cref{sec:pvPhaseDiffObsExample}.
We first consider the fast pressure, which established a robust phase relative to the transpiration across all example configurations (second row in \cref{fig:phaseHistrograms}).
This robust phase relation is an immediate consequence of the control and the structure of the fast pressure solution:
The opposition control scheme, \cref{eq:varyingPhaseOc}, relates the wall transpiration to the sensor measurement and thus constrains the wall-normal profile of $\four{v}_{\bkappa}$ between $y_w$ and $y_d$.
The phase of the fast pressure at the wall corresponds to a weighted average of $\angle \four{v}_{\bkappa}$ according to \cref{eq:phaseFastSlowPressure}.
Since the weighting function decays quickly away from the wall, the phase of the fast wall pressure is determined by the wall-normal velocity in the constrained region, and a robust phase relation $\Delta \theta_{\bkappa,f}$ can be established.

\begin{figure}
    \centering
    \begin{subfigure}[b]{0.48\textwidth}
        \caption{$\phaseShift = - 3\pi / 4, \bkappa_s$}
        \includegraphics[width=\textwidth]{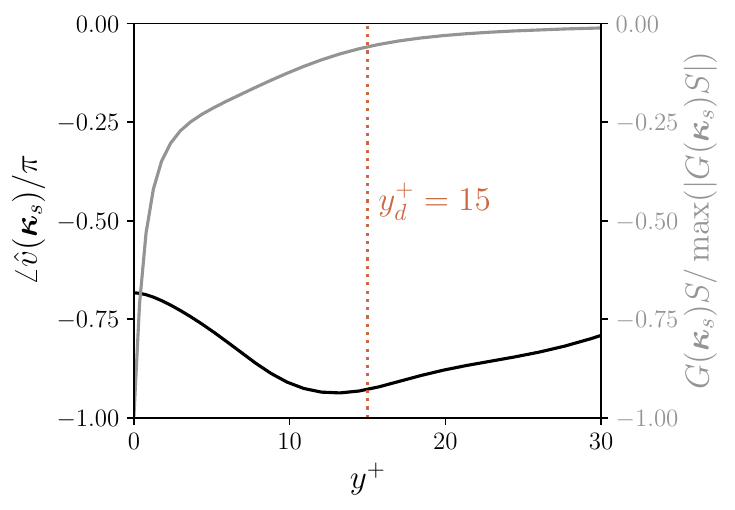}
    \end{subfigure}
    \hfill
    \begin{subfigure}[b]{0.48\textwidth}
        \caption{$\phaseShift = - \pi / 4, \bkappa_s$}
        \includegraphics[width=\textwidth]{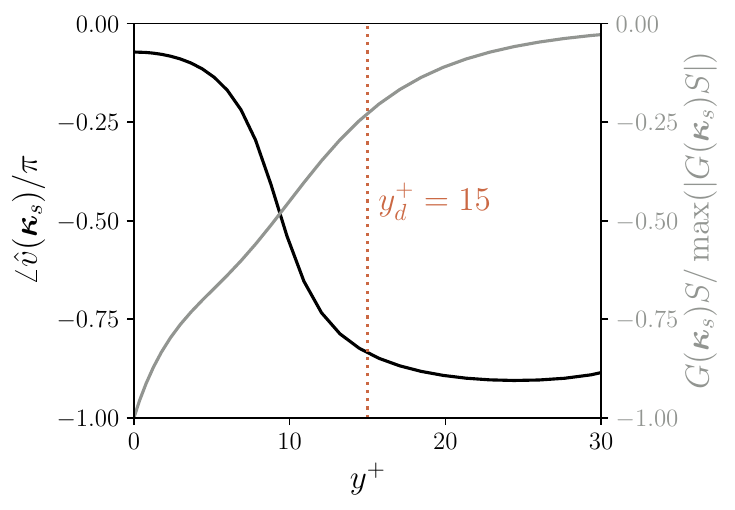}
    \end{subfigure}
    \caption{Representative instantaneous phase profile $\angle \four{v}$ (solid black line) and product of Green's function kernel $G$ and mean shear $S$ normalized by the maximum value (solid gray line). The two plots show the same spatial scale $\bkappa_s$ at different $\phaseShift$.}
    \label{fig:phaseGreensFuncFastPres}
\end{figure}

This point is illustrated in \cref{fig:phaseGreensFuncFastPres} for scale $\bkappa_s$ and two different phase shifts:
$\phaseShift = -3 \pi / 4$, which resulted in a small spread about the circular mean, and $\phaseShift = -\pi / 4$, which had a larger variance.
For $\phaseShift = - 3 \pi / 4$, the normalized weighting function (gray curve) is indeed close to zero for $y>y_d$, demonstrating that the phase of the fast pressure at the wall is determined by $y \in [y_w, y_d]$ where $\angle \four{v}_{\bkappa}$ is closely constrained.
In contrast, the normalized weighting function for $\phaseShift = -\pi / 4$ is larger for $y>y_d$, allowing the fast pressure at the wall to acquire phase from regions where $\angle \four{v}_{\bkappa}$ is unconstrained.
This leads to the larger observed spread in \cref{fig:phaseHistrograms}.
It is important to note that both plots in \cref{fig:phaseGreensFuncFastPres} show the same spatial scale $\bkappa_s$ at two different $\phaseShift$.
The Green's function kernel is identical for the two cases and the difference in the weighting function comes from the mean shear, which is greater in the drag-increasing case.
The phase relation between fast wall pressure and transpiration is thus particularly robust for small scales (due to the fast decay of the Green's function kernel) or drag-increased flows (due to the large mean shear close to the wall), as confirmed by the histogram for scale $\kappa_r$ and $\phaseShift = +\pi / 2$.

The source term of the slow pressure is the divergence of the nonlinear advection term.
The nonlinearity can further be split into an irrotational and solenoidal part, each of which has a distinct dynamical role \cite{ChorinMarsden1993MathematicalIntroductionFluid}.
Only the irrotational component enters the source term of the slow pressure and determines the nonlinear effects in the pressure field.
The solenoidal part on the other hand governs the nonlinear effects in the evolution of the velocity field and represents the forcing in input-output formulations of the Navier-Stokes equations \citep{McKeonSharma2010criticallayerframework}.
Due to its quadratic nonlinearity, the
source term of the slow pressure depends on all scales that are triadically consistent with $\bkappa$, but not on $\bkappa$ itself (see \cref{eq:pressurePoissonSplit}).
With disjoint spectral domains of dependence, it is unlikely that the weighted integral of the source term is strongly correlated with $\four{v}_{\bkappa}(y_w)$ and that a robust phase relation can be established.
The large variance in $\Delta \theta_{\bkappa, s}$, as shown in the histograms of \cref{fig:phaseHistrograms}, supports this interpretation.
The phase difference between transpiration and slow pressure can possibly be modeled as a stochastic process, but the sample size is too small to draw any meaningful conclusion about a suitable probability distribution.

\begin{figure}
    \centering
    \begin{subfigure}[t]{0.47\textwidth}
        \caption{$\phaseShift = - \pi / 4, \bkappa_s$}
        \includegraphics[width=\textwidth]{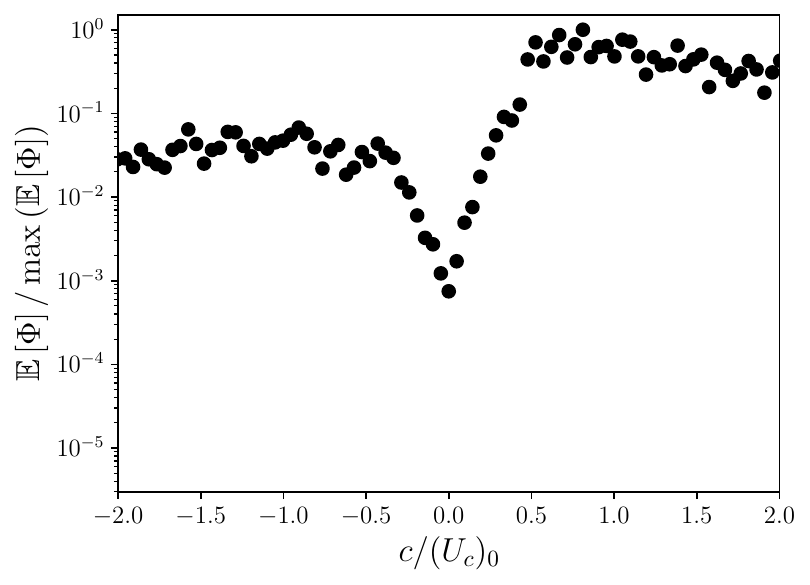}
    \end{subfigure}
    \hfill
    \begin{subfigure}[t]{0.47\textwidth}
        \caption{$\phaseShift = +\pi / 2, \bkappa_r$}
        \includegraphics[width=\textwidth]{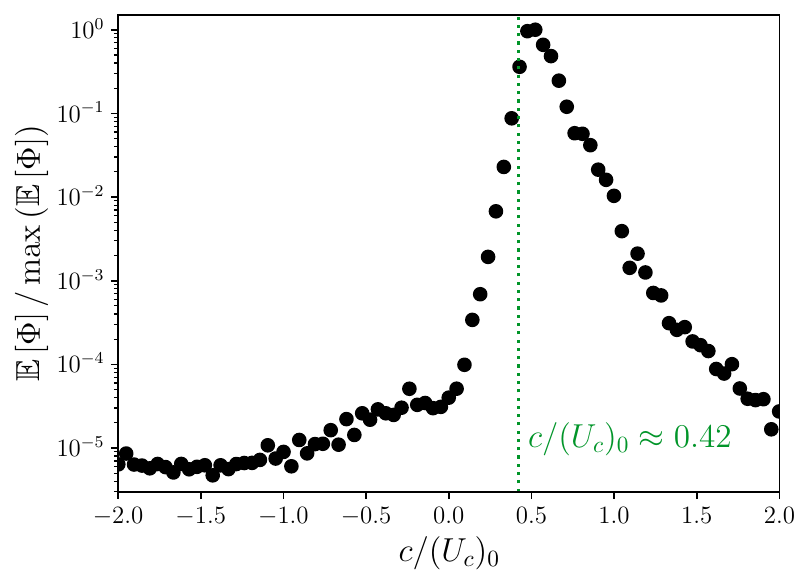}
        \label{fig:timeSpectrumP05pi}
    \end{subfigure}
    \caption{Power spectral density estimate $\mathbb{E} \left[ \Phi \right]$ of $\partial \four{v}_{\bkappa} / \partial t$ for two different scales and phase shifts. Both spectra are normalized by their respective maximum values. The dotted vertical line in \cref{fig:timeSpectrumP05pi} indicates the wavespeed of the amplified eigenmode reported in \cite{ToedtliYuMcKeon2020origindragincrease}.}
    \label{fig:timeSpectra}
\end{figure}

Finally, the phase difference between Stokes pressure and transpiration in \cref{fig:phaseHistrograms} is clustered about the circular mean for $\phaseShift = +\pi / 2$, with greater variance in the other cases.
\Cref{eq:phaseStokesPressure} suggests that $\Delta \theta_{\bkappa, s}$ is approximately determined by the temporal frequency content of $\partial \four{v}_{\bkappa} / \partial t$ and that a robust phase relation can be established when the frequency content is sparse.
To validate this, we analyze the temporal frequency content of two example control configurations.
In this specific instance, the DNS is run at a constant time step to enforce a uniform sampling rate and the temporal frequency content is estimated using Welch's method.
Each time series is divided into segments of $M$ samples, which are multiplied by a Hamming window
\begin{equation}
        w[n] = 0.54 - 0.46 \cos \left( \frac{2 \pi n}{M-1} \right), \quad 0 \leq n \leq M-1
\end{equation}
to enforce periodicity.
Consecutive segments have a $50\%$ overlap and the robustness of the results was assessed through a convergence study.
The results for the two example configurations is shown in \cref{fig:timeSpectra}:
the left figure shows the temporal frequency content of $\partial \four{v}_{\bkappa} / \partial t$ at scale $\bkappa_s$ for $\phaseShift = - \pi / 4$ (larger variance in $\Delta \theta_{\bkappa, St}$), while the right figure presents scale $\bkappa_r$ and $\phaseShift = + \pi / 2$ (small variance).
Each of the power spectra is normalized with its maximum value and only a subset of the resolved phase speeds $c = \omega / k_x$ is shown for clarity.

For controller N25, the spectral content of $\partial \hat{v}_{\bkappa} / \partial t$ is distributed over a large range of positive phase speeds.
The rate of change of the wall transpiration is not restricted to wave speeds below the centerline velocity, as is typically the case for the velocity fluctuations in a canonical turbulent channel flow \citep{BourguignonTroppSharmaEtAl2014Compactrepresentationwall}.
This difference can likely be attributed to the time derivative, which amplifies higher frequencies, and to the closed loop control, which introduces additional dynamics.
The temporal frequency content is broad-band and $\angle \four{p}_{St}$ therefore depends on multiple temporal frequencies.
This can decrease the correlation to $\four{v}_{\bkappa}$ and leads to the broader range of phase differences observed in \cref{fig:phaseHistrograms}.

In contrast, the temporal frequency content of $\partial \hat{v}_{\bkappa} / \partial t$ for configuration P50 and scale $\bkappa_r$ has a pronounced peak.
\Cref{eq:phaseStokesPressure} suggests that the phase difference between Stokes pressure and transpiration is $-\pi / 2 + n 2\pi$ when a single frequency dominates, consistent with the histogram in \cref{fig:phaseHistrograms}.
The dominant wave speed roughly aligns with the phase speed of the amplified eigenvalue (green dotted line), which \cite{ToedtliYuMcKeon2020origindragincrease} computed for this configuration under idealized conditions (uncontrolled mean, single mode in isolation).
The temporal frequency content thus provides further evidence that the amplified eigenmode drives the flow response at $\bkappa_r$ and $\phaseShift = +\pi / 2$ and leaves a clear imprint not only in the velocity field, but also in the Stokes pressure.

\subsection{Phase difference at all scales}  \label{sec:pvPhaseDiffAllScales}
\begin{figure}
    \centering
    \begin{subfigure}[t]{0.45\linewidth}
        \caption{$\phaseShift = -3 \pi / 4$}
        \includegraphics[width=\linewidth]{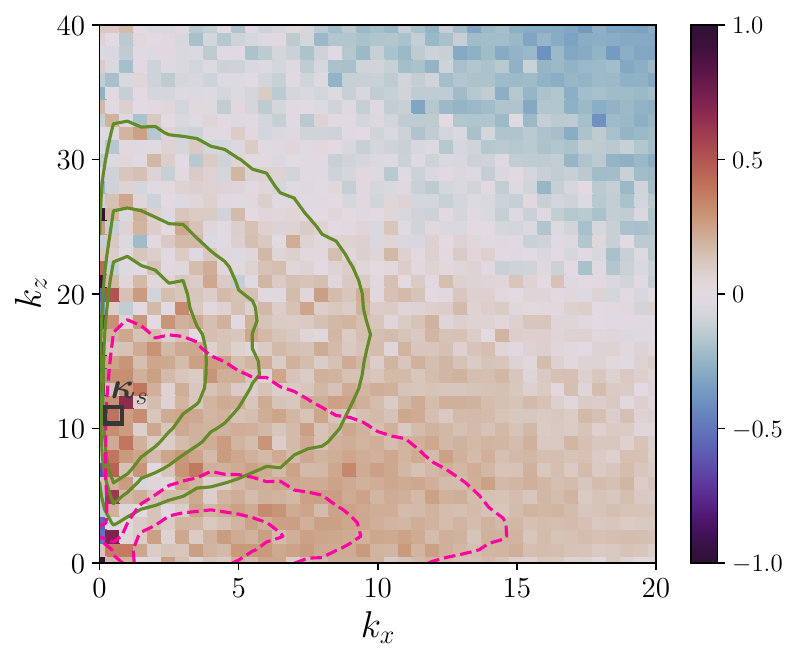}
        \label{fig:phaseDiffN075piAllScales}
    \end{subfigure}
    \begin{subfigure}[t]{0.45\linewidth}
        \caption{$\phaseShift = - \pi / 4$}
        \includegraphics[width=\linewidth]{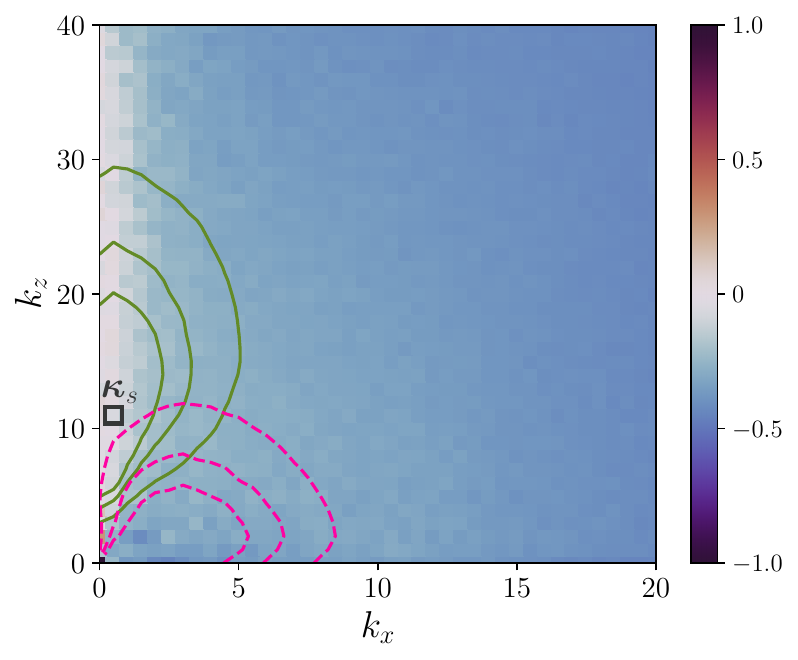}
        \label{fig:phaseDiffN025piAllScales}
    \end{subfigure}
    \begin{subfigure}[t]{0.45\linewidth}
        \caption{$\phaseShift = + \pi / 2$}
        \includegraphics[width=\linewidth]{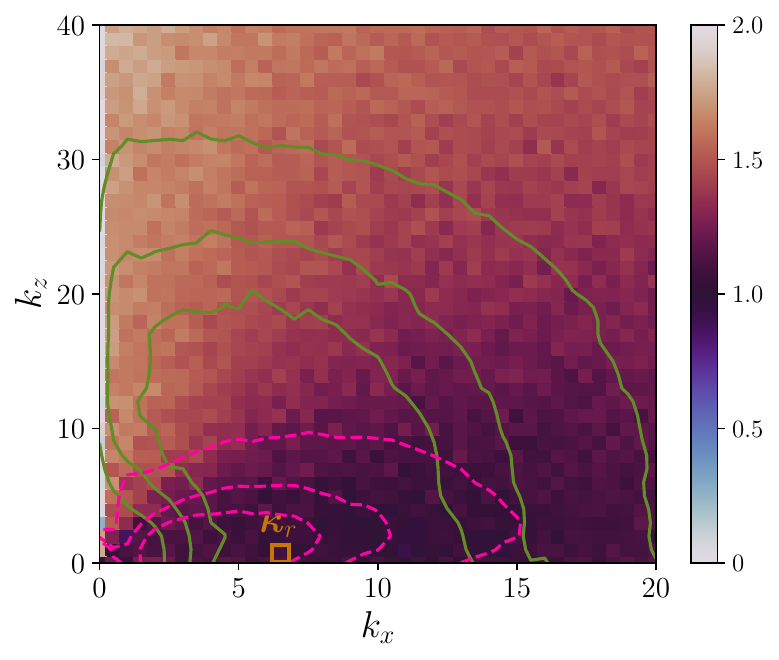}
        \label{fig:phaseDiffP05piAllScales}
    \end{subfigure}
    \caption{Phase difference spectra $\mean{\Delta \theta}_{\bkappa} / \pi$ of the example control configurations. The contour lines are the same as in \cref{fig:vSpatialSpectraAct,fig:pSpatialSpectra} and label $\Phi_{qq}^+ / \max(\Phi_{qq}^+) = (0.15, 0.3, 0.45)$ for $v$ (solid green lines) and $p$ (dashed pink lines).}
    \label{fig:spectraPhaseAllScales}
\end{figure}

With a better understanding of how individual wall pressure components establish their phase relative to the transpiration and how the relative magnitude of the components determines the overall phase of the wall pressure, we next discuss the mean phase difference across all scales.
\Cref{fig:spectraPhaseAllScales} shows the mean phase difference $\mean{\Delta \theta}_{\bkappa}$ between total pressure and transpiration for the example control configurations, over the same range of scales as the spectra in \cref{fig:vSpatialSpectraAct,fig:pSpatialSpectra}.
The interpretation of these spectra requires care:
The phase of a Fourier coefficient with (close to) zero magnitude is ill-defined, which implies that the phase difference at scale $\bkappa$ is only meaningful if both $\four{p}_{\bkappa}$ and $\four{v}_{\bkappa}$ are non-zero.
The non-zero regions have to be obtained from the corresponding spectra, and are indicated by the solid green ($v$) and dashed pink contour lines ($p$) introduced earlier.

The average phase difference for the drag-reducing controller N25 is shown in \cref{fig:phaseDiffN025piAllScales}.
The highlighted entry at scale $\bkappa_s$ corresponds to the circular average of the histogram shown in \cref{fig:phaseHistogramsN025pi},
and similar histograms could be constructed for all other scales.
Analogous to scale $\bkappa_s$, the phase difference at scales with large control input is approximately zero, indicating that transpiration and pressure at the streak scales are in-phase when $\phaseShift = - \pi / 4$.
On the other hand, the phase relation between $v$ and $p$ for controller N75 (\cref{fig:phaseDiffN075piAllScales}) is less clear.
The average phase difference for the energetic streak scales is between $0$ and $+\pi / 2$, with significant variability from scale to scale.
As shown in \cref{fig:phaseHistogramsN075pi}, the circular variance for this control configuration is larger compared to the other ones, which makes the value of the circular mean less conclusive and possibly contributes to the observed spread across $\bkappa$.
Finally, transpiration and pressure at the roller scales are approximately out-of-phase for $\phaseShift = + \pi / 2$, as can be seen in \cref{fig:phaseDiffP05piAllScales}.
This is again consistent with the earlier observations for example scale $\bkappa_r$.

The above discussion highlights that the example scales $\bkappa_s$ and $\bkappa_r$ are representative of their scale families also in terms of the mean phase difference.
This suggests that the relative importance of the pressure components and the robustness of the phase relations through the domain of dependence apply broadly and underlie the observed phase difference spectra of each scale family.

\subsection{Connection to kinetic energy}  \label{sec:connectionKinEn}
The scale-by-scale phase difference between wall pressure and transpiration, $\Delta \theta_{\bkappa}$, is further related to the energetics of the actuation, as will be shown next.
The starting point is the rate of change of the volume-integrated kinetic energy
\begin{equation}  \label{eq:kineticEnergy}
    \td{}{t} \int_{-1}^{1} \avgxz{e} \, \mathrm{d}y
        = - 2 G_x U_b
        - \frac{1}{\reBulk} \int_{-1}^{1} \avgxz{\left( \frac{\partial u_i}{\partial x_j} \right)^2} \mathrm{d}y
        - \left[ \frac{\avgxz{v^3}}{2} + \avgxz{pv} \right]_{y=-1}^{y=1}
\end{equation}
where $e = u_i u_i / 2$.
The first two terms on the right-hand side are analogous to a canonical turbulent channel flow and represent the work done by the mean pressure gradient and the viscous dissipation.
The last two terms in square brackets are the contributions from the wall transpiration and quantify the injection of kinetic energy (first term) and the work done against the local pressure (second term).

The transpiration-pressure work at each wall can further be written in terms of Fourier modes
\begin{equation}  \label{eq:pvWork}
\avgxz{pv}(y_w) = 2
    \sum_{l} \sum_{m \in \mathcal{M}(l)} \absVal{\four{p}_{\bkappa}(y_w)} \, \absVal{\four{v}^{\ast}_{\bkappa}(y_w)} \, \cos(\Delta \theta_{\bkappa})
\end{equation}
where the superscript $\ast$ denotes a complex-conjugate and the index set $\mathcal{M}$ contains all integers for $l \neq 0$ and only positive integers for $l = 0$.
\Cref{eq:pvWork} shows that the instantaneous scale-by-scale phase difference between velocity and pressure determines whether a specific scale injects or extracts kinetic energy from the flow through the pressure work term.
Specifically at the lower wall, scales with $ \absVal{\Delta \theta_{\bkappa}} \leq \pi / 2$ inject energy into the flow, while scales with $\absVal{\Delta \theta_{\bkappa}} > \pi / 2$ extract energy from it.
The net effect results from the sum over $\bkappa$ and the energetic behavior can vary in spectral space, so that cancellations across scales are in principle permissible.
Depending on the spread of $\Delta \theta_{\bkappa}$ over time, it is further possible that a specific scale acts as an energy sink and source for different parts of the time series.
\Cref{fig:phaseHistogramsN075pi} shows that scale $\bkappa_s$ in case N75 is an example of this behavior.

\Cref{fig:phaseHistrograms,fig:spectraPhaseAllScales} further demonstrate that the streak scales inject energy in configuration N25, while the roller scales extract energy through the pressure term in configuration P50.
However, the sign of the pressure work term does not say much about the overall effect of the transpiration on the flow.
Configuration P50 has increased turbulence intensity and a higher drag relative to the uncontrolled flow, even though the pressure work term extracts kinetic energy.
Conversely, configuration N25 with positive pressure work term has less turbulence intensity and lower drag.
Two comments are important to interpret the disconnect between the sign of the pressure work and the effect on the flow:
First, the pressure work is only one of two transpiration-related terms in \cref{eq:kineticEnergy} and their sum may have the opposite sign.
The transpiration-pressure phase relation does therefore not fully characterize the energetics of the actuation.
Second, the data only consider the steady-state of the controlled flow and the energetics of individual scales may be different during the transient stage.
For example, the energetic roller scales in configuration P50 are associated with an amplified eigenvalue of the linearized Navier-Stokes operator.
Their amplification in the transient stage, where the structural flow changes occur, is not part of the above discussion and beyond the scope of this work.

\section{Scale response to control and connection to pv phase}  \label{sec:comparisonPhaseDrag}
This section relates the phase difference between $p$ and $v$ to scale suppression or amplification under control.
\Cref{sec:scaleContribCf} defines a scale-by-scale drag contribution metric and \cref{sec:scaleByScaleDragChange} analyzes how this metric changes across the example control configurations.
The change in scale-by-scale drag contribution provides a basis to objectively define scale suppression or amplification and \cref{sec:connectionDragPvPhase} shows that both possible scale responses coincide with distinct mean phase differences between wall pressure and transpiration.

\subsection{Scale-by-scale drag contribution}  \label{sec:scaleContribCf}
The friction coefficient $c_f$ (see \cref{eq:defDr}) of the present flow configuration can be expressed as the sum of a laminar ($c_{f,l}$) and a turbulent contribution ($c_{f,t}$) \citep[see][for details]{FukagataIwamotoKasagi2002ContributionReynoldsstress}.
The turbulent contribution results from a weighted wall-normal integral of the Reynolds stresses and can further be decomposed into scale-by-scale contributions $c_{f,t\bkappa}$
\begin{equation}  \label{eq:frictionCoeffFIK}
\begin{aligned}
    c_f &= c_{f,l} + c_{f,t} = \frac{12}{\reBulk} + 12 \int_{-1}^{1} y (\mean{\fluct{u} \fluct{v}}) \, \mathrm{d} y \\
    &= \frac{12}{\reBulk} + \sum_{k_x} \sum_{k_z} \underbrace{12 \int_{-1}^1 y \, \Re \{ \four{u}_{\bkappa} \four{v}^{\ast}_{\bkappa} \} \, \mathrm{d}y}_{=c_{f,t \bkappa}}
\end{aligned}
\end{equation}
This flow identity is valid as long as the assumptions outlined in \cite{FukagataIwamotoKasagi2002ContributionReynoldsstress} are met and the velocities are nondimensionalized by twice the bulk velocity.
The sign of $c_{f, t \bkappa}$ results from the phase relation between $\four{u}_{\bkappa}$ and $\four{v}_{\bkappa}$ across $y$.
In a canonical turbulent channel flow, $\four{u}_{\bkappa}$ and $\four{v}_{\bkappa}$ are out-of-phase on average for $y\leq 0$ (and vice-versa for $y>0$), so that the scale-by-scale turbulent drag contribution is positive.
Control alters $\four{u}_{\bkappa}$ and $\four{v}_{\bkappa}$ and possibly their relative arrangement, so that $c_{f, t \bkappa}$ of a controlled flow can in principle have either sign.
A negative sign would imply that control fundamentally alters the phase relation between $\four{u}_{\bkappa}$ and $\four{v}_{\bkappa}$ so that the Reynolds stresses become an accelerating force.

\begin{figure}
    \centering
    \begin{subfigure}[b]{0.45\linewidth}
        \captionsetup{width=\extraCaptionWidth}
        \caption{Uncontrolled ($\frac{\sum_{\Gamma} \tau_{t \bkappa}}{\tau_{t}} = 93.3 \% $)}
        \includegraphics[width=\textwidth]{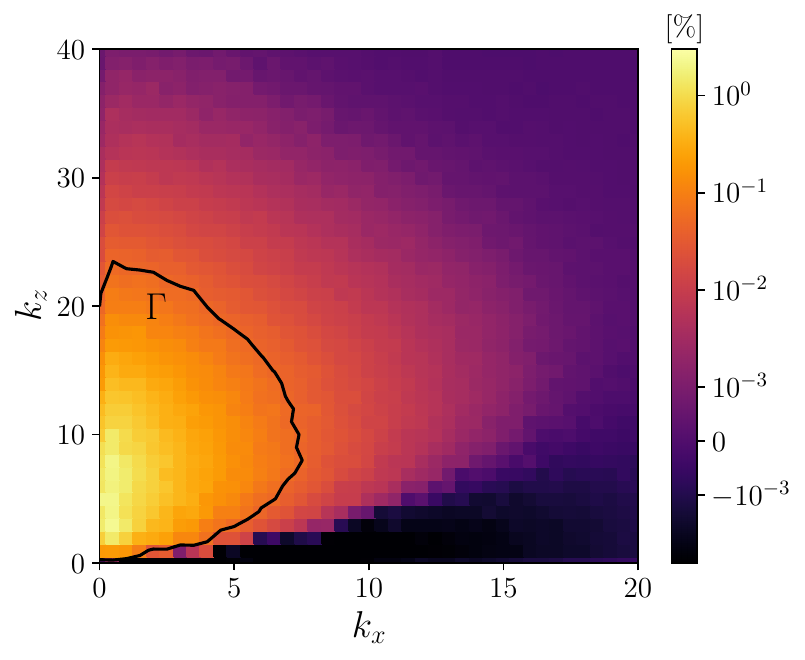}
        \label{fig:turbulentDragContribRef}
    \end{subfigure}
    \hspace{0.1in}
    \begin{subfigure}[b]{0.45\linewidth}
        \captionsetup{width=\extraCaptionWidth}
        \caption{$\phaseShift = - 3 \pi / 4$ ($\frac{\sum_{\Gamma} c_{f, t \bkappa}}{c_{f,t}} = 84.8 \%$)}
        \includegraphics[width=\textwidth]{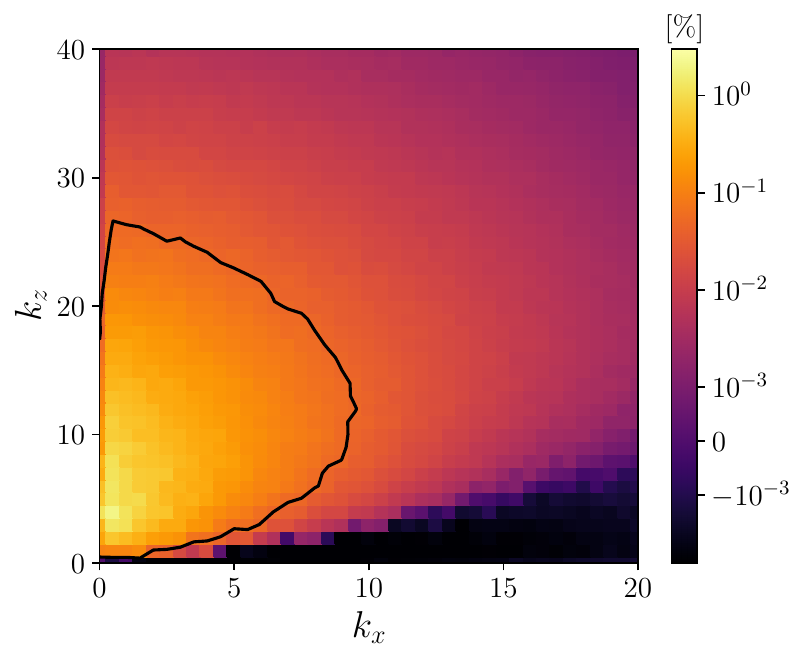}
        \label{fig:turbulentDragContribN075pi}
    \end{subfigure}
    \begin{subfigure}[b]{0.45\linewidth}
        \captionsetup{width=\extraCaptionWidth}
        \caption{$\phaseShift = -\pi / 4$ ($\frac{ \sum_{\Gamma} c_{f, t \bkappa}}{c_{f,t}} = 94.4 \%$)}
        \includegraphics[width=\linewidth]{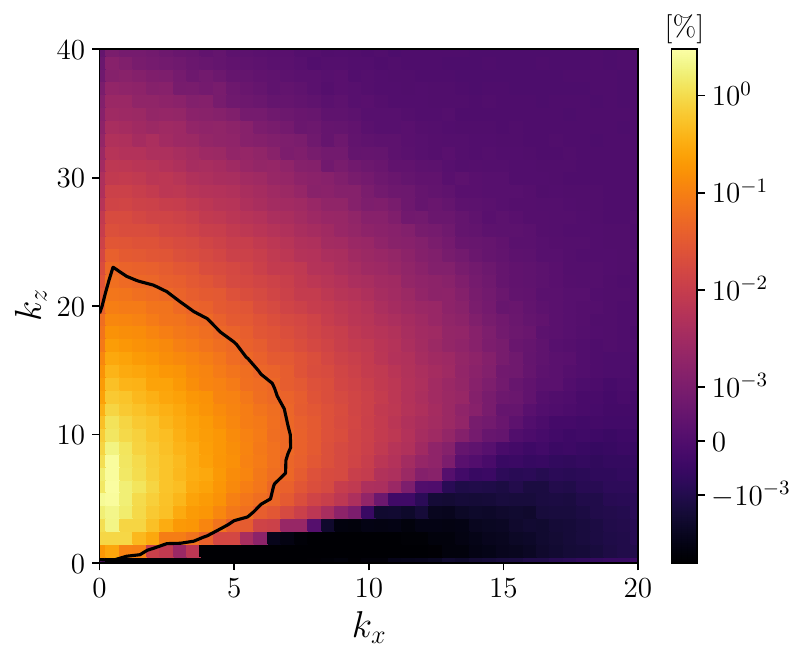}
        \label{fig:turbulentDragContribN025pi}
    \end{subfigure}
    \hspace{0.1in}
    \begin{subfigure}[b]{0.45\linewidth}
        \captionsetup{width=\extraCaptionWidth}
        \caption{$\phaseShift = +\pi / 2$ ($\frac{\sum_{\Gamma} c_{f, t \bkappa}}{c_{f,t}} = 80.2 \%$)}
        \includegraphics[width=\linewidth]{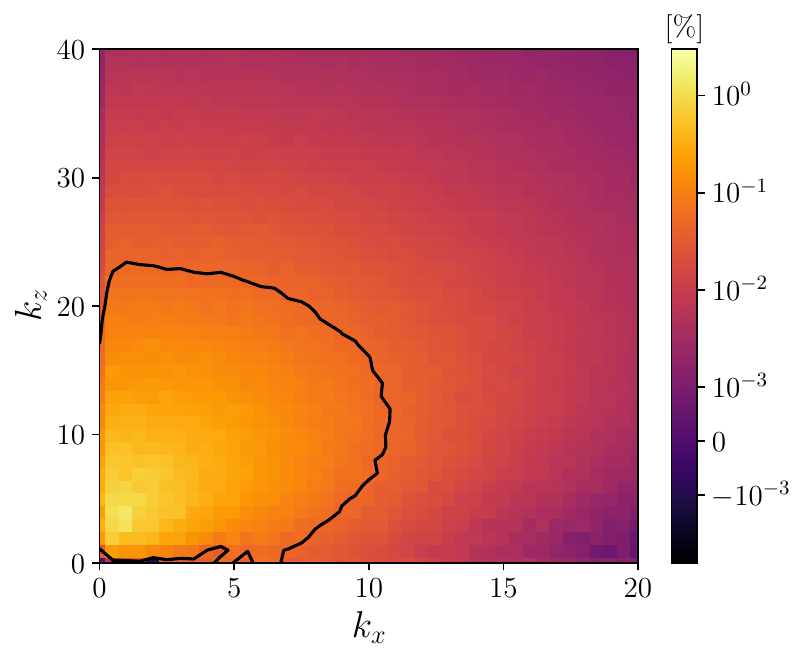}
        \label{fig:turbulentDragContribP05pi}
    \end{subfigure}
    \caption{Relative turbulent drag contribution for the uncontrolled reference flow (\cref{fig:turbulentDragContribRef}) and various controlled flows (\cref{fig:turbulentDragContribN075pi,fig:turbulentDragContribN025pi,fig:turbulentDragContribP05pi}). The spatial scales contained within the black solid contour $\Gamma$ contribute at least $0.05\%$ of the turbulent drag and sum to the total value indicated in brackets.}
    \label{fig:turbulentDragContrib}
\end{figure}

The mass flow rate (or, equivalently, $Re_{b}$) of the present channel flow is fixed, which implies that $c_{f,l}$ is constant across configurations.
Any change in friction coefficient (and wall shear stress) due to control is therefore reflected in the turbulent contribution and the discussion can be limited to $c_{f,t \bkappa}$.
\Cref{fig:turbulentDragContrib} shows the relative turbulent drag contribution $c_{f,t\bkappa} / c_{f,t}$ for an uncontrolled reference configuration and the three example controlled flows.
The wavenumber range is again limited to the subset shown earlier in \cref{fig:vSpatialSpectraAct}.
Large streamwise scales with spanwise wavenumbers $0 \lesssim k_{z} \lesssim 10$ contribute most to the friction coefficient in all configurations, but the range of active $\bkappa$ varies.
In the uncontrolled (\cref{fig:turbulentDragContribRef}) and drag-reduced flow N25 (\cref{fig:turbulentDragContribN025pi}) the active scales are contained in a compact region, which is very similar in both cases.
In contrast, more and smaller scales contribute to the friction coefficient of the drag-increased flows (\cref{fig:turbulentDragContribN075pi,fig:turbulentDragContribP05pi}) and it is apparent that spanwise-constant scales ($k_{z} = 0$) only substantially contribute to the wall shear stress in case P50.
We also note that statistically significant scale-by-scale drag contributions are positive across all flow configurations (the small negative numbers are most likely statistical errors due to the finite sample size).
The Reynolds stresses therefore act as a decelerating force in the controlled cases as well.

\Cref{sec:scaleByScaleDragChange} will quantify the difference in $c_{f,t\bkappa}$ between configurations, with a focus on scales that contribute significantly to the wall-shear stress and drag.
In this context, a scale will be considered drag-contributing if it accounts for at least for $0.05\%$ to the turbulent friction coefficient.
These drag-contributing scales lie close in spectral space and are contained within a closed contour, which is shown by the black contour line with label $\Gamma$ in \cref{fig:turbulentDragContrib}.
The contours of the example configurations share similarities, but their exact shape and extent varies from case to case.
The scales within $\Gamma$ account for at least $80\%$ of the turbulent wall shear stress, with the exact amount indicated above each figure.
We note that the threshold choice at $0.05\%$ is somewhat arbitrary, but does not affect the conclusions: the contour (and total drag captured) would uniformly increase or decrease across all configurations if a different threshold were chosen.

\subsection{Scale-by-scale drag change}  \label{sec:scaleByScaleDragChange}
\Cref{eq:defDr} defined the drag change between two configurations as the normalized difference between the corresponding friction coefficients.
The decomposition of $c_{f}$ according to \cref{eq:frictionCoeffFIK} naturally induces a scale-by-scale drag change metric
\begin{equation}
    \Delta c_{f,t \bkappa} = (c_{f, t \bkappa})_0 - (c_{f, t \bkappa})_c
\end{equation}
To highlight the effect of control, we consider a normalized form of this metric
\begin{equation}
    \Delta \xi_{t \bkappa} = \frac{\Delta c_{f,t \bkappa}}{\lvert (c_{f, t \bkappa})_0 \rvert}
\end{equation}
where the absolute value is needed to preserve the sign change in case of a negative scale-by-scale contribution in the uncontrolled flow.
The scale-by-scale drag change can be used to objectively define scale suppression and amplification:
A scale is said to be suppressed under control if $\Delta \xi_{t \bkappa} > 0$ (i.e. the turbulent drag contribution decreases) and amplified if $\Delta \xi_{t \bkappa} < 0$ (turbulent drag contribution increases).
However, $\Delta \xi_{t, \bkappa}$ is ill-defined if a scale contributes close to zero wall-shear stress in the canonical flow, as is typically the case for large $k_{x}$ and $k_{z}$.
This issue can be prevented if the scale-by-scale turbulent drag contribution is only evaluated for scales inside the contours $\Gamma$ introduced in \cref{fig:turbulentDragContrib}.
In order to focus on scales that are active under control, we choose the contours of the controlled flows.
It is important to note that the controlled contours are somewhat larger than the canonical one, but the additional scales are typically also energetic enough in the uncontrolled flow to make the metric well-defined.

\begin{figure}
    \centering
    \begin{subfigure}[t]{0.45\linewidth}
        \caption{$\phaseShift = - 3 \pi / 4$}
        \includegraphics[width=\linewidth]{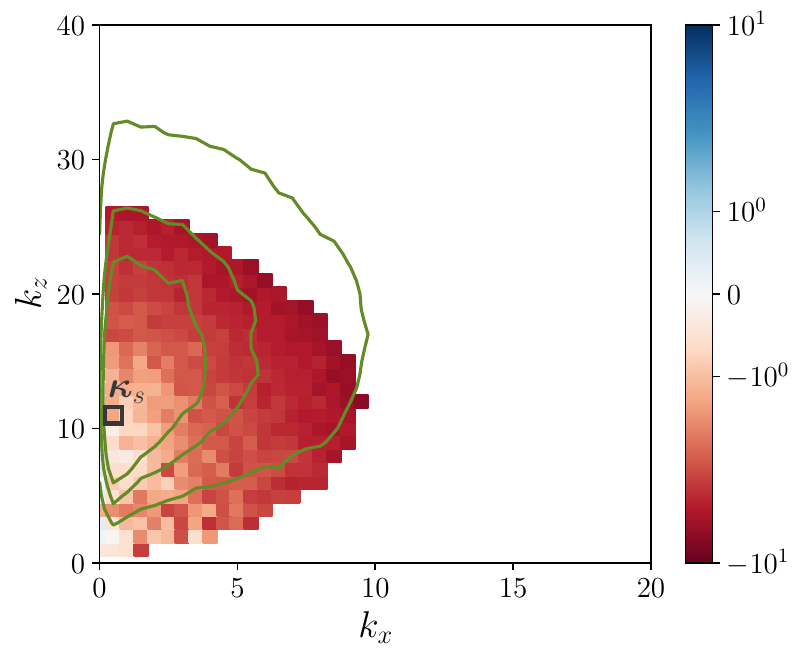}
        \label{fig:relDragChangeN075pi}
    \end{subfigure}
    \begin{subfigure}[t]{0.45\linewidth}
        \caption{$\phaseShift = -3 \pi / 4$}
        \includegraphics[width=\linewidth]{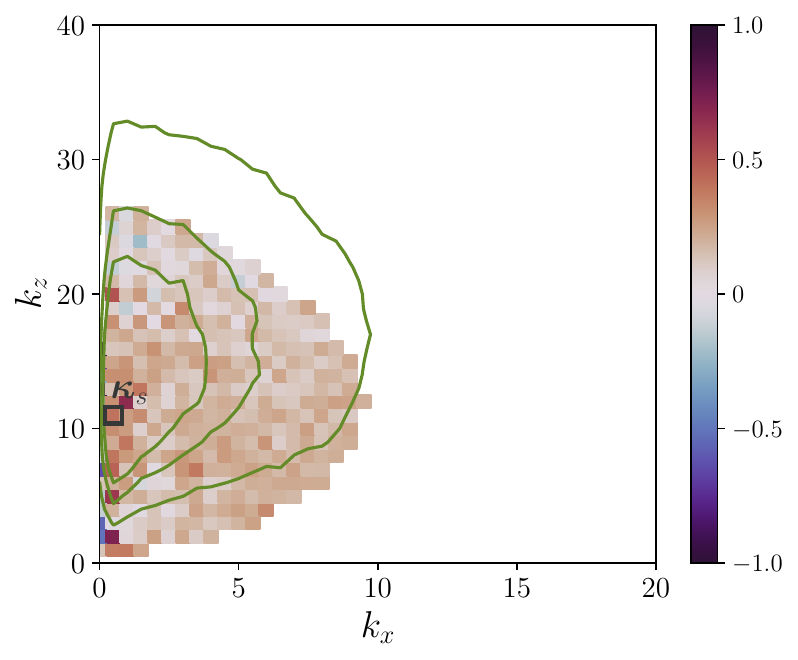}
        \label{fig:phaseDiffN075pi}
    \end{subfigure}
    \begin{subfigure}[t]{0.45\linewidth}
        \caption{$\phaseShift = - \pi / 4$}
        \includegraphics[width=\linewidth]{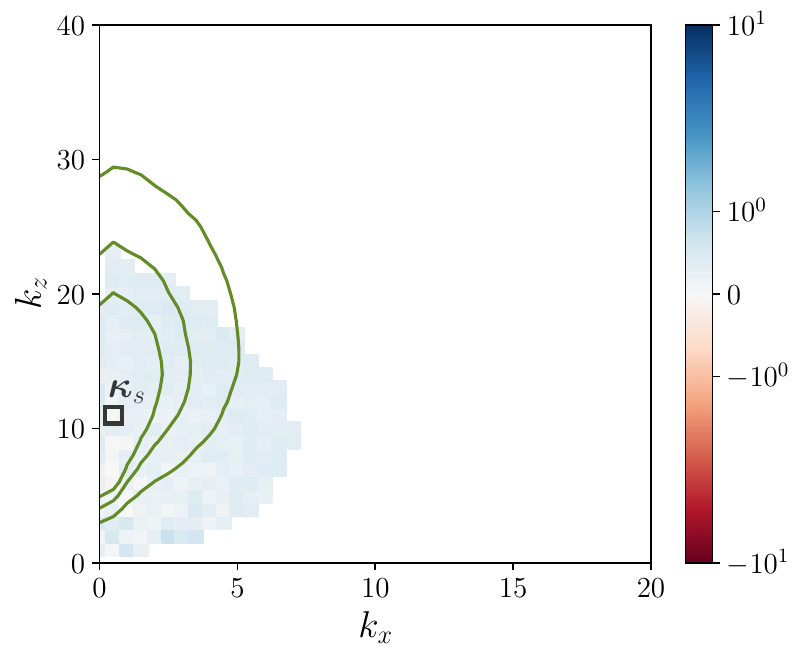}
        \label{fig:relDragChangeN025pi}
    \end{subfigure}
    \begin{subfigure}[t]{0.45\linewidth}
        \caption{$\phaseShift = - \pi / 4$}
        \includegraphics[width=\linewidth]{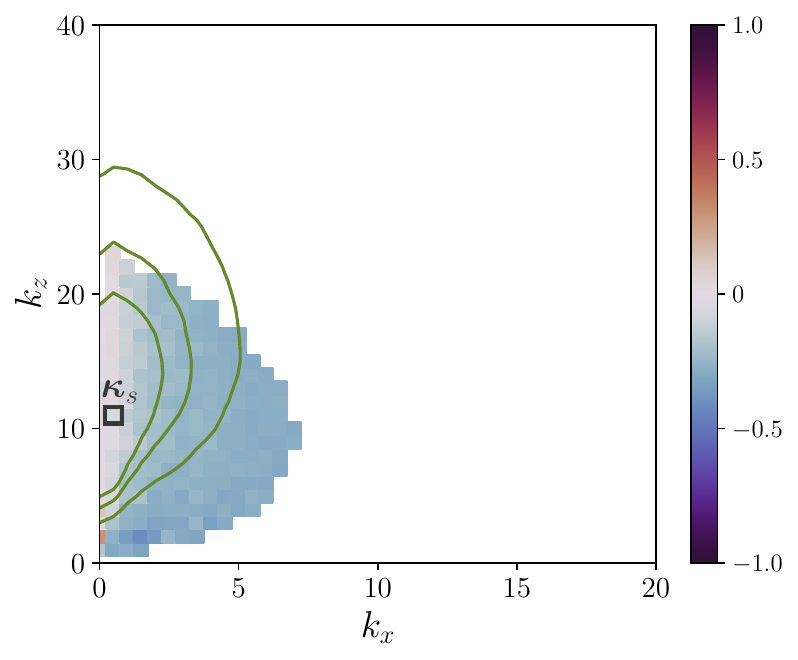}
        \label{fig:phaseDiffN025pi}
    \end{subfigure}
    \begin{subfigure}[t]{0.46\linewidth}
        \caption{$\phaseShift = + \pi / 2$}
        \includegraphics[width=\linewidth]{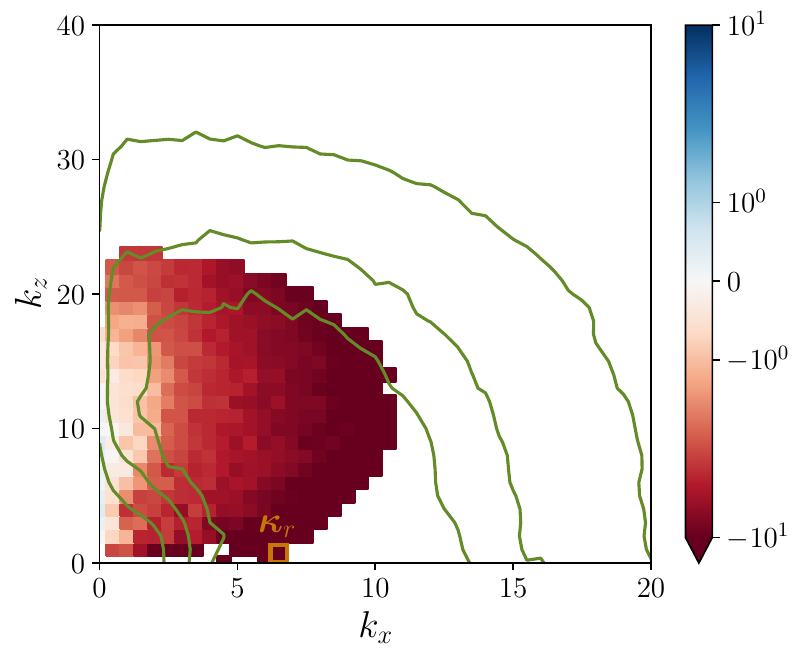}
        \label{fig:relDragChangeP05pi}
    \end{subfigure}
    \begin{subfigure}[t]{0.45\linewidth}
        \caption{$\phaseShift = + \pi / 2$}
        \includegraphics[width=\linewidth]{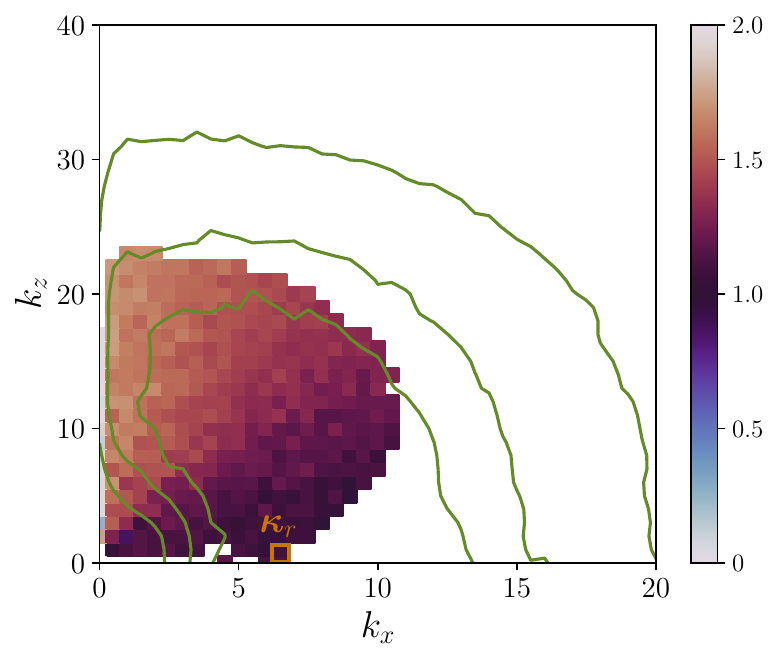}
        \label{fig:phaseDiffP05pi}
    \end{subfigure}
    \caption{Left column: relative change in turbulent drag contribution $\Delta \xi_{t \bkappa}$, right column: phase difference spectra $\mean{\Delta \theta}_{\bkappa}/\pi$. Both quantities are only shown for the scales within the black contour $\Gamma$ defined in \cref{fig:turbulentDragContrib}. The green contour lines are the same as in \cref{fig:vSpatialSpectraAct} and label $\Phi_{vv}^+ / \max(\Phi_{vv}^+) = (0.15, 0.3, 0.45)$.}
    \label{fig:spectraDrPhase}
\end{figure}

The left column of \cref{fig:spectraDrPhase} shows the change in turbulent drag contribution within the contours $\Gamma$.
In addition, the green lines outline the spectral content of the actuation signal.
The overlap of the two spectral regions indicates that the transpiration acts on the drag-carrying scales and the scale-restricted control experiments (\cref{sec:scaleFamilies,sec:scaleRestrictedControllers}) suggest that the amplification and suppression of each scale $\bkappa$ results mostly from transpiration and $\phaseShift$ at that $\bkappa$.
All active scales of the drag-reducing configuration N25 (\cref{fig:relDragChangeN025pi}) are suppressed due to the control.
The relative drag reduction is uniformly less than one, which indicates that the controller reduces or decorrelates the Reynolds stress at individual $\bkappa$, but does not alter the phase relation between $\four{u}_{\bkappa}$ and $\four{v}_{\bkappa}$ enough to make the Reynolds stress an accelerating force.
In contrast, the active scales are amplified in the drag-increased flows N75 and P50 (\cref{fig:relDragChangeN075pi,fig:relDragChangeP05pi}).
The strength of the amplification is not uniform in spectral space and can be a multiple of the contribution in the canonical flow (note that the colorbar is logarithmic and saturated).
The amplification is especially pronounced in case P50, where the actuation energizes the roller scales.
A comparison between configurations N75 and N25 further shows that control inputs with similar spectral signature can have opposite effects on the controlled scales.
The only difference between the two configurations is $\phaseShift$, which provides further evidence that the phase shift (streamwise position relative to the background flow) of the transpiration determines the control effect.

\Cref{fig:spectraDrPhase} also provides insights into the difficulty of drag reduction:
The present transpiration reduces or at best annihilates the scale-by-scale Reynolds stress contribution (all $\Delta \xi_{t \bkappa} \leq 1$) which limits the achievable drag reduction.
On the other hand, there is no apparent limit on how much control can amplify a scale in drag-increasing configurations and the amplification factors observed in \cref{fig:relDragChangeN075pi,fig:relDragChangeP05pi} are typically much larger than one.
It is therefore in general much more difficult to reduce drag than to increase it.
In addition, the effect of control is often non-uniform in spectral space (see e.g. \cref{fig:relDragChangeN075pi}) and a highly amplified scale can cancel the effect of many weakly suppressed scales.

\subsection{Connection to pv phase}  \label{sec:connectionDragPvPhase}
The results thus far indicate that the scale response to control and the phase difference between $p$ and $v$ at the wall depend on $\phaseShift$.
We explore next if scale amplification and suppression coincide with specific $\mean{\Delta \theta}_{\bkappa}$.
The right column of \cref{fig:spectraDrPhase} displays the mean phase difference of $\mean{\Delta \theta}_{\bkappa}$, which was shown earlier in \cref{fig:spectraPhaseAllScales}, but is now restricted to the contours $\Gamma$.
A comparison between the left and right column indicates that there is a correlation between scale amplification and velocity-pressure phase difference.
In the drag-reducing configuration N25, the streak scales are suppressed and transpiration is in-phase with the wall pressure.
In contrast, the drag-increasing configuration P50 amplifies the roller scales and transpiration is out-of-phase with the wall pressure.
The data for configuration N75 are less conclusive: the streak scales are amplified and the phase difference between transpiration and wall pressure is positive, but different from $\pi$ and shows considerable scatter.
It is also important to keep in mind that the variance in the phase difference was largest in case N75, which reduces the confidence in the exact value of the circular mean.
One may speculate that the phase relation between $\four{v}_{\bkappa}$ and $\four{p}_{\bkappa}$ at the streak scales varies smoothly with $\phaseShift$.{}
The two are in-phase at $\phaseShift = - \pi / 4$ when the drag is maximally reduced and out-of-phase at $\phaseShift = \pm \pi$ when the drag increase is largest and the DNS diverges.
From this perspective, configuration N75 could represent an intermediate state of positive phase shift as the scales approach the out-of-phase relation.

These observations suggest that the phase difference between transpiration and wall pressure is an important parameter of the problem.
Drag reduction with streak scales coincides with actuation and wall pressure being in-phase, while drag increase with roller scales occurs if the two are out-of-phase.
It is therefore plausible, that the wall pressure encodes the state of the background flow, relative to which the transpiration is shifted by means of $\phaseShift$.
The non-local domain of dependence (\cref{sec:domainDependencePwPhase}) is further consistent with an interpretation of the wall pressure as an averaged background flow state.

\section{Summary and implications for control and modeling}
\label{sec:summary}
This section summarizes our results and puts them in broader perspective.
A brief overview of the varying-phase opposition control results is given in \cref{sec:resultSummary}.
\Cref{sec:connectionTailoredSurf} relates these results to the tailored surfaces and discusses implications for control and drag reduction.
Possible avenues to extend the present results to higher Reynolds numbers are presented in \cref{sec:relevanceHigherRe}.

\subsection{Summary of results}  \label{sec:resultSummary}
In this study, we conduced direct numerical simulation of low Reynolds number turbulent channel flows with wall transpiration and investigated the wall pressure field and its phase relation to the transpiration.
The analysis considered the Fourier domain in the wall-parallel $\{x, z\}$ directions and focused on distinct spatial scales identified by the wavenumber combination $\bkappa = (k_x, k_z)^{\intercal}$.
The wall-normal and temporal coordinate were retained in physical space so that the Fourier coefficients depended on $y$ and $t$.

The transpiration was generated by the varying-phase opposition control scheme, which measures the Fourier coefficient of the wall-normal velocity $\four{v}_{\bkappa}(y_d, t)$ at a distance $y_d$ from the wall, and generates a wall transpiration $\four{v}_{\bkappa}(y_w, t) = - \four{A}_d \four{v}_{\bkappa}(y_d, t)$.
The complex-valued controller gain $\four{A}_d$ allowed shifting the phase of the transpiration ($\angle \four{v}_{\bkappa}(y_w, t)$) relative to the sensor measurement and the background flow.
The gain was carefully defined so that $\phaseShift$ corresponded to a shift in the streamwise direction.
The sensor location was fixed at $y_d^+ = 15$ and $\phaseShift$ was varied over the entire parameter range $[-\pi, +\pi]$.
Our analysis focused on three example controllers: N75 ($\phaseShift = -3 \pi / 4$), N25 ($\phaseShift = -\pi / 4$) and P50 ($\phaseShift = +\pi / 2$).

The structure and magnitude of the wall-pressure as well as the relative importance of its components (fast, slow and Stokes) changed with $\phaseShift$ of the transpiration.
Notably, the transpiration added an inertial forcing term to the Stokes pressure and made its magnitude comparable to the fast and slow pressure.
The Stokes pressure is therefore not necessarily negligible in flows with wall transpiration.

We further analyzed the structure of the pressure Green's function solution.
Specifically, we showed how the phase of the wall pressure is established and formulated conditions for robust phase relations between $\four{p}_{\bkappa}$ and $\four{v}_{\bkappa}$.
The phase relation between transpiration and fast and slow pressure was determined by a weighted wall-normal integral and a robust relation resulted if the integrand was correlated to $\four{v}_{\bkappa}(y_w)$ over the wall-normal extent where the weight function was non-zero.
This was typically the case for the fast pressure but not for the slow pressure.
The difference is due to the spectral dependence of the two source terms: the source term of the fast pressure depends on $\four{v}_{\bkappa}$ at the same wavenumber, which is constrained by the varying-phase opposition control scheme.
On the other hand, the nonlinear source term of the slow pressure depends on all wavenumbers that are triadically consistent with $\bkappa$ and is typically not strongly correlated to $\four{v}_{\bkappa}(y_w)$.
The phase relation of the Stokes pressure depended on the temporal frequency content of the transpiration dynamics at the wall.
A robust phase relation resulted for approximately sparse frequency content, which was observed when the eigenspectrum of the linearized Navier-Stokes operator had an amplified eigenvalue.

Most of our subsequent analysis focused on the phase difference (relative streamwise arrangement) between wall pressure and transpiration, which changed as a function of $\phaseShift$.
We considered scales at distinct $\bkappa$, which belonged to one of two families:
streak scales ($k_x$ small and $k_x < k_z$, associated with the near-wall cycle at the current $\reTau$) and roller scales ($k_z$ small and $k_x > k_z$).
A significant contribution from the roller scales was only observed in configuration P50.
For such positive $\phaseShift$, wall pressure and transpiration were approximately out-of-phase ($\mean{\Delta \theta}_{\bkappa} \approx \pi$) and control amplified the roller scales, which lead to a large drag increase.
Earlier studies showed that the drag-increase coincides with the appearance of an amplified eigenvalue in the linearized Navier-Stokes operator and the presence of spanwise rollers in the DNS \citep{ToedtliYuMcKeon2020origindragincrease}.
On the other hand, the streak scales were most active for $\phaseShift \lesssim 0$.
For configuration N25, wall pressure and transpiration at the streak scales were in-phase ($\mean{\Delta \theta}_{\bkappa} \approx 0$) and transpiration suppressed the near-wall cycle and resulted in maximum drag reduction.
The phase difference for more negative $\phaseShift$ was positive, with large variance at each $\bkappa$ and significant variation across scales.
In this case, transpiration amplified the near-wall cycle and lead to a drag increase.

To our knowledge, this study presented the first direct evidence that drag reduction correlates with $\Delta \theta_{\bkappa} \approx 0$ in the full non-linear flow, confirming the hypothesis of \citet{XuRempferLumley2003Turbulenceovercompliant}.
Similarly, the present study provided evidence that $\Delta \theta_{\bkappa} \approx \pm \pi$ correlates with drag increase and the appearance of spanwise rollers, which to our knowledge had not been appreciated thus far.
It is possible that $\Delta \theta_{\bkappa} \approx \pm \pi$ can induce other flow structures as well (we hypothesized that this phase relation could also be established at $\phaseShift = \pm \pi$) but this aspect remained speculative.
At its core, the present study made a case that the phase relation between wall transpiration and pressure deserves more attention in the flow control community.

\subsection{Connection to tailored surfaces and implications for modeling and control}  \label{sec:connectionTailoredSurf}
We next connect the varying-phase opposition control results to tailored surfaces.
As described in the introduction, the occurrence of spanwise rollers is well-documented for a range of tailored surfaces including porous \citep{JimenezUhlmannPinelliEtAl2001Turbulentshearflow} and permeable walls \citep[for example,][]{BreugemBoersmaUittenbogaard2006influencewallpermeability,EfstathiouLuhar2018Meanturbulencestatistics,Gomez-de-SeguraGarcia-Mayoral2019Turbulentdragreduction}
and riblets past the viscous breakdown \citep{Garcia-MayoralJimenez2011Hydrodynamicstabilitybreakdown}.
The appearance of spanwise rollers coincides with a significant drag increase and, in many cases, with the presence of an amplified eigenvalue in the spectrum of the linearized Navier-Stokes operator.
A number of pressure-driven tailored surfaces further impose a phase relation between pressure and transpiration at the wall.
For example, a porous wall can be modeled as transpiration boundary condition \citep{JimenezUhlmannPinelliEtAl2001Turbulentshearflow}
\begin{equation}  \label{eq:porousWall}
    v(x, y_w , z, t) = -\beta p^{\prime}(x, y_w, z, t)
\end{equation}
where $\beta \geq 0$ is a material parameter and $p^{\prime}$ denotes the pressure fluctuations above the background pressure and mean gradients.
By construction, wall pressure and transpiration are out-of-phase in this configuration.

The flow response to varying-phase opposition control with roller scales and positive $\phaseShift$ leads to the same observations:
the transpiration induces spanwise rollers near the wall and causes a large drag increase; the onset of the rollers and their wall-normal structure can be predicted from the eigenspectrum of the linearized Navier-Stokes equations; and wall transpiration and pressure are out-of-phase.
The close correspondence of flow responses suggest that transpiration with the roller scales and $\phaseShift > 0$ is ``dynamically equivalent'' to this class of tailored surfaces.
We use the term dynamical equivalence to indicate that the overall flow response is analogous, even if finer details are different.
For example, \cref{fig:wallPressureP05pi} suggests that wall pressure and transpiration of configuration P50 are not simply related by a constant, which is different from the porous wall boundary condition (\ref{eq:porousWall}).
To further test the dynamical equivalence, it would be interesting to verify if $\four{v}_{\bkappa}$ and $\four{p}_{\bkappa}$ are out-of-phase when other configurations, such as riblets, generate spanwise rollers.
As mentioned in the introduction, the notion of transpiration in this case would apply to a plain inside the flow domain and the suitable definition of this plain is clearly challenging.

Tailored surfaces also interact with the near-wall cycle.
For example, riblets in the viscous regime suppress the near-wall cycle and reduce drag, while riblets with spacings past the optimum can amplify these scales and increase the drag \citep{ChavarinLuhar2020ResolventAnalysisTurbulent}.
Varying-phase opposition control with streak scales, which are associated with the near-wall cycle at the present $\reTau$, leads to the same observations:
A transpiration with small negative $\phaseShift$ suppresses the near-wall cycle and leads to drag reduction, analogous to riblets in the viscous regime.
On the other hand, transpiration with large negative or positive $\phaseShift$ amplifies the near-wall cycle and leads to a drag increase, analogous to the flow response to riblets past the viscous regime.
The close correspondence of flow responses again suggests that  transpiration with streak scales is dynamically equivalent to tailored surfaces interacting with the near-wall cycle.
It would be interesting to verify also in this case if wall pressure and transpiration are in-phase when riblets suppress the near-wall cycle.
A further analogy between the two control methods occurs for transpiration with $\phaseShift > 0$ and the riblets past the viscous breakdown: both configurations amplify the near-wall cycle and induce spanwise rollers, but the amplification of the rollers is much stronger so that they dominate the flow response and the drag increase (see \cref{fig:dragReduction,fig:actSpectrumP05pi} and \citet{Garcia-MayoralJimenez2011Hydrodynamicstabilitybreakdown}).

It is also important to highlight that varying-phase opposition control and at least a subset of tailored surfaces, namely the pressure-driven ones, have a complimentary perspective on how the $pv$ phase relation is established:
the varying-phase opposition control scheme applies a transpiration with a phase shift $\phaseShift$, but does not explicitly depend on the pressure.
Once the transpiration is generated, the pressure field responds and the phase relation is established according to the discussion in \cref{sec:wallPressure,sec:pvPhaseWall}.
In contrast, a pressure-driven tailored surface generates a transpiration that explicitly depends on the pressure and the phase difference between $p$ and $v$ may be imposed by the boundary condition (see e.g. \cref{eq:porousWall}).
In this case, the flow has to adjust the wall-normal profiles of the pressure source term and the spectral content of the Stokes pressure boundary condition to satisfy this phase relation (see \cref{sec:pvPhaseDiffStrucExample}).
This is in some sense a complementary problem to our configuration, but the tools we presented can still be used to analyze the flow changes in detail.

The observed dynamical equivalence suggests that the two scale families together with the phase difference $\Delta \theta_{\bkappa}$ (or, equivalently, $\phaseShift$) build essential building blocks for flows over tailored surfaces.
From a physical perspective, this suggests that the two flow families can provide a unifying framework to analyze tailored surfaces and understand their dynamical effect.
From a modeling perspective, this provides further evidence that tailored surfaces can be represented by velocity boundary conditions and suggests that transpiration has to be a key component of any such dynamical representation.

We conclude this section with a few general remarks regarding control:
First, the results of this study suggest that drag reduction with passive pressure-driven tailored surfaces are difficult to achieve.
The spectral content of the wall pressure and transpiration are reasonably aligned when the drag is increased (see \cref{fig:wallPressureP05pi}), but not when the drag is reduced (\cref{fig:wallPressureN025pi}).
The pressure therefore likely activates the ``wrong'' scales if the tailored surface is scale-agnostic.
Moreover, a passive configuration induces an out-of-phase relation between $v$ and $p$, which based on the results presented herein does not lead to scale suppression and drag reduction.
In order to reduce drag, a passive pressure-driven tailored surface would therefore need to impose 1) a scale-dependent response, and 2) a different phase relation between $p$ and $v$.
This could perhaps be achieved through an array of suitably placed Helmholtz resonators (see, e.g., \citet{DacomeSiebolsBaars2023InnerscaledHelmholtz})
or a still to be developed meta-material.
A permeable substrate with complex-valued transpiration coefficient would be another interesting test case.
\citet{Gomez-de-SeguraGarcia-Mayoral2019Turbulentdragreduction} considered real-valued transpiration coefficients, which induced an out-of-phase relation between $p$ and $v$ at the substrate boundary.
Drag reduction was observed for small values of the wall-normal permeability, which prevented the formation of spanwise rollers.
It would be interesting to see if the spanwise rollers can alternatively be suppressed with complex-valued transpiration coefficients, which induce a different phase relation between $p$ and $v$.

The second comment regards the pressure as a sensor measurement for control:
the present study indicates that the wall pressure signal encodes the background flow and can inform a suitable transpiration.
However, this pressure information has to come from the controlled flow itself, since the pressure changes significantly in the presence of transpiration.
For example, the Stokes pressure is negligible in the canonical case but can become leading-order in the presence of transpiration (see \cref{sec:wallPressure}).
Control policies or flow estimators based on data of canonical wall-bounded flows are therefore likely not very effective once control is applied.
An additional challenge with basing the control input on the wall pressure is the direct coupling between $v$ and $p$, especially through the Stokes pressure.
This is different from, say, opposition control, where the sensor and actuator are separated by a viscous flow region that reduces direct coupling.

\subsection{Relevance at higher Reynolds numbers}  \label{sec:relevanceHigherRe}
The results presented in this study were for a low Reynolds number turbulent channel flow, which corresponded to $(\reTau)_0 \approx 180$ in the absence of transpiration.
Most practical applications operate at significantly higher Reynolds number.
An obvious question is whether the scale families and the $pv$ phase relation are relevant at higher $\reGeneral$ as well.
We offer a few different perspectives on this question.

The first perspective is a phenomenological one: the two scale families were associated with distinct flow structures.
The persistence of these flow structures at higher Reynolds numbers may be taken an indication that the scale families remain relevant.
Following this reasoning, results in the literature suggest that the streak and roller scales and their phase relations remain relevant for at least another decade in $\reTau$.
For example, an experimental investigation by \citet{EfstathiouLuhar2018Meanturbulencestatistics} found evidence for the existence of spanwise rollers over a porous wall at $\reTau \approx 1700$.
Similarly, the DNS study of \citet{DengHuangXu2016Origineffectivenessdegradation} suggests that classical opposition control can reduce drag until at least $\reTau \approx 1000$.
Classical opposition control is closely related to $\phaseShift = 0$, at which the streak scales lead to drag reduction.
A resolvent study on compliant walls by \citet{LuharSharmaMcKeon2015frameworkstudyingeffect} at $\reTau \approx 2000$ further suggests that the (fast) wall pressure and transpiration have to be in-phase to suppress the near-wall cycle, consistent with the results presented earlier.
However, the $\reGeneral$-scaling of the spatial scales remains an open question.

As the Reynolds number increases, large-scale flow structures in the logarithmic region become more energetic and their imprint reaches all the way down to the wall \citep[see, e.g.,][]{HutchinsMarusic2007Largescaleinfluences}.
A second question is how the results apply to these structures.
Substantiated answers for this question are sparse, because effective flow manipulation schemes for large-scale structures are still rare.
There has been recent success in suppressing flow structures in the logarithmic region by means of spanwise traveling waves \citep{MarusicChandranRouhiEtAl2021energyefficientpathway}, but it remains inconclusive if transpiration-based control can suppress these structures as well \citep{AbbassiBaarsHutchinsEtAl2017Skinfrictiondrag}.
Opposition control in the logarithmic region has so far lead to drag increase and the generation of spanwise rollers for some configurations \citep{GusevaJimenez2022Linearinstabilityresonance}, which  suggests that the roller scales represent a general sensitivity of the flow to a relaxed no-throughflow condition.
On the other hand, a different phase relation between $v$ and $p$ may be needed to suppress the large-scale motions in the logarithmic layer.
A recent resolvent study suggests that wall pressure and transpiration no longer need to be in-phase to optimally suppress these motions \citep{LuharSharmaMcKeon2015frameworkstudyingeffect}, but the caveat is again that this analysis only captures the fast pressure component.
A future study may consider a modified version of the present analysis, with $\four{A}_d$ according to \cref{eq:controllerGainAllScales} and sensors located in the logarithmic region, to clarify if and how transpiration-based control can increase or decrease the drag and what the $vp$ phase relation in each case is.

Finally, the present work offers a fresh perspective on flow control that can also be applied at higher Reynolds numbers.
The phase difference $\Delta \theta_{\bkappa}$ is a wall quantity and its strong correlation with drag reduction suggests that flow control can perhaps be understood from the perspective of the wall.
Control based on the wall pressure naturally selects scales that have an imprint at the wall and removes the somewhat unsatisfying necessity of identifying target structures inside the flow, at the expense of not being able to target scales that cannot be sensed at the wall.
In addition, a wall transpiration with a specific phase relation to the wall pressure systematically alters the scale-by-scale vorticity flux, which represents the dynamical effect of the wall and is perhaps the most natural control target at any Reynolds number.
The wall vorticity flux may therefore deserve more attention in the flow control community, as envisioned early on by \citet{Koumoutsakos1999Vorticityfluxcontrol}.

\bigskip
\noindent
\textbf{Acknowledgements.} The authors are grateful to Drs. Nicholas Hutchins, Daniel Chung, Dan Meiron, Adrian Lozano-Duran and Petros Koumoutsakos for many fruitful discussions.

\medskip
\noindent
\textbf{Funding.} The authors acknowledge funding from the Office of Naval Research through ONR Grant Number N00014-17-1-3022.
The National Center for Atmospheric Research is sponsored by the National Science Foundation of the United States.

\medskip
\noindent
\textbf{Declaration of interest.} The authors report no conflict of interest.

\medskip
\noindent
\textbf{Author ORCIDs.}\\
Simon Toedtli: \url{https://orcid.org/0000-0001-9371-9572}\\
Anthony Leonard: N/A \\
Beverley McKeon: \url{https://orcid.org/0000-0003-4220-1583}

\appendix
\section{Scale-restricted controllers}  \label{sec:scaleRestrictedControllers}
A scale-restricted controller is formally defined by \cref{eq:varyingPhaseOc} with a different controller gain
\begin{equation}
    \four{A}_{d, \bkappa} = \begin{cases}
        \four{A}_d & \text{if } \bkappa \in \bkappa_c \\
        0 & \text{else}
    \end{cases}
\end{equation}
In the above expression, $\bkappa_c$ denotes the set of controlled scales, for which the controller gain is identical to \cref{eq:controllerGainAllScales}.
All scales not contained in $\bkappa_c$ remain uncontrolled, which means $\four{v}_{\bkappa}(y_w) = 0$ for these scales.
The scale-restricted controllers thus generate a wall transpiration with very limited spectral content.
Two different choices of $\bkappa_c$ will be considered subsequently:
Single-scale controllers, for which $\bkappa_c = \lbrace (k_{xc}, k_{zc})^{\intercal} \rbrace$ contains a single spatial scale, and single-wavenumber controllers, for which $\bkappa_c = \lbrace (k_{xc}, k_z)^{\intercal} | \, k_z = 0, \pm 1, \dots \rbrace$ contains all spanwise wavenumbers at a given $k_{xc}$.

The control scales in $\bkappa_c$ are an additional parameter of the scale-restricted scheme, besides the sensor location $y_d$ and phase shift $\phaseShift$.
We consider a scale-restricted controller for each scale family (streaks and rollers) in isolation, which results in two different sets $\bkappa_c$.
The scale selection within each family is somewhat arbitrary, but ideally the controlled scales are energetic enough so that their effect is detectable in the flow in a mean sense.
Suitable energetic scales can thus be identified from the actuation spectra shown in \cref{fig:vSpatialSpectraAct}.

The roller scales are particularly active at large positive phase shifts, and $\bkappa_c$ can thus be selected from the transpiration spectrum of configuration P50 in \cref{fig:actSpectrumP05pi}.
The requirement of a strong flow response suggests selecting the most energetic scale, which was previously identified at $\bkappa_r = (k_x = 6.5, k_z = 0)^{\intercal}$.
A suitable scale-restricted controller for the roller family is therefore given by
\begin{equation}
    \bkappa_c = \left \lbrace \bkappa_r = (k_{xr} = 6.5, k_{zr} = 0)^{\intercal} \right \rbrace
\end{equation}
and the red squares in \cref{fig:dragReduction} show the drag change due to control with this $\bkappa_c$.
The flow response to this single-scale controller is strong enough to be statistically significant for positive phase shifts, but no drag change is observed for negative phase shifts.
We confirmed that increasing the control input by incorporating additional roller scales does not lead to a drag change either, which suggests that the roller scales do not contribute to the flow response in this parameter regime.
This interpretation is also consistent with the transpiration spectra at negative $\phaseShift$ (\cref{fig:actSpectrumN075pi,fig:actSpectrumN025pi}).

The streak scales on the other hand are most active for $\phaseShift \leq 0$ and suitable control scales $\bkappa_c$ can be selected from the transpiration spectrum of configuration N25 (\cref{fig:actSpectrumN025pi}).
The most energetic scale of this configuration occurs at $\bkappa_s = (k_x = 0.5, k_z = 11)^{\intercal}$, but a controller based on this single scale generates a weak flow response that is not statistically significant.
The control input can be amplified by adding more spatial scales.
A suitable scale-restricted controller for the streak scales is therefore given by the single-wavenumber controller
\begin{equation}  \label{eq:streamwiseElongatedSw}
    \bkappa_c = \left \lbrace \lbrace \bkappa \rbrace_{k_{xs}} = (k_{xs} = 0.5, k_z)^{\intercal} | \, k_z = 0, \pm 1, \dots \right \rbrace
\end{equation}
and the blue triangles in \cref{fig:dragReduction} show the drag change due to control with this $\bkappa_c$.
\Cref{eq:streamwiseElongatedSw} is preferred over other possible choices of $\bkappa_c$, because it contains many energetic scales (see \cref{fig:actSpectrumN025pi}) and ensures that no three scales are triadically consistent, since all have the same $k_{xs}$.
It is also important to point out that even though the single-wavenumber controller acts on all spanwise wavenumbers, they only correspond to approximately 1\% of all the Fourier modes.

\bibliographystyle{jfm}
\bibliography{research_bibliography_sst}

\end{document}